%% file: final_v2.tex
\documentclass[referee]{raa}

\usepackage{graphicx,times}
\usepackage{natbib}
\usepackage{amssymb,amsmath}
\usepackage{multirow}
\usepackage{rotating}
\usepackage{pdflscape} 
\usepackage{enumitem}
\usepackage{booktabs} 
\usepackage{float}
\bibpunct{(}{)}{;}{a}{}{,}

\usepackage[pagebackref=true]{hyperref}

\begin{document}

\title{Not All Who Wander Are Lost: Early Excess Demographics in the Volume-limited ZTF DR2 SN Ia Sample}

 \volnopage{ {\bf 20XX} Vol.\ {\bf X} No. {\bf XX}, 000--000}
   \setcounter{page}{1}

   \author{C\'esar Rojas-Bravo
   \inst{1,2,*}
\and Ning-Chen Sun
   \inst{1,2,3,*}
\and Mathew Smith
\inst{4}
\and Chun Chen
   \inst{5,6, 7}
\and Xiaohan Chen
   \inst{2,8}
\and Zexi Niu
   \inst{1,2}
\and Anyu Wang
   \inst{1,2}
\and Zi-Yang Wang
   \inst{1,2}
\and Yi-Han Zhao
   \inst{1,2}
\and Jifeng Liu
   \inst{1,2,3,9}
\footnotetext{$*$Corresponding authors:  {cesar.rojasbravo@ucas.ac.cn, sunnc@ucas.ac.cn}}
}

\institute{School of Astronomy and Space Science, University of Chinese Academy of Sciences, Beijing 100049, People's Republic of China\\
       \and
       National Astronomical Observatories, Chinese Academy of Sciences, Beijing 100101, China\\
       \and
       Institute for Frontiers in Astronomy and Astrophysics, Beijing Normal University, Beijing 102206, People's Republic of China\\
       \and
       Department of Physics, Lancaster University, Lancaster LA1 4YB, UK \\
       \and
       School of Physics and Astronomy, Sun Yat-sen University, Zhuhai 519082, China\\
       \and
       CSST Science Center for the Guangdong-Hong Kong-Macau Greater Bay Area, Sun Yat-sen University, Zhuhai 519082, China\\
       \and
       Dipartimento di Fisica, Universit\`a di Napoli ``Federico II'', Compl. Univ. di Monte S. Angelo, Via Cinthia, I-80126, Napoli, Italy\\
       \and
       School of Physics and Astronomy, China West Normal University, Nanchong 637002, China\\
       \and New Cornerstone Science Laboratory, National Astronomical Observatories, Chinese Academy of Sciences, Beijing 100012, People’s Republic of China \\     
\vs \no
   {\small Received 20XX Month Day; accepted 20XX Month Day}
}

\abstract{Early-time flux excesses in Type Ia supernovae (SNe~Ia) offer a unique insight into their progenitor systems and explosion mechanisms. Although individual early-excess events and larger searches have been reported, demographic studies remain limited by sample size. We present a systematic search for early-time excess emission in a volume-limited sample ($z<0.06$) of SNe~Ia based on the Zwicky Transient Facility Data Release 2 (ZTF DR2). Using ZTF $g$- and $r$-band light curves, we identify candidates showing early-excesses shortly after the explosion time, and we apply conservative coverage and quality requirements to build reliable ``excess'' and ``no-excess'' bump and no-bump catalogs.
From an initial sample of 1547 SNe~Ia, our final catalogs contain 42 early-excess and 110 no-excess events. We compare the two populations using SN and host environment parameters from ZTF DR2 and quantify the differences using two-sample statistical tests. We find the strongest differences are in SN light-curve properties: early-excess events have larger SALT2 stretch $x_1$ ($7.91\sigma$) and larger $r$-band secondary-maximum flux $\mathcal{F}_{r_2}$ ($6.25\sigma$), while differences in SALT2 color $c$ are weak ($0.57\sigma$). Early-excess events also favor bluer $(g-z)_{\rm local}$ ($3.41\sigma$) and lower $\log_{10} (M_*/M_\odot)_{\rm local}$ ($2.73\sigma$). Our results connect early excesses with SNe~Ia diversity, and motivate further analyses of upcoming larger samples.
\keywords{supernovae: general --- transients: supernovae
}
}

\authorrunning{C. Rojas-Bravo et al.}
\titlerunning{Demographics of Early-time Excess Emission in ZTF DR2 SNe Ia}
   \maketitle

%
\section{Introduction} \label{sec:introduction}

Type Ia supernovae (SNe Ia)—the thermonuclear explosions of white dwarf stars \citep{Hoyle1960}— are essential tools in modern cosmology \citep{Brout2022,descosmo}, enabling precision measurements of the accelerating expansion of the Universe \citep{Phillips1993,Riess1998,Perlmutter1999}. As we now have thousands of these supernovae observed across a wide range of distances, it has become critical to better account for astrophysical sources of error—such as the diversity in progenitor and explosion physics \citep{Wang2012,Maoz2014,Jha2019,Ruiter2025}, and correlations with host-galaxy environments \citep{Sullivan2006,Kelly2010,Sullivan2010,Lampeitl2010,Childress2014,Jones2018,Rose2019,Rigault2020,Brout2021,Thorp2022,Kelsey2023,Wiseman2023,Padilla2025,Ramaiya2025,Wojtak2025,ztfdr2c,ztfdr2x1,ztfdr2gal,Toy2025}. However, these fundamental questions still remain unclear despite extensive efforts (e.g., \citet{Dimitriadis2019, Tucker2020, Liu2023}).

In recent years, numerous SNe~Ia have shown early flux above a simple power-law rise (e.g., SN 2017cbv \citep{Hosseinzadeh2017}, SN 2018oh \citep{Dimitriadis2019,Li2019,Shappee2019}, SN 2023bee \citep{Wang2024}; for a full compilation, see \citet{Ye2024}). Several physical scenarios have been proposed to explain early-time excess emission in SNe~Ia (see \citet{Liu2023} for a recent review), such as (i) ejecta-companion interaction, in which the collision of the SN ejecta with a non-degenerate companion produces a viewing-angle-dependent shock-heated emission, strongest at blue/UV wavelengths \citep{Kasen2010}; (ii) shallow radioactive heating (e.g., surface $^{56}$Ni or outward mixing of radioactive material), which can increase the earliest optical flux with different early colors \citep[e.g.,][]{Piro2014,Piro2016}; (iii) circumstellar material (CSM) interaction, adding early luminosity with color/timescale signatures depending on the CSM mass, opacity and other parameters \citep{Piro2016,Moriya2023}; and (iv) double detonations in sub-$M_{\rm Ch}$ progenitors (detonation of the Carbon-Oxygen white dwarf caused by a previous detonation of a thin surface layer of helium) \citep{Polin2019,Piro2025}. Therefore, the early excess emission serves as a powerful tool to investigate the progenitor system and explosion mechanism of SNe~Ia.

However, distinguishing these scenarios in individual events is challenging, since their early-time signatures can overlap, and many predicted observables are viewing-angle-dependent or affected by survey limitations such as cadence and filter coverage \citep[e.g.,][]{Kasen2010,deckers2022,Dhawan2022,Ye2024}. To this end, it is essential to build large, homogeneous samples to measure how often early-time excesses occur and what they correlate with. Recent analyses have systematically tried to identify early-time bumps in modern datasets \citep[e.g.,][]{Olling2015,Jiang2018,Fausnaugh2021,Burke2022,deckers2022,Fausnaugh2023,Hoogendam2024,Ye2024,Ni2025,Wu2025,Wu2026}. For example, \citet{Ye2024} performed a search for early excesses in the Zwicky Transient Facility (ZTF) first data release DR \citep{Yao2019,Dhawan2022} using multi-band power-law and Gaussian fits. Their study allowed quantitative classification of bumps and their possible correlation with some light curves and host-galaxy parameters, such as the global host mass.

Building on these investigations and recognizing the need for large, homogeneous samples to measure how often early-time excesses occur and their correlations, we here present, to our knowledge, the largest sample to date of SNe~Ia with early-time excesses. A central aspect of our analysis is the use of the ZTF DR2 volume-limited subset \citep{ztfdr2overview} to minimize selection biases and support demographic comparisons using uniformly derived light-curve and host-galaxy parameters provided in ZTF DR2. 

This paper is organized as follows. In Section \ref{sec:data} we describe the ZTF DR2 sample, the volume-limited selection, and the SN/host parameters used in this work. In Section \ref{sec:methodology}, we present our early-time fitting framework, model comparison strategy, and detection/robustness criteria used to define the bump and no-bump catalogs. In Section \ref{sec:results}, we present the resulting bump and no-bump catalogs and report demographic comparisons between the two populations. In Section \ref{sec:discussion}, we discuss the implications, limitations, and connections to prior work. Finally, we conclude in  Section \ref{sec:conclusions} with a summary of this work and thoughts on future directions.

\section{Data} \label{sec:data}

\subsection{ZTF DR2} \label{subsec:ztfdr2}

The data used in this work were obtained from the ZTF Data Release 2 (ZTF DR2), as described fully in \cite{ztfdr2overview}.
This data release consists of $gri-$band light curves of 3628 spectroscopically-classified SNe Ia up to redshift $z\lesssim0.2$, observed between March 2018 and December 2020 by the ZTF camera on the P48 Schmidt telescope at Mount Palomar Observatory, California, with a $5\sigma$ depth of $20.5$ mag in 30 s exposures. The light curves have different sampling rates, with overall cadences of 2.9, 2.5, and 6 days for the $g$, $r$, and $i$ bands, respectively. For further details on the ZTF DR2 and their photometry, we refer the reader to \cite{ztfdr2overview}. The ZTF DR2 data were downloaded directly from the dedicated ZTF SN Ia DR2 server\footnote{\url{https://ztfcosmo.in2p3.fr/download/data}}.

ZTF DR2 also includes light-curve parameters and host-galaxy properties. All the ZTF DR2light-curve and host-galaxy  parameters used in this work are described in Table \ref{tab:param_inventory} and references therein, but we provide a summary here:
\begin{itemize}

\item ZTF DR2 light curves are fit with SALT2 \citep{Guy07,Guy10,Betoule14} model retrained by \cite{Taylor2021} in the time window $[-10,+40]$ days (rest-frame) \citep{ztfdr2overview,rigaultfits}. The parameters used here are the stretch $x_1$ parameter (how fast the SN brightens and fades after peak, associated with the amount of $^{56}$Ni produced in the explosion \citep{Phillips1993,Kasen2007}), the color $c$ (associated with intrinsic/extrinsic reddening), and time of maximum ($t_0$). These parameters are described in detail in \cite{ztfdr2overview,ztfdr2x1,ztfdr2c}.

\item Redshifts come from public galaxy catalogs, galaxy emission lines, and \texttt{SNID} \citep{Blondin07} SN Ia template matching, and are fully described in \cite{ztfdr2overview}. All redshifts are in the heliocentric frame.

\item The derivation of host-galaxy properties is described in detail in \cite{ztfdr2overview}: the host galaxy for each SN was identified with the directional light radius method (DLR, \cite{Sullivan2006,Gupta2016}  ); SNe with no host galaxy within  DLR $<7$ were removed by the ZTF team. Global photometry was obtained with \texttt{HostPhot} \citep{hostphot} on public PS1 DR2 images. In contrast, local photometry was obtained within a 2 kpc radius. Stellar masses and rest-frame colors were obtained with the galaxy spectral code \texttt{PEGASE2} \citep{pegase2}. Throughout the paper, we refer to the local and global stellar masses as $\log_{10} (M_*/M_\odot)_{\rm local}$ and $\log_{10} (M_*/M_\odot)_{\rm global}$, and to the local and global rest-frame colors as $(g-z)_{\rm local}$ and $(g-z)_{\rm global}$, respectively.

\item The absolute peak magnitudes, decline rates, increase rates, and $g-r$ colors are provided in \cite{ztfdr2diversity}. To compare these parameters for SNe Ia at different redshifts, we use their K-corrected values \citep{Kim96, Nugent02}.

\item The time of the secondary maximum in the $r$-band and the integrated flux under the secondary maximum in this same band are described in detail in \cite{ztf2max}. The secondary maximum, observed $\approx 2$–$3$ weeks after peak flux in the optical $i$ band and at near-infrared wavelengths \citep{Elias1981}, with a shoulder in the optical $r$ band, provides information on the $^{56}$Ni mass \citep{Dhawan2016}, the total ejecta mass \citep{Papadogiannakis2019}, and can be used to standardize SNe~Ia \citep{Nobili2005,Shariff2016}.  \citet{ztf2max} analyze this secondary maximum in the ZTF $r$-band (caused by the ionization transition of iron-group elements from double to single ionized below $\sim 7000$ K) by quantifying its strength: they normalize the flux to the $r$-band peak, integrate the flux from $+15$ to $+40$ days after peak, and divide by the interval length ($25$ days). They find that the strength of the $r$-band secondary maximum correlates with $x_1$ in a slightly non‑linear way and that it is a better standardization parameter than $x_1$. 

\item Finally, the galaxy type and galaxy component in which the SN lies are all provided by \cite{ztfdr2gal}.

\end{itemize}

\begin{landscape}
\begin{table*}
\centering
\begin{minipage}{\textwidth}
\caption{Parameters used in this work, their units, sources, and availability in the ZTF DR2 SN~Ia volume-limited sample ($z<0.06$).\label{tab:param_inventory}}
\end{minipage}
\setlength{\tabcolsep}{3pt}
\small
\begin{tabular}{l l l l l r r}
\hline\noalign{\smallskip}
Parameter & ZTF DR2 name & Unit & Source & Comment & Total $N_{\rm with}$ & Total $N_{\rm without}$ \\
\hline\noalign{\smallskip}
$z$ & \texttt{z} & dimensionless & \cite{ztfdr2overview} & redshift, heliocentric & 1547 & 0 \\
$x_1$ & \texttt{x\_1} & dimensionless & \cite{ztfdr2overview} & SALT2 stretch & 1532 & 15 \\
$c$ & \texttt{c} & dimensionless & \cite{ztfdr2overview} & SALT2 color & 1532 & 15 \\
$E(B-V)_{\mathrm{MW}}$ & \texttt{mwebv} & dimensionless & \cite{ztfdr2overview} & Assumed Milky Way dust with $R_V=3.1$ & 1547 & 0 \\
$\log_{10} (M_*/M_\odot)_{\rm local}$ & \texttt{localmass} & dex & \cite{ztfdr2overview} & Local host stellar mass, SED-fit; 2 kpc radius & 1515 & 32 \\
$\log_{10} (M_*/M_\odot)_{\rm global}$ & \texttt{globalmass} & dex & \cite{ztfdr2overview} & Global host stellar mass, SED-fit & 1528 & 19 \\
$(g-z)_{\rm local}$ & \texttt{localrestframe\_gz} & mag & \cite{ztfdr2overview} & Local host color $(g-z)$ * & 1515 & 32 \\
$(g-z)_{\rm global}$ & \texttt{globalrestframe\_gz} & mag & \cite{ztfdr2overview} & Global host color $(g-z)$ * & 1528 & 19 \\
$d_\mathrm{DLR}$ & \texttt{d\_dlr} & dimensionless & \cite{ztfdr2overview} & normalized direct light distance & 1528 & 19 \\
$M_{\rm{g,max}}$ & \texttt{abs\_mag\_peak\_g\_ks} & mag & \cite{ztfdr2diversity} & Absolute peak brightness, $g$-band * & 525 & 1022 \\
$M_{\rm{r,max}}$ & \texttt{abs\_mag\_peak\_r\_ks} & mag & \cite{ztfdr2diversity} & Absolute peak brightness, $r$-band * & 525 & 1022 \\
$\Delta \rm{m}_{{\rm15,g}}$ & \texttt{dm\_p15\_g\_ks} & mag & \cite{ztfdr2diversity} & Decline rate in 15 days from peak, $g$-band * & 525 & 1022 \\
$\Delta \rm{m}_{{\rm15,r}}$ & \texttt{dm\_p15\_r\_ks} & mag & \cite{ztfdr2diversity} & Decline rate in 15 days from peak, $r$-band * & 525 & 1022 \\
$\Delta \rm{m}_{{\rm-5,g}}$ & \texttt{dm\_m5\_g\_ks} & mag & \cite{ztfdr2diversity} & Increase rate from 5 days before peak, $g$-band * & 525 & 1022 \\
$\Delta \rm{m}_{{\rm-5,r}}$ & \texttt{dm\_m5\_r\_ks} & mag & \cite{ztfdr2diversity} & Increase rate from 5 days before peak, $r$-band * & 525 & 1022 \\
$(\rm{g-r})_{\rm{max}}$ & \texttt{gr\_peak\_ks} & mag & \cite{ztfdr2diversity} & $g-r$ colour at peak * & 525 & 1022 \\
$(\rm{g-r})_{+15}$ & \texttt{gr\_p15\_ks} & mag & \cite{ztfdr2diversity} & $g-r$ colour at $+15$ days * & 525 & 1022 \\
$(\rm{g-r})_{-5}$ & \texttt{gr\_m5\_ks} & mag & \cite{ztfdr2diversity} & $g-r$ colour at $-5$ days * & 525 & 1022 \\
$t^{r-g}_{\rm{}max}$ & \texttt{tr\_tg\_ks} & days & \cite{ztfdr2diversity} & Time difference of peak $r$-band and peak $g$-band * & 525 & 1022 \\
$t_{\rm{2,r}}$ & \texttt{t2r} & days & \cite{ztf2max} & Timing of secondary maximum in the $r$-band * & 795 & 752 \\
\multirow{2}{*}{$\mathcal{F}_{r_2}$} & \multirow{2}{*}{\texttt{F2\_r}} & \multirow{2}{*}{dimensionless} & \multirow{2}{*}{\cite{ztf2max}} & Integrated flux under the secondary maximum in $r$-band, & \multirow{2}{*}{876} & \multirow{2}{*}{671} \\
& & & & normalised to peak & & \\
Galaxy type & \texttt{gal\_type} & --- & \cite{ztfdr2gal} & best fit model to the galaxy & 337 & 1210 \\
SN-component & \texttt{SN\_component} & --- & \cite{ztfdr2gal} & which galaxy component the SN lies in & 337 & 1210 \\
\noalign{\smallskip}\hline
\end{tabular}
\tablecomments{\textwidth}{Parameters marked with * have been K-corrected with SALT2 in the respective paper.}
\end{table*}
\end{landscape}

\subsection{Redshift Cut}
\label{subsec:sample}

We adopt the redshift cut ($z<0.06$) used by the ZTF collaboration to define their ``volume-limited'' SN~Ia sample \citep{ztfdr2sims}. For non-peculiar SNe~Ia, \citet{ztfdr2sims} show that the sample up to $z=0.06$ is free from Malmquist bias (and other non-random selection effects), such that the observed distributions of light-curve parameters are representative of the underlying population. After applying this cut, our parent sample contains 1547 SNe~Ia. 37 non-peculiar SNe~Ia in ZTF DR2 with $z<0.015$ are not included in our sample since their light curves are not released by the ZTF team.

\subsection{Cosmology Population Cuts}
\label{subsec:cosmocuts}

Before we fit the data and identify SNe with and without potential early excesses, we apply several cuts aligning with previous SN~Ia cosmology studies \citep{foundation,pplus,descosmo} and similar early excess studies \citep{deckers2022,Ye2024}: 

\begin{enumerate}[label=\roman*)]
    \item SN is spectroscopically normal or 1991T-like (according to the ZTF DR2 classification \citep{ztfdr2overview,ztfdr2diversity}).
    \item Milky Way dust with $R_V=3.1$ is $\le 0.2$.
    \item $x_1$ uncertainty: $\sigma_{x_1} \le 1.0$.
    \item $c$ uncertainty: $\sigma_c \le 0.1$.
    \item Uncertainty of time of peak brightness is $\le0.8$.
    \item $-3.0 \le x_1 \le 3.0$.
    \item $-0.3 \le c \le 0.3$.
    \item SALT2 fit probability $\texttt{fitprob}$ is $ > 10^{-7}$.
    
\end{enumerate}

\section{Methodology} \label{sec:methodology}

\subsection{Photometry quality cuts}

After downloading the full ZTF DR2 sample and applying redshift and cosmology population cuts (see sections \ref{subsec:sample} and \ref{subsec:cosmocuts}), we clean the light curves according to the ZTF DR2 quality control guidelines (private communication), by removing datapoints with unphysical flux errors, poor PSF fits, processing failures, clouds, invalid baseline corrections and images in secondary grid (bitmask flags 1, 2, 4, 8, 16, and 32 in the ZTF DR2 \texttt{flag} column). We also remove all $i$-band data due to its lower cadence, and focus solely on the $g$ and $r$ bands.

\begin{figure}[!htbp]
\centering
\includegraphics[width=0.95\linewidth]{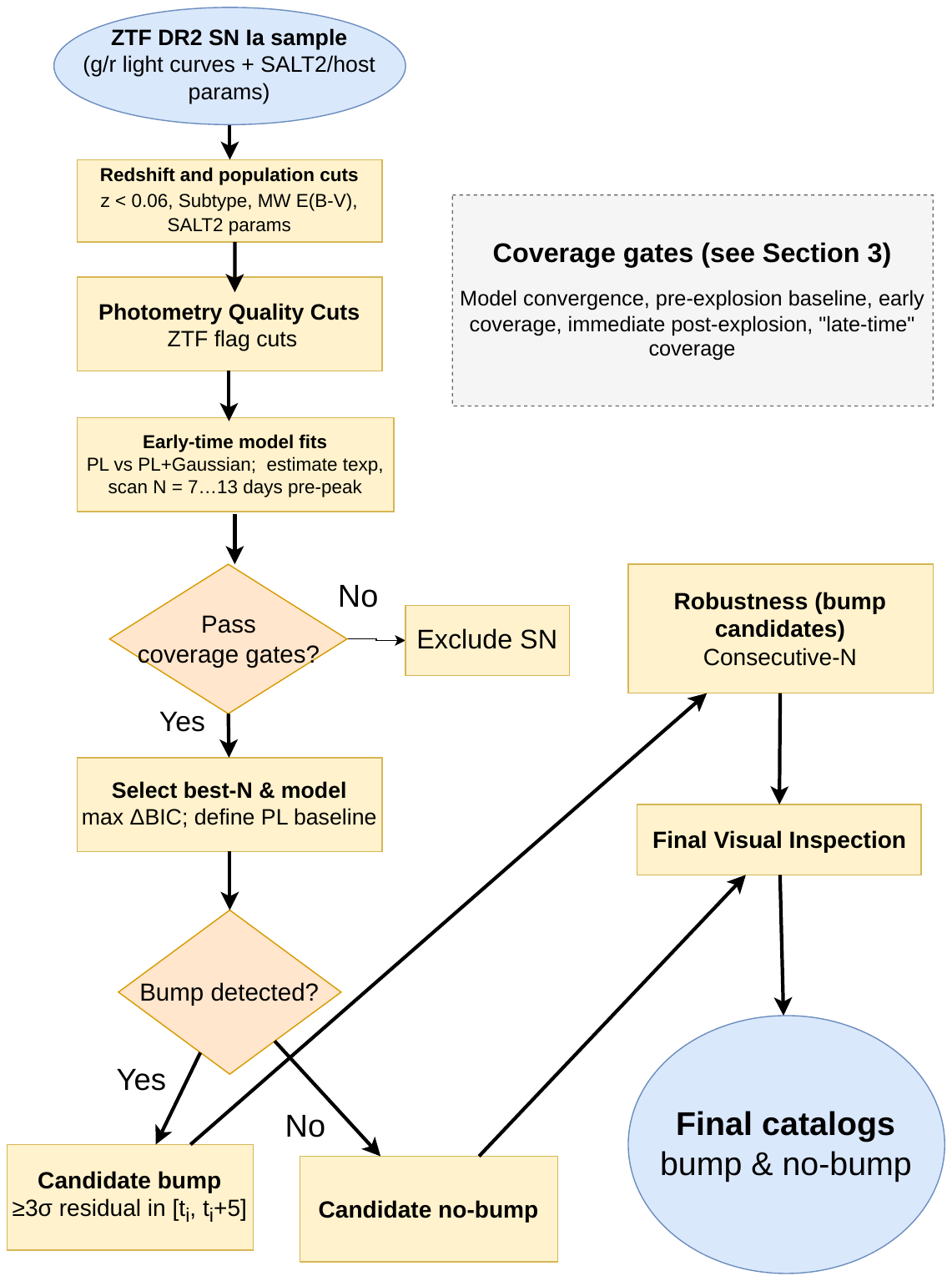}
\caption{Overview of our pipeline to create the final bump and no-bump catalogs from the ZTF DR2 SN~Ia sample. After initial photometry quality control, a $z<0.06$ redshift cut and population cuts, we fit early-time $g$ and $r$ light curves with a power-law (PL) and a PL+Gaussian (PL+G) model while scanning the pre-peak fit window ($N=7$--13~days; Section~\ref{sec:methodology}). After selecting objects which pass the coverage gates (Section~\ref{subsec:steps}), we select the best-$N$ and preferred model via the maximum $\Delta$BIC and then classify candidates based on statistically significant early-time excesses. We apply a robustness (consecutive-$N$) cut to the bump candidates. Finally, after a final visual inspection, we produce the final bump and no-bump catalogs. The exact order of the final cuts and number of SNe after each cut is given in Table~\ref{tab:waterfall_full}.}
\label{fig:flowchart}
\end{figure}

\subsection{Fitting Models} \label{subsec:fittingmodels}

Following \cite{Ye2024}, we perform joint two-band ($g,r$) fits to describe the early-time emission and quantify statistically significant excesses (``bumps'') in the ZTF DR2 light curves. For each band, we model the early-time flux as a rising power law (PL), 
\begin{equation}
f_{\rm PL}(t) = B + A\,(t - t_{\rm i})^{\alpha}\,H(t - t_{\rm i}),
\label{eq:fpl}
\end{equation}
where $t_{\rm i}$ is the explosion time parameter, $A$ is the power-law scale, $\alpha$ is the power-law index, $B$ is the baseline offset, and $H$ is the Heaviside step function.

In addition, we also fit a power-law + Gaussian model (PL+G) to the early light curves
\begin{equation}
f_{\rm PLG}(t) = f_{\rm PL}(t) + \frac{C}{\sigma\sqrt{2\pi}}
\exp\!\left[-\frac{(t-\mu)^2}{2\sigma^2}\right],
\label{eq:fplg}
\end{equation}
where $C\ge 0$ is the Gaussian normalization factor, and $\sigma$ and $\mu$ are the standard deviation and location of the Gaussian component. We allow $\mu$ to vary between $t_{\rm i}$ and $t_{\rm i} +5$ days, since we only search for bumps in the early light curve phases. We share $t_{\rm i}$, $\sigma$ and $\mu$ across both bands, but keep $A$, $B$, $C$, and $\alpha$ independent across each band.

We adopt the early-excess fitting parameter constraints with the same values presented in Table 2 in \cite{Ye2024}, with two differences to account for differences in our datasets: first, since the data in \cite{Ye2024} have different zeropoints than the ZTF DR2 dataset, we scale their $A$ and $C$ constraints by a factor of $33.62$ (the mean flux ratio between overlapping SNe in both samples); second, unlike the ZTF DR2 data, the data in \cite{Ye2024} are not baseline-corrected, so we tighten our bounds on $B$ to $[-0.5,0.5]$, without scaling. We present the final constraint values in Table~\ref{tab:early_excess_constraints}.

\begin{table}[htbp]
\centering
\caption{Constraints on Early-excess Fitting Parameters}
\label{tab:early_excess_constraints}
\begin{tabular}{l r}
\toprule
Parameter & Constraint \\
\midrule
$B_{g,r}$         & $[-0.5,\,0.5]$ \\
$A_{g,r}$         & $[0.0,\,3.362\times10^{4}]$ \\
$t_{\rm{i}}$     & $[t_0-45,\,t_0-10]$ \\
$\alpha_{g,r}$    & $[1.0,\,3.0]$ \\
$\mu$             & $[t_{\mathrm{i}},\,t_{\mathrm{i}}+5]$ \\
$\sigma$          & $[0.5,\,4.0]$ \\
$C_{g,r}$         & $[0.0,\,1.681\times10^{4}]$ \\
\bottomrule
\end{tabular}
\end{table}

\subsection{Fitting Steps} \label{subsec:steps}

We fit each $g,r$ light curve using all individual photometric points 
with both models from $t_0-40$ to $t_0-N$, where $t_0$ is the time of maximum light (provided in the ZTF DR2 metadata tables) and $N$ ranges from $7$ to $13$ days in integer values (larger $N$ means fewer datapoints are used in the fit), choosing these values of $N$ to have the most data possible but keeping the PL/PL+G approximations valid. 

We apply ``coverage gates'' to the ZTF DR2 light curves to ensure reliable model fits. After several iterations, we determined cuts that guarantee good coverage but are not so aggressive as to exclude potential early excesses:

\begin{enumerate}
    \item \textbf{Model convergence:} Both PL and PL+G fits converge successfully.
    \item \textbf{Pre-explosion baseline:} At least 1 epoch per band between $t_{\rm i} - 5$ days and $t_{\rm i}-0.5$ days.
    \item \textbf{Immediate post-explosion:} At least 1 epoch per band between $t_{\rm i}-0.5$ days and $t_{\rm i}+2.5$ days.
    \item \textbf{Early coverage:} At least 2 epochs per band between $t_{\rm i}-0.5$ days and $t_{\rm i}+5.5$ days. 

    \item \textbf{Late-time coverage:} At least 1 epoch per band between $t_{\rm i}+5$ and the end of the fit.
\end{enumerate}

If a SN does not meet any of these requirements, it is discarded from our analysis.

Next, for each $N$, we compute the Bayesian Information Criterion (BIC) \citep{kass1995bayes}

\begin{equation}
\mathrm{BIC} = \chi^2 + k \ln n,
\label{eq:bic}
\end{equation}
where $k$ is the number of free parameters in the model, $n$ the number of datapoints, and 

\begin{equation}
\chi^2 = \sum_{i=1}^{n} \frac{[f_i - f_{\mathrm{model}}(t_i)]^2}{\sigma_i^2}
\label{eq:chisq}
\end{equation}
is the chi-squared statistic, with $f_i$ and $\sigma_i$ are the observed flux and uncertainty at time $t_i$.

We define the statistic

\begin{equation}
\Delta\mathrm{BIC}(N) = \mathrm{BIC}_{\mathrm{PL}}(N) - \mathrm{BIC}_{\mathrm{PLG}}(N),
\label{eq:dbic}
\end{equation}
where positive values favor the PL+G model. For each SN~Ia, we choose the “best-$N$” as the integer which maximizes $\Delta{\rm BIC}(N)$. We adopt PL+G as the preferred model if $\Delta\mathrm{BIC}_{\mathrm{max}} > 2$, following \citet{kass1995bayes}.

\subsection{Early-excess Flagging} \label{subsec:flags}

We flag potential bumps by first defining the baseline model (different to ZTF DR2 light curve baseline where flux = 0): if the preferred model is PL+G, the baseline model is the power-law component of the PL+G fit; if the preferred model is PL, the baseline model is the power-law fit itself. Then, we bin observations by night (using inverse-variance-weighted averages) to calculate residuals relative to this baseline model. We then define the potential bumps as $3\sigma$ residuals relative to this baseline model, from $t_{\rm i}$ to $t_{\rm i}+5$ days. We also allow a one-day tolerance window before $t_{\rm i}$ for visualization purposes to account for minor uncertainties in the determination of the explosion time, but we classify the SN as having a bump only if at least one $3\sigma$ excess epoch occurs at $t\geq t_{\rm i}$.

In a small number of objects, statistically significant excess residuals appear at a negative phase (i.e., before the fitted $t_{\rm i}$). We point out that this is not unreasonable since in the PL+G model $t_{\rm i}$ is not the explosion epoch but simply the start of the PL component. Because $t_{\rm i}$ is inferred from model fits and is degenerate with the early-time rise parameters \citep{Ye2024}, we interpret these negative-phase bumps as reflecting the uncertainty in the fitted explosion time—especially for light curves with sparse sampling around first light—rather than as evidence for physical emission prior to explosion. 

\subsection{Final cuts} 
\label{subsec:samplecuts}

Finally, we add two further quality cuts based on iterative testing to ensure fit reliability:

\begin{enumerate}[label=\roman*)]

    \item \textbf{Consecutive-N test:}  For each bump candidate (best-$N$), we test $N \pm 1$ and $N \pm 2$ (clipped to $7 \le N \le 13$).  A bump is considered robust (and kept in the final sample) if flagged in $\ge 2$ different $N$. We exclude cases in which different $N$ values share the same data points during the fitting process (due to sparse sampling). This test is introduced to ensure robustness against single data points that could influence the detection of the early excess.
    
    \item \textbf{Visual inspection:} We visually inspect the data quality and fits of all bump and no-bump candidates. We remove four bump candidates (ZTF18abtfvsk, ZTF20aaeuxqk, ZTF20aayxldg, ZTF20acucbek) and two no-bump candidates (ZTF18abbikrz and ZTF18abjgyyr) due to the following reasons (plots for the removed bump objects are shown in Figure \ref{fig:removed}):
\end{enumerate}


\begin{enumerate}
    \item ZTF18abtfvsk: located $4.8''$ from the center of its host galaxy, the light curve shows possible galaxy contamination/nuclear variability, which may create spurious fits and is therefore excluded from the analysis.
     \item ZTF20aaeuxqk: shows a very clear early excess epoch (only) in $r$-band at MJD 58850.49, a $\sim3\sigma$ single-epoch spike relative to its 1-day neighbors, suggesting a spurious point, even though it passes the ZTF catalog flag/quality mask cuts. We therefore remove it from the analysis.
     \item ZTF20aayxldg: some epochs present large scatter, which probably causes spurious fits. We therefore exclude this SN from the analysis.
     \item ZTF20acucbek: this SN only shows an apparent early bump in $g$-band (no corresponding valid epoch in $r$-band); the remaining early-time data are consistent with a smooth rise. We therefore exclude it as a non-robust bump candidate.
     \item ZTF18abbikrz and ZTF18abjgyyr (no early excesses): both SNe show pre-explosion data well-below the baseline, suggesting imperfect subtraction or background systematics. Therefore, we remove them from our final no-bump sample.
\end{enumerate}
A full schematic of our methodology is presented in Figure \ref{fig:flowchart}. We provide example light curves of bump and no-bump candidates in Figure \ref{fig:examplelcs}, and light curve of all the bump candidates in Appendix \ref{sec:app_tables}. The light curves of all our 110 no-bump candidates are available upon request.

\begin{figure}[!htbp]
    \centering
    \begin{subfigure}{0.49\textwidth}
        \centering
        \includegraphics[width=\linewidth]{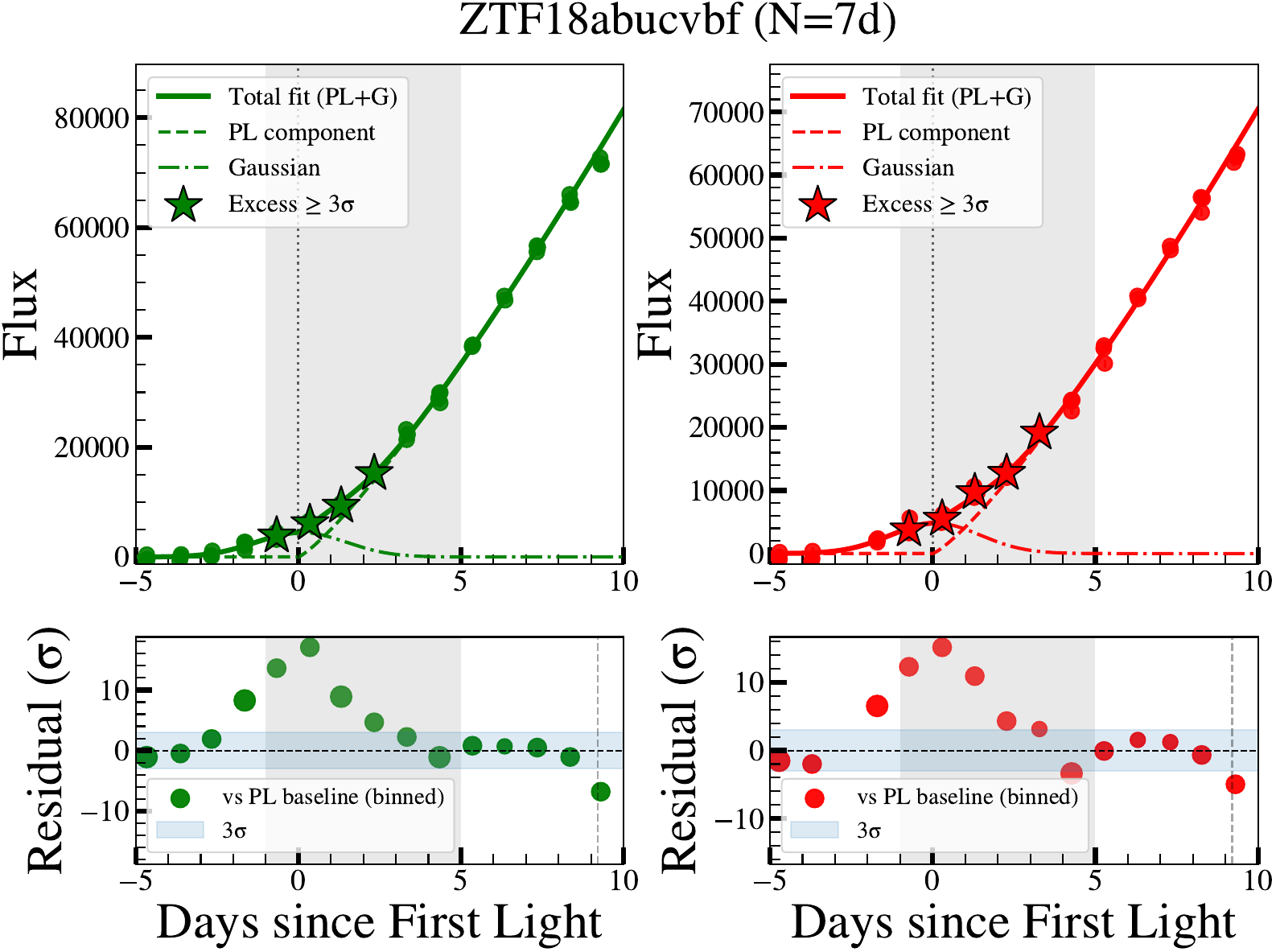}
        \caption{}
        \label{subfig:a}
    \end{subfigure}
    \hfill
    \begin{subfigure}{0.49\textwidth}
        \centering
        \includegraphics[width=\linewidth]{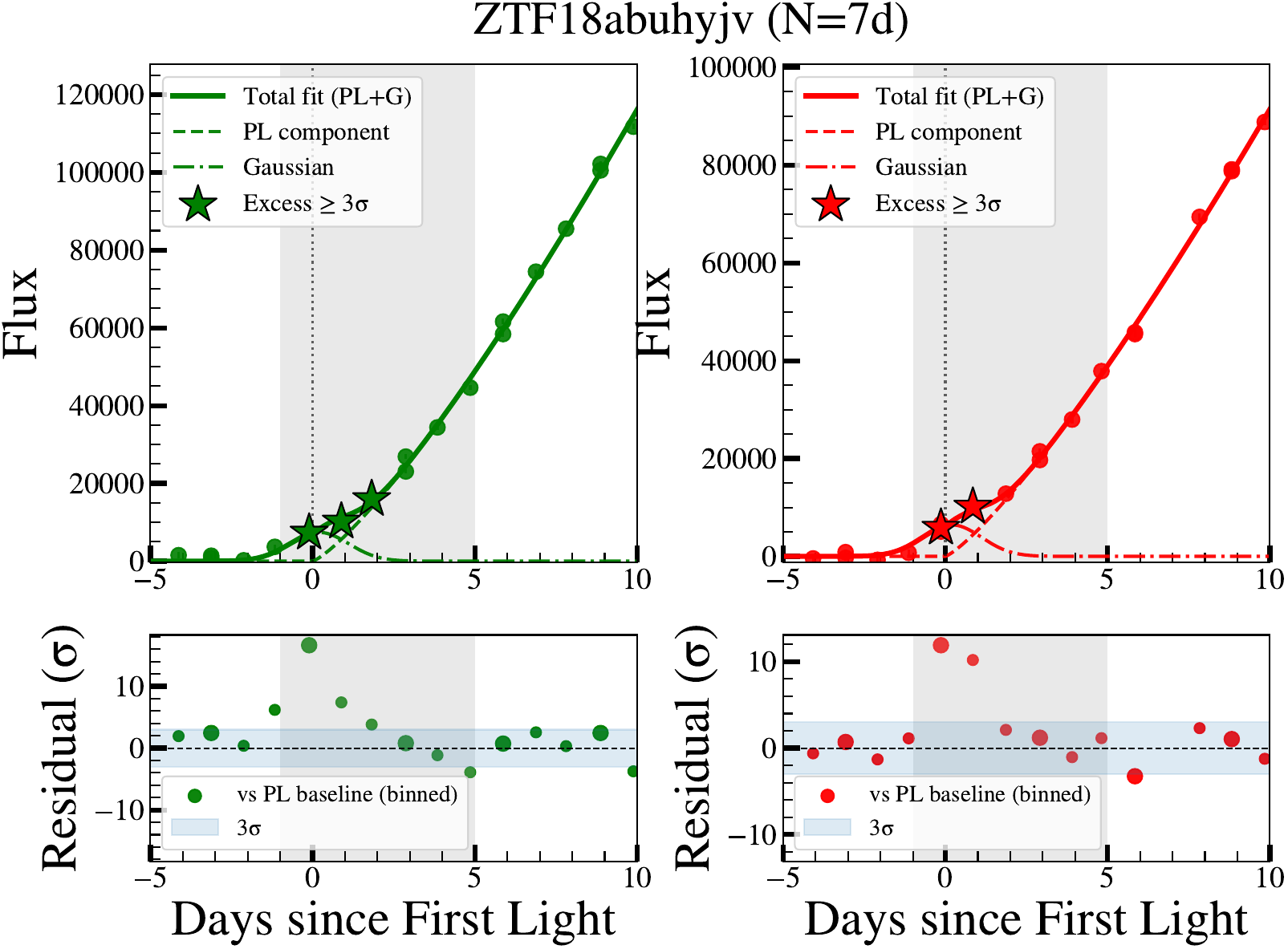}
        \caption{}
        \label{subfig:b}
    \end{subfigure}
    
    \vspace{0.5cm} 
    
    \begin{subfigure}{0.49\textwidth}
        \centering
        \includegraphics[width=\linewidth]{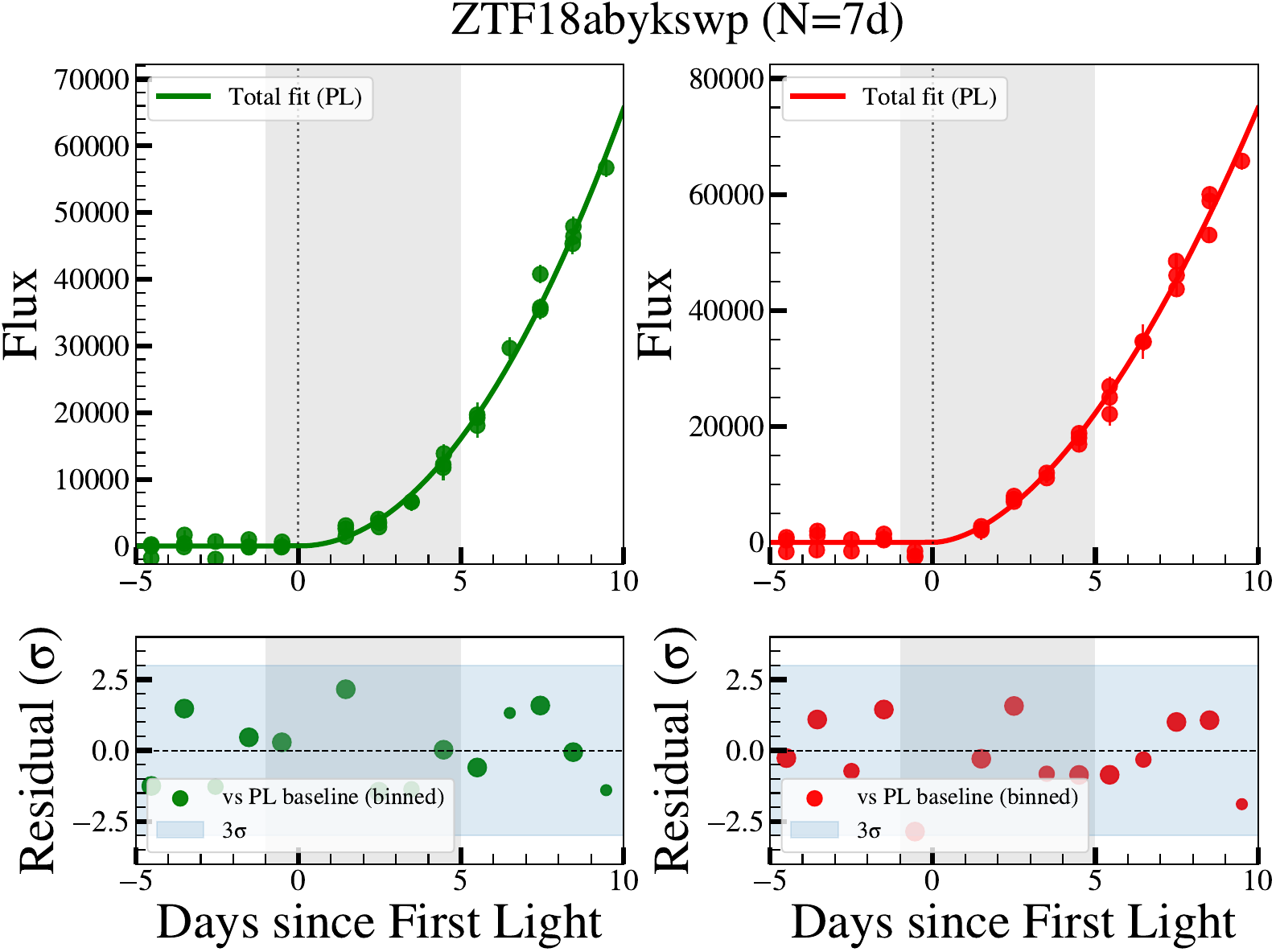} 
        \caption{}
        \label{subfig:c}
    \end{subfigure}
    \hfill
    \begin{subfigure}{0.49\textwidth}
        \centering
        \includegraphics[width=\linewidth]{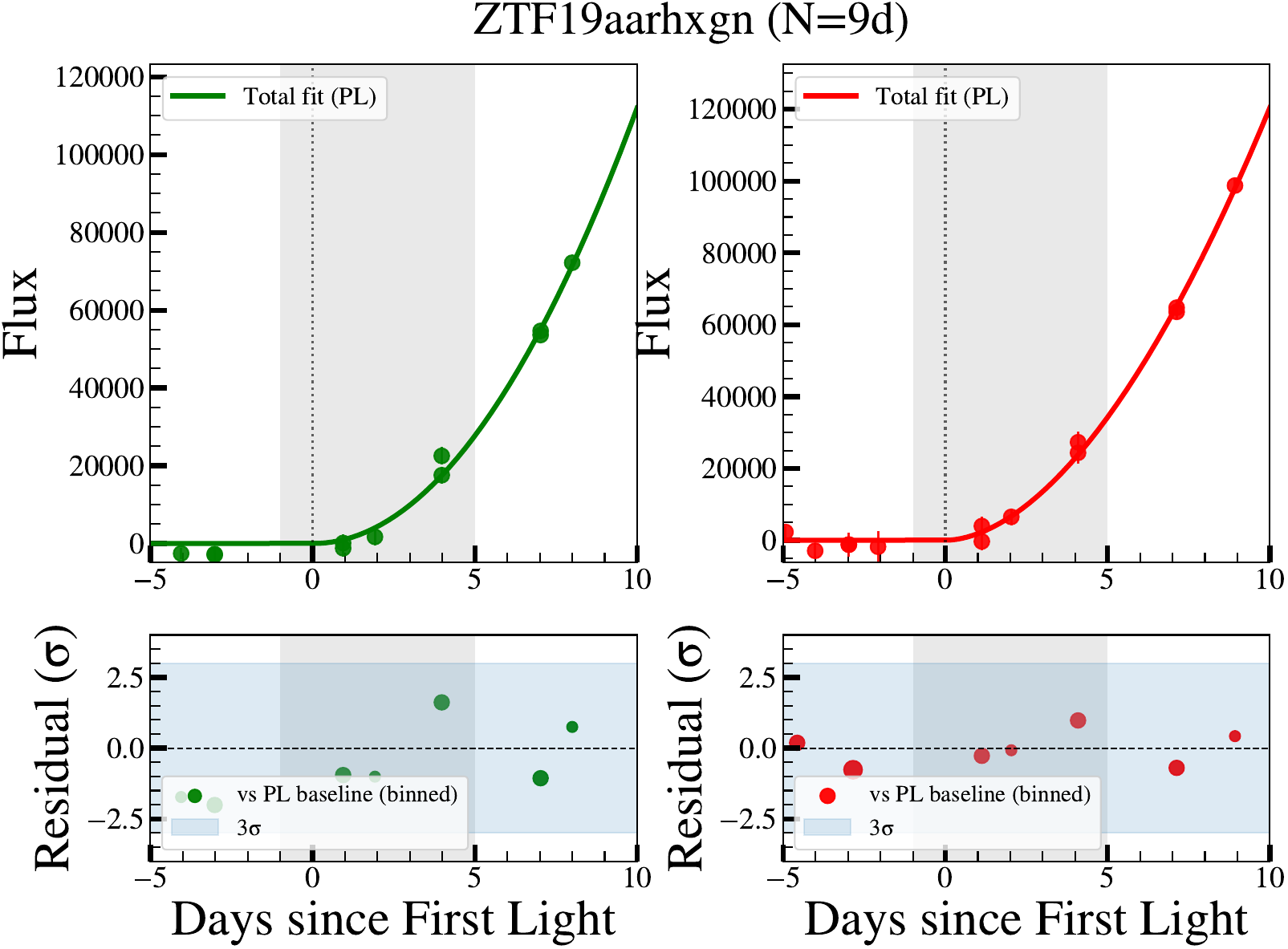} 
        \caption{}
        \label{subfig:d}
    \end{subfigure}
    
 \caption{Example ZTF DR2 light curves from the final bump and no-bump subsamples. Each panel shows observed $g$ and $r$ photometry for a single SN~Ia, together with the best-fit early-time model used by our bump-finding procedure (Section~\ref{sec:methodology}) and the residuals relative to the baseline model (PL component in the PL+G model, PL in the PL model). Early excesses are flagged as bumps if they are $3\sigma$ from the baseline within the search window (gray shaded region, from $t_{\rm i}$ to $t_{\rm i}+5$ days -- we allow a one-day tolerance window before $t_{\rm i}$ to account for explosion‑time uncertainties, but we classify the SN as having a bump only if at least one $3\sigma$ excess epoch occurs at $t\geq t_{\rm i}$). The light curves are zoomed in from $-5$ days to $+10$ days to highlight the bump region. The top row shows two SNe classified as with possible early-excesses, while the bottom row shows two SNe classified as no-bump. The light curves for the remaining bump candidates are shown in Appendix \ref{sec:app_tables}}.

    \label{fig:examplelcs}
\end{figure}

\section{Results} \label{sec:results}

\subsection{Final sample} \label{sec:finalsample}

From the initial published ZTF DR2 sample (3591 SNe~Ia), 1547 are below redshift $z<0.06$. From this subsample, 820 pass our cosmology population cuts, and 716 SNe~Ia converge successfully in both our PL and PL+G models. After applying our early-time data-quality requirements (pre-explosion baseline, early coverage around explosion time, and immediate post-explosion sampling, see Section \ref{subsec:steps}), our sample is reduced to 200 SNe Ia suitable for an early-excess search according to our criteria. The sample selection summary is presented in Table \ref{tab:waterfall_full}.

We apply our early-excess search algorithm (see Section \ref{sec:methodology}) to this subsample, preliminarily identifying 88 SNe~Ia candidates with bumps, and 112 SNe~Ia candidates without bumps. Next, we apply several final cuts to each cohort, such as the consecutive-N cut and manual vetting. Notably, the consecutive-N cut, applied only to the bump candidates, removes the most objects. We remove four bump candidates and two no-bump candidates, see section \ref{subsec:samplecuts}. The cumulative effect of each cut on the bump and no-bump samples is shown in Table~\ref{tab:waterfall_full}. The final bump sample consists of 42 candidates, while the final no-bump sample consists of 110 candidates. We provide lists of all the final candidates in Appendix \ref{sec:app_tables}.

\begin{table*}
\centering
\begin{minipage}{\textwidth}
\caption{Sample selection summary and final cuts.\label{tab:waterfall_full}}
\end{minipage}
\setlength{\tabcolsep}{4pt}
\small
\begin{tabular}{l r r}
\hline\noalign{\smallskip}
Selection step & $N$ removed & $N$ remaining \\
\hline\noalign{\smallskip}
\multicolumn{3}{l}{\textbf{A. Initial ZTF DR2 and population cuts}} \\
Full ZTF DR2 SN~Ia sample & --- & 3591 \\
Redshift cut ($z < 0.06$) & 2044 & 1547 \\
Subtype $\in \{\mathrm{norm}, \mathrm{91T}\}$ & 356 & 1191 \\
$E(B-V)_{\mathrm{MW}} \le 0.2$ & 94 & 1097 \\
$\sigma_{x_1} \le 1.0$ & 73 & 1024 \\
$\sigma_c \le 0.1$ & 31 & 993 \\
$\sigma_{t_0} \le 0.8$ & 50 & 943 \\
$-3.0 \le x_1 \le 3.0$ & 4  & 939 \\
$-0.3 \le c \le 0.3$ & 88 & 851 \\
$\mathrm{fitprob} > 10^{-7}$ & 31 & 820 \\
\multicolumn{3}{l}{\textbf{B. Model Coverage Gates}} \\
Model convergence (PL and PL+G) & 104 & 716 \\
Pre-explosion baseline, early coverage & 476 & 240 \\
Late-time coverage & 40 & 200 \\
\multicolumn{3}{l}{\textbf{C. Preliminary bump/no-bump classification}} \\
Preliminary bump candidates (early excess) & --- & 88 \\
Preliminary no-bump candidates (no early excess) & --- & 112 \\
\hline
\multicolumn{3}{l}{\textbf{D. Final cuts}} \\
\multicolumn{3}{l}{\small\textit{Bump cohort}} \\
$\;$Consecutive-$N$ robustness cut & 42 & 46 \\
$\;$Final visual inspection & 4 & \textbf{42} \\
\multicolumn{3}{l}{\small\textit{No-bump cohort}} \\
$\;$Final visual inspection & 2 & \textbf{110} \\
\noalign{\smallskip}\hline
\end{tabular}
\tablecomments{\textwidth}{The consecutive-$N$ robustness cut is applied only to bump candidates. The no-bump cohort passes directly from preliminary classification to visual inspection. Starting numbers for each cohort are the preliminary counts from Section C. Final numbers: 42 bump and 110 no-bump events.}
\end{table*}

\subsection{Light-curve and host-galaxy parameter differences}\label{subsec:param_differences}

We compare the light-curve and host-galaxy parameters between the final bump and no-bump subsamples using Welch’s $t$-tests \citep{Welch1947}. Our results are presented in Table~\ref{tab:welch_global_allrows}. We present the sample sizes, means, standard deviations, Welch $p$-values, and equivalent two-sided Gaussian significances, $\sigma_{\rm gauss}$, for all parameters. We convert each two-sided Welch $p$-value to a Gaussian-equivalent (two-sided) significance via
\begin{equation}
\sigma_{\rm gauss}\equiv \Phi^{-1}(1-p_t/2),
\end{equation}
where $\Phi^{-1}$ is the inverse cumulative distribution function of the standard normal distribution.

\begin{figure}[!htbp]
\centering
\includegraphics[width=\linewidth]{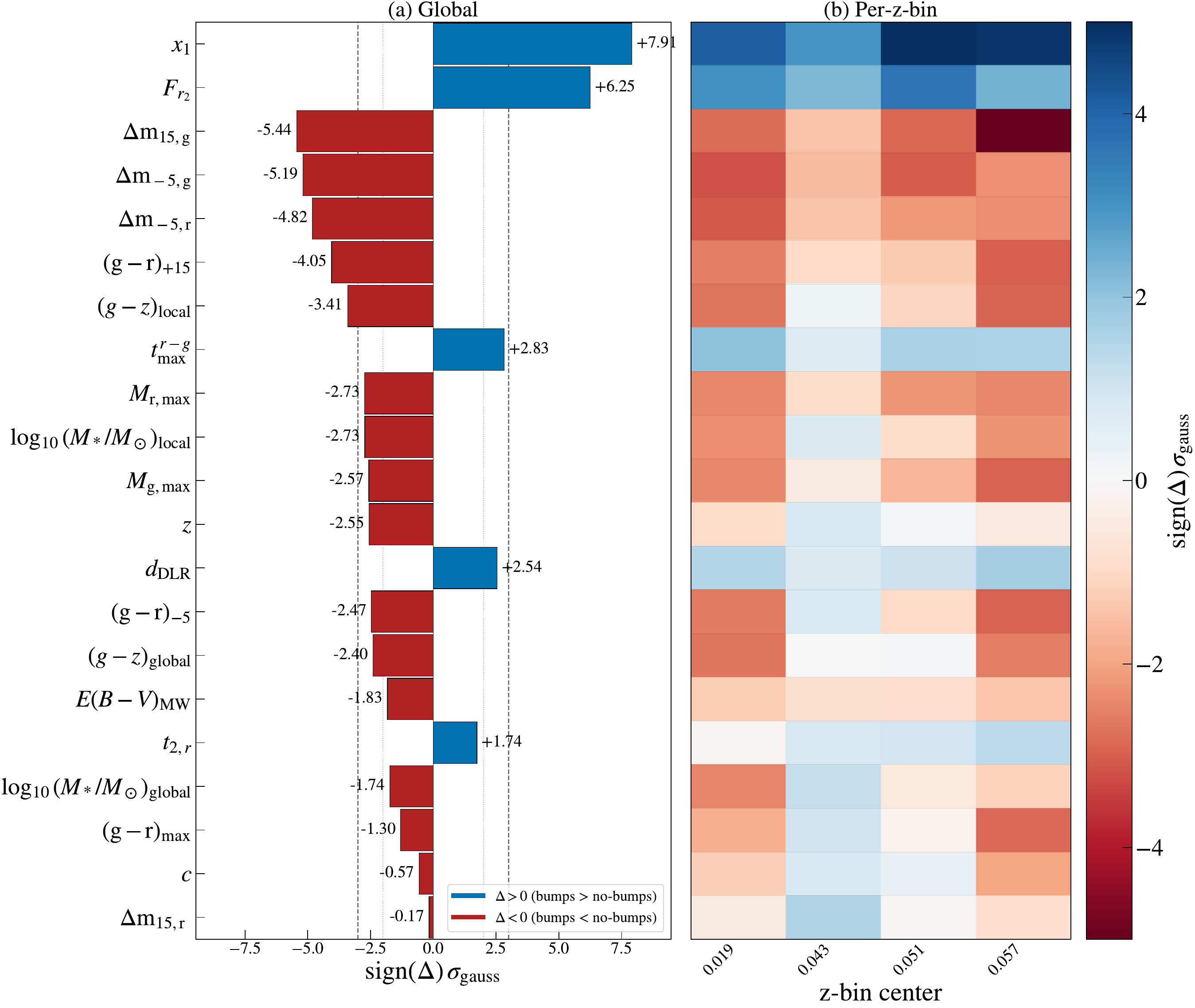}
\caption{Summary of Gaussian-equivalent significances for all tested light curve and host-galaxy parameters. \emph{Left:} global signed significance, ${\rm sign}(\Delta\langle x\rangle)\,\sigma_{\rm gauss}$, where $\Delta\langle x\rangle\equiv \langle x\rangle_{\rm bump}-\langle x\rangle_{\rm nobump}$ (Table~\ref{tab:welch_global_allrows}). \emph{Right:} the same signed significance calculated independently within four redshift bins (Section~\ref{sec:zdependence}). The most significant differences (e.g., $x_1$ and $\mathcal{F}_{r_2}$) persist across redshift bins and are not driven by a single bin.}

\label{fig:allsigma}
\end{figure}

We visually compare all the tested parameters in Figure~\ref{fig:allsigma}. The left panel shows the Gaussian-equivalent significance, ${\rm sign}(\Delta\langle x\rangle)\,\sigma_{\rm gauss}$, for each parameter in Table~\ref{tab:welch_global_allrows}. The most significant positive offsets (bump $>$ no-bump) occur in $x_1$ and $\mathcal{F}_{r_2}$, whereas several decline-rate/rise-rate parameters show significant negative offsets (bump $<$ no-bump). The host-environment properties ($(g-z)_{\rm local}$, $\log_{10} (M_*/M_\odot)_{\rm local}$, $(g-z)_{\rm global}$, $\log_{10} (M_*/M_\odot)_{\rm global}$) also show significant but smaller differences than the leading light-curve trends, with $(g-z)_{\rm local}$ and $\log_{10} (M_*/M_\odot)_{\rm local}$ more significant than the corresponding global parameters.

\begin{table*}
\centering
\begin{minipage}{\textwidth}
\caption{Global Welch comparisons between the bump and no-bump cohorts. \label{tab:welch_global_allrows}}
\end{minipage}
\setlength{\tabcolsep}{2.5pt}
\footnotesize
\begin{tabular}{l r r r r r r r r r}
\hline\noalign{\smallskip}
Parameter &
$N_{\rm bump}$ &
$N_{\rm nobump}$ &
$\langle x\rangle_{\rm bump}$ &
$\langle x\rangle_{\rm nobump}$ &
$\Delta\langle x\rangle$ &
SD$_{\rm bump}$ &
SD$_{\rm nobump}$ &
$\sigma_{\rm gauss}$ &
$p_{\rm t}$ \\
\hline\noalign{\smallskip}
$x_1$                   & 42 & 110 &  0.542 & $-$0.602 &  1.144 & 0.556 & 0.983 & \textbf{7.91} & $<10^{-6}$ \\
$\mathcal{F}_{r_2}$      & 37 &  98 &  0.390 &  0.346 &  0.044 & 0.025 & 0.049 & \textbf{6.25} & $<10^{-6}$ \\
$\Delta m_{15,\mathrm{g}}$ & 29 &  69 &  0.762 &  0.917 & $-$0.156 & 0.098 & 0.153 & \textbf{5.44} & $<10^{-6}$ \\
$\Delta m_{-5,\mathrm{g}}$ & 29 &  69 &  0.146 &  0.189 & $-$0.043 & 0.028 & 0.047 & \textbf{5.19} & $<10^{-6}$ \\
$\Delta m_{-5,\mathrm{r}}$ & 29 &  69 &  0.146 &  0.194 & $-$0.048 & 0.037 & 0.048 & \textbf{4.82} & $1.0\times10^{-6}$ \\
$(g-r)_{+15}$            & 29 &  69 & $-$0.033 &  0.132 & $-$0.165 & 0.149 & 0.222 & \textbf{4.05} & $5.1\times10^{-5}$ \\
$(g-z)_{\rm local}$      & 39 & 110 &  0.762 &  1.075 & $-$0.312 & 0.478 & 0.429 & \textbf{3.41} & $6.5\times10^{-4}$ \\
$t^{r-g}_{\mathrm{max}}$ & 29 &  69 & $-$0.004 & $-$0.446 &  0.442 & 0.751 & 0.354 & \textbf{2.83} & $4.7\times10^{-3}$ \\
$\log_{10} (M_*/M_\odot)_{\rm local}$ & 39 & 110 &  8.604 &  9.042 & $-$0.438 & 0.799 & 0.940 & \textbf{2.73} & $6.3\times10^{-3}$ \\
$M_{\mathrm{r,max}}$      & 29 &  69 & $-$18.988 & $-$18.809 & $-$0.179 & 0.297 & 0.240 & \textbf{2.73} & $6.3\times10^{-3}$ \\
$M_{\mathrm{g,max}}$      & 29 &  69 & $-$19.106 & $-$18.900 & $-$0.207 & 0.361 & 0.319 & \textbf{2.57} & $1.0\times10^{-2}$ \\
$z$                       & 42 & 110 &  0.041 &  0.047 & $-$0.006 & 0.012 & 0.010 & 2.55 & $1.1\times10^{-2}$ \\
$d_\mathrm{DLR}$          & 41 & 109 &  1.226 &  0.677 &  0.549 & 1.226 & 0.878 & 2.54 & $1.1\times10^{-2}$ \\
$(g-r)_{-5}$              & 29 &  69 & $-$0.113 & -0.059 & $-$0.054 & 0.093 & 0.102 & 2.47 & $1.4\times10^{-2}$ \\
$(g-z)_{\rm global}$     & 41 & 109 &  0.910 &  1.052 & $-$0.142 & 0.312 & 0.326 & 2.40 & $1.6\times10^{-2}$ \\
$E(B-V)_{\mathrm{MW}}$    & 42 & 110 &  0.042 &  0.053 & $-$0.011 & 0.030 & 0.043 & 1.83 & $6.8\times10^{-2}$ \\
$\log_{10} (M_*/M_\odot)_{\rm global}$ & 41 & 109 &  9.710 &  9.955 & $-$0.246 & 0.708 & 0.895 & 1.74 & $8.2\times10^{-2}$ \\
$t_{2,\mathrm{r}}$        & 37 &  96 & 20.042 & 19.170 &  0.872 & 2.193 & 3.341 & 1.74 & $8.2\times10^{-2}$ \\
$t^{r-g}_{\mathrm{max}}$  & 29 &  69 & $-$0.121 & -0.093 & $-$0.029 & 0.095 & 0.104 & 1.30 & $1.9\times10^{-1}$ \\
$c$                       & 42 & 110 &  0.021 &  0.030 & $-$0.009 & 0.086 & 0.099 & 0.57 & $5.7\times10^{-1}$ \\
$\Delta m_{15,\mathrm{r}}$ & 29 &  69 &  0.677 &  0.679 & $-$0.002 & 0.059 & 0.070 & 0.17 & $8.7\times10^{-1}$ \\
\noalign{\smallskip}\hline
\end{tabular}
\tablecomments{\textwidth}{Means and standard deviations are listed for each cohort. $\Delta\langle x\rangle \equiv \langle x\rangle_{\rm bump}-\langle x\rangle_{\rm nobump}$. $\sigma_{\rm gauss}$ is the Gaussian-equivalent significance derived from the two-sided Welch $p_{\rm t}$ value. We mark in bold the most significant differences.}
\end{table*}

\textit{Light curve properties.}
Table~\ref{tab:welch_global_allrows} shows that the strongest difference between the bump and no-bump samples occurs in the SALT2 stretch parameter $x_1$: the bump sample has systematically larger $x_1$ than the no-bump sample ($\langle x_1\rangle_{\rm bump}=0.542$ vs.\ $\langle x_1\rangle_{\rm nobump}=-0.602$, $\Delta\langle x_1\rangle=1.144$), with a $\sigma_{\rm gauss}=7.91$ significance. We also find a highly significant difference in the $r$-band secondary-maximum integrated flux $\mathcal{F}_{r_2}$, with the bump sample having higher values on average ($\langle \mathcal{F}_{r_2} \rangle_{\rm bump}=0.390$ vs.\ $\langle \mathcal{F}_{r_2} \rangle_{\rm nobump}=0.346$, $\Delta\langle \mathcal{F}_{r_2} \rangle=0.044$; $\sigma_{\rm gauss}=6.25$).

Furthermore, several additional light-curve shape and color-evolution parameters have significant differences at the $\gtrsim 3\sigma$ level, such as the decline rate at $15$ days post peak in $g$-band, the $r$ and $g$-band rise rates at $-5$ days pre peak, the $g-r$ color at $15$ days post peak, and the time difference of peak between $r$ and $g$-bands ( $\Delta \rm{m}_{{\rm15,g}}$, $\Delta \rm{m}_{{\rm-5,g}}$, $\Delta \rm{m}_{{\rm-5,r}}$, $(\rm{g-r})_{+15}$, and $t^{r-g}_{\rm{}max}$, respectively). Peak absolute magnitudes in both $g$ and $r$-band  show modest differences at the $\sim2.5-2.7\sigma$ level. In contrast, the SALT2 color parameter $c$  ($\Delta\langle c\rangle=-0.009$; $\sigma_{\rm gauss}=0.57$).

The SALT2 fitting in ZTF DR2 was performed only in the time window $[-10,+40]$ days (rest-frame), which does not include any possible bumps; therefore, the values of $x_1$ and $c$ should not be affected by the bumps themselves. To further test whether bumps influence these parameters, we repeat the Welch comparisons using alternative fits obtained from the ZTF team via private communication: SALT2 fits in the wider window $[-20,+50]$ days, and SALT3 fits in both $[-10,+40]$ and $[-20,+50]$ days. 

To make sure we perform a homogeneous comparison, we remove ZTF18ABFHRYC (bump) and ZTF18ACAHUPH (no-bump) from our final samples, since they lack a successful SALT3 fit in $[-20,+50]$ days and a SALT2 fit in $[-20,+50]$ days, respectively. After removing these objects, the SALT2 $[-10,+40]$ fits gives $\sigma_{\rm gauss}=7.85$ for $x_1$ and $\sigma_{\rm gauss}=0.62$ for $c$. For the alternative fits we find: SALT2 $[-20,+50]$ yields $x_1$ $\sigma_{\rm gauss}=6.73$ and $c$ $\sigma_{\rm gauss}=1.09$; SALT3 $[-10,+40]$ yields $7.98$ and $1.06$; SALT3 $[-20,+50]$ yields $6.03$ and $1.29$. In all cases, the large $x_1$ difference between bump and no-bump populations persists, while the $c$ differences remain negligible, in agreement with the original trend.

\textit{Host-galaxy and local environment properties.}
Host-galaxy and local-environment parameters also show statistically significant, though generally smaller, differences than the most prominent light-curve trends. The bump subsample is bluer on average in both local and global rest-frame color, with $\langle (g-z)_{\rm local}\rangle_{\rm bump}=0.762$ versus $\langle (g-z)_{\rm local}\rangle_{\rm nobump}=1.075$ ($\Delta\langle g-z\rangle_{\rm local}=-0.312$; $\sigma_{\rm gauss}=3.41$). These differences are larger than those in global rest-frame color ($\Delta\langle g-z\rangle_{\rm global}=-0.142$; $\sigma_{\rm gauss}=2.40$).

We also find that bump SNe occur in lower-mass local environments on average ($\langle \log_{10} (M_*/M_\odot)_{\rm local}\rangle_{\rm bump}=8.604$ vs.\ $\langle \log_{10} (M_*/M_\odot)_{\rm local}\rangle_{\rm nobump}=9.042$), while the difference in global host mass is weaker ($\Delta\langle \log_{10} (M_*/M_\odot)_{\rm global}\rangle=-0.246$, $\sigma_{\rm gauss}=1.74$). Additionally, the bump subgroup has a larger directional light radius $d_{\rm DLR}$, but with less significant difference ($\sigma_{\rm gauss}=2.54$). Milky Way reddening $E(B-V)_{\mathrm{MW}}$ differs only marginally ($\sigma_{\rm gauss}=1.83$).

Additionally, we studied the distribution of our bump and no-bump cohorts with respect to galaxy type and a SN location component label presented in \citet{ztfdr2gal}. However, these parameters are available for only a small fraction of our final cohorts: bump vs no-bump (13/42 bumps, 22/110 no-bumps). Within this limited subset, the bump subgroup appears more frequently associated with disk-like environments than the no-bump cohort,  qualitatively consistent with our local color and local mass trends, since late-type/disk systems are statistically bluer than early-type/bulge systems \citep{Strateva2001}. However, since the total fraction of final objects with this classification is very sparse, we treat this as an illustrative cross-check and do not draw quantitative conclusions from these labels.

\begin{figure}[!htbp]
\centering
\includegraphics[width=0.55\linewidth]{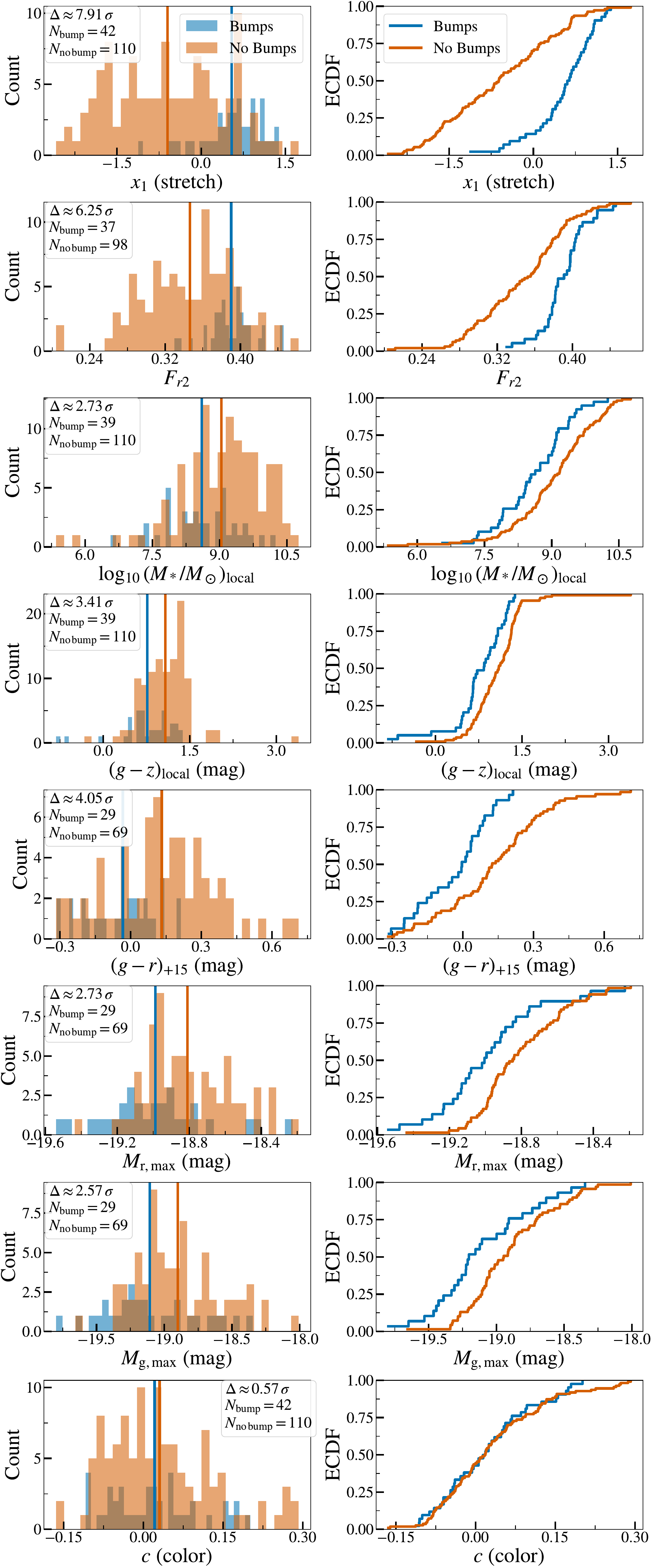}
\caption{Histograms (left column) and empirical cumulative distribution functions (ECDFs; right column) for the final bump (blue) and no-bump (orange) subgroups. The parameters displayed, from top to bottom, are: the SALT2 stretch $x_1$, the $r$-band secondary-maximum integrated flux ratio $\mathcal{F}_{r_2}$, the local stellar mass $\log_{10}(M_\ast/M_\odot)_{\rm local}$, the local rest-frame $(g-z)$ color, the color at +15~days, $(g-r)_{+15}$, the maximum absolute magnitudes in $r$- and $g$-bands ($M_{\rm{r,g,max}}$), and the SALT2 color $c$. Vertical lines in the histograms indicate parameter means. Each row is annotated with the Gaussian-equivalent significance $\sigma_{\rm gauss}$ (computed from the two-sided Welch $p_t$ value) and the sample sizes used for that parameter (Table~\ref{tab:welch_global_allrows}).}

\label{fig:hists_ecdf}
\end{figure}

Figure~\ref{fig:hists_ecdf} presents histograms and ECDFs of the most significant light-curve and host-environment parameters: $x_1$, $\mathcal{F}_{r_2}$, $\log_{10} (M_*/M_\odot)_{\rm local}$, $(g-z)_{\rm local}$, $(g-r)_{+15}$, $M_{\rm{r,max}}$ and $M_{\rm{g,max}}$ (the decline rate is directly related to $x_1$ \citep{ztfdr2diversity}, and therefore is not included in our analysis). For $x_1$ and $\mathcal{F}_{r_2}$, the histograms and ECDFs show a clear shift of the bump subgroup toward larger values (smaller for $g-r$ at $+15$\,d), consistent with their $\sigma_{\rm gauss}$ values in Table~\ref{tab:welch_global_allrows}. The host-environment parameters $(g-z)_{\rm local}$ and $\log_{10} (M_*/M_\odot)_{\rm local}$ also show a shift (though less pronounced) of the bump subgroup towards smaller values, again consistent with their $\sigma_{\rm gauss}$ values. The peak absolute magnitudes $M_{\rm{r,max}}$ and $M_{\rm{g,max}}$ also show differences in these distributions, although with smaller significance. Finally, we include the color parameter $c$ due to its importance in SN cosmology: consistent with its $\sigma_{\rm gauss}$ value, the differences in distributions between the bump and no‑bump cohorts are negligible.

We additionally apply two-sample Kolmogorov-Smirnov (KS) tests to four of our primary parameters of interest ($x_1$, $\mathcal{F}_{r_2}$, $\log_{10} (M_*/M_\odot)_{\rm local}$, and $(g-z)_{\rm local}$). Consistent with the ECDF separations in Fig. \ref{fig:hists_ecdf}, two-sample KS tests show that the bump and no-bump distributions differ for $x_1$ ($D=0.579$, $p=5.2\times10^{-10}$) and $\mathcal{F}_{r_2}$ ($D=0.532$, $p=1.6\times10^{-7}$), and show smaller but significant differences for $\log_{10} (M_*/M_\odot)_{\rm local}$ ($D=0.304$, $p=7.3\times10^{-3}$) and $(g-z)_{\rm local}$ ($D=0.307$, $p=6.5\times10^{-3}$).

\begin{figure}[!htbp]
\centering
\includegraphics[width=\linewidth]{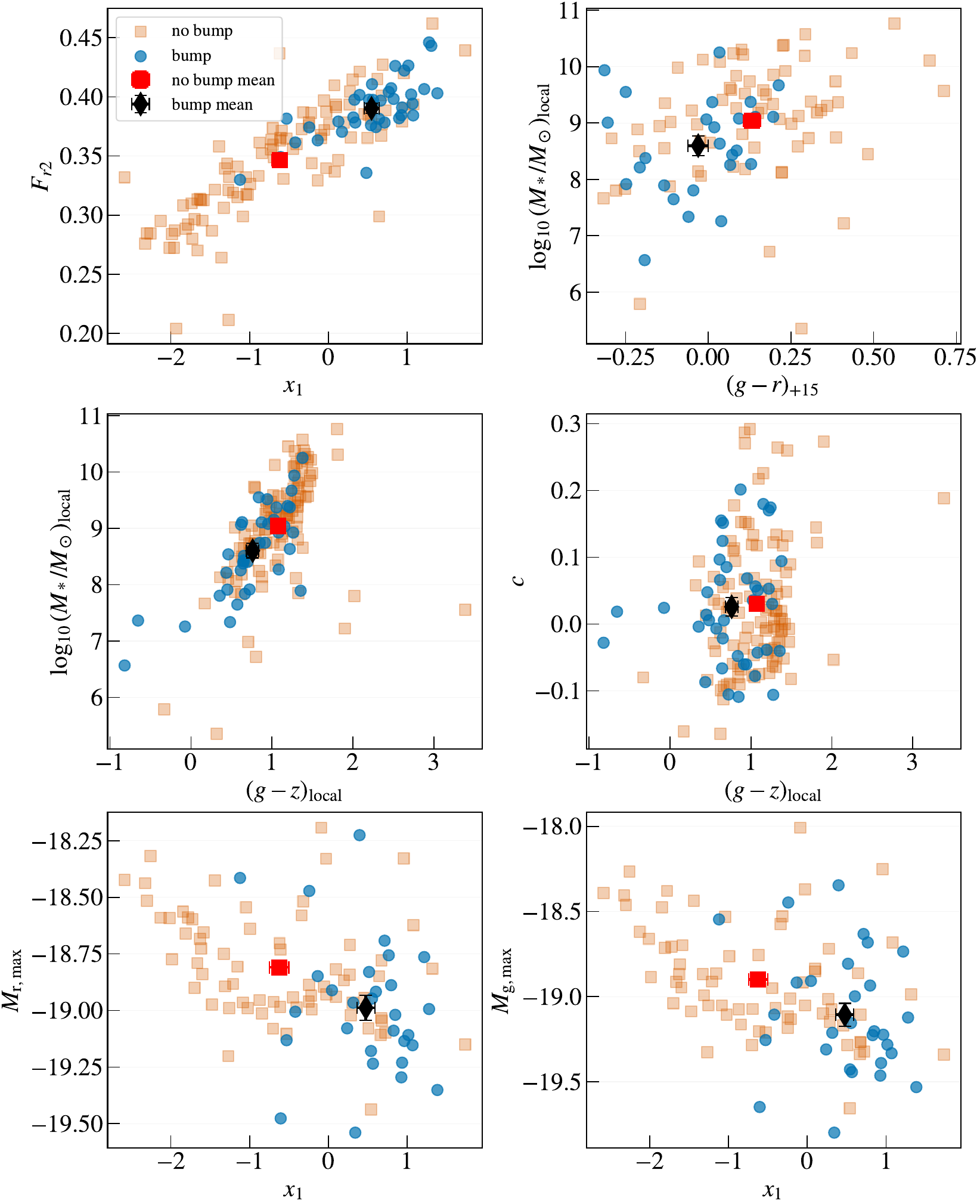}
\caption{Two-dimensional distributions for key light-curve and host-environment parameters for the final bump (blue circles) and no-bump (orange squares) subgroups. The different panels show: $\mathcal{F}_{r_2}$ vs.\ $x_1$ (top left), $(g-r)_{+15}$ vs.\ $\log_{10} (M_*/M_\odot)_{\rm local}$ (top right), $\log_{10} (M_*/M_\odot)_{\rm local}$ vs.\ $(g-z)_{\rm local}$ (middle left), SALT2 color $c$ vs.\ $(g-z)_{\rm local}$ (middle right), and $M_{\rm{r,max}}$ and $M_{\rm{g,max}}$ vs.\ $x_1$ (bottom). The large black diamond and red square represent the mean values of the bump and no-bump subgroups, respectively. The bump candidates generally show higher $x_1$ and $\mathcal{F}_{r_2}$, and are, on average, in bluer, lower-mass local environments, while the $c$--$(g-z)_{\rm local}$ plane shows significant overlap between both populations, consistent with the negligible global difference in $c$.}

\label{fig:planes}
\end{figure}

Finally, Figure~\ref{fig:planes} shows two-dimensional projections for some of the most important parameters. The $x_1$--$\mathcal{F}_{r_2}$ and $(g-r)_{+15}$--$\log_{10} (M_*/M_\odot)_{\rm local}$ planes show a clear difference between the distributions and mean values of the bump and no-bump populations, with the bump population shifted toward higher $x_1$ and $\mathcal{F}_{r_2}$ values and toward lower-mass local environments on average. Similarly, the $\log_{10} (M_*/M_\odot)_{\rm local}$--$(g-z)_{\rm local}$ planes show that bump SNe lie in bluer, lower-mass environments on average. On the other hand, the $c$--$(g-z)_{\rm local}$ distributions show substantial overlap between the populations in $c$ at fixed local color, consistent with the negligible global difference in $c$. Finally, the absolute magnitudes--$x_1$ planes also show significant differences in their mean values and distributions, though more pronounced in the $x_1$ direction.

\subsection{Redshift dependence} \label{sec:zdependence}

Table~\ref{tab:welch_global_allrows} shows a small but non-negligible difference in redshift between the bump and no-bump subsamples ($\Delta\langle z\rangle=-0.006$; $\sigma_{\rm gauss}=2.55$). Although our full sample is restricted to the ZTF DR2 volume-limited sample with $z<0.06$ (Section~\ref{sec:data}), some parameters may correlate with redshift due to selection effects, S/N, or parameter systematics. Therefore, we explore the bump/no-bump differences as a function of redshift.

We divide the final bump/no-bump subsamples into four redshift bins, each with approximately equal numbers of SNe, and repeat the bump/no-bump comparisons within each bin. The right-hand panel of Figure~\ref{fig:allsigma} summarizes the significance per bin for all parameters: the differences in the most significant parameters persist across all redshift bins and do not appear to be driven by a single bin.   

\begin{figure}[!htbp]
\centering
\includegraphics[width=\linewidth]{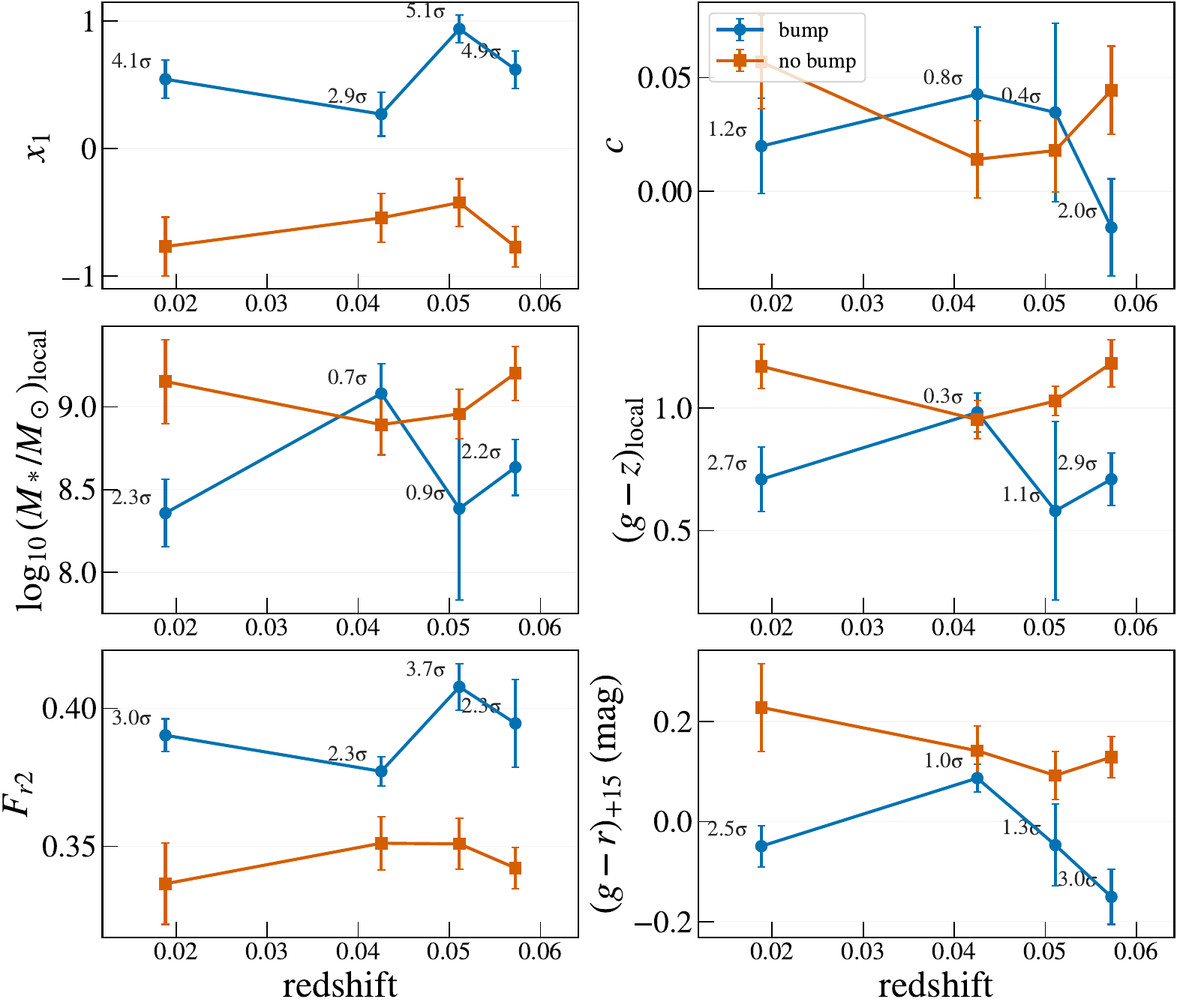}
\caption{Redshift-binned mean comparisons for representative parameters in the final bump and no-bump subgroups. Each panel shows the mean value in each of the four redshift bins (bin centers on the horizontal axis; bin sizes not shown for clarity), with uncertainties calculated as the standard errors of the mean. From top to bottom / left to right: $x_1$, $c$, $\log_{10} (M_*/M_\odot)_{\rm local}$, $(g-z)_{\rm local}$, $\mathcal{F}_{r_2}$, and $(g-r)_{+15}$. We also indicate the Gaussian-equivalent significance of the bump--no-bump difference in each bin. The strongest light-curve parameter differences ($x_1$ and $\mathcal{F}_{r_2}$) remain significant across all bins, while the host-galaxy parameters ($\log_{10} (M_*/M_\odot)_{\rm local}$ and $(g-z)_{\rm local}$) are still significant in the first and fourth bins, and remain within errorbars in the remaining bins. These trends possibly hint that these offsets are not driven by redshift-dependent selection effects (Section~\ref{sec:zdependence}).}

\label{fig:zbinned}
\end{figure}

In more detail, Figure~\ref{fig:zbinned} presents the redshift-binned means for key parameters ($x_1$, $c$, $\mathcal{F}_{r_2}$, $(g-z)_{\rm local}$, $\log_{10} (M_*/M_\odot)_{\rm local}$, and $(g-r)_{+15}$), with uncertainties shown as standard errors on the mean. The trends in $x_1$ and $\mathcal{F}_{r_2}$ are strong and consistent across all bins (and in a slightly lower measure for $(g-r)_{+15}$), supporting the interpretation that these differences arise from an intrinsic population difference rather than a redshift-driven effect. As expected from the global trends, the SALT2 $c$ shows very low significance in all bins.

The bin-by-bin host-environment parameters ($(g-z)_{\rm local}$, $\log_{10} (M_*/M_\odot)_{\rm local}$) present a slight variation and significance consistent with their smaller global $\sigma_{\rm gauss}$ values. However, the difference between the bump and no-bump subgroups is consistent within the uncertainties for bins 1, 3 and 4, and is larger than 2.7$\sigma$ for bins 1 and 4 for $(g-z)_{\rm local}$, and larger than 2.2$\sigma$ for bins 1 and 4 for $\log_{10} (M_*/M_\odot)_{\rm local}$.

Our main results are insensitive to the visual vetting of the four manually-removed bump-SNe (Section \ref{subsec:samplecuts}). When including them in the analysis, the $x_1$, $\mathcal{F}_{r_2}$, $(g-z)_{\rm local}$, $\log_{10} (M_*/M_\odot)_{\rm local}$, and $c$ $\sigma$ significances are 7.42, 5.60, 3.28, 2.78 and 0.61, respectively, consistent with the values reported throughout this work when excluding these objects.

\section{Discussion} \label{sec:discussion} 

\subsection{Early-time excesses in ZTF DR2}

Using our large, homogeneously selected sample of early-excess candidates from ZTF DR2, we find potential evidence that the bump and no-bump subgroups differ in several parameters. The strongest differences are in light-curve shape and the secondary maximum: the bump subgroup shows significantly larger $x_1$ and larger $\mathcal{F}_{r_2}$. These trends persist across redshift bins, which supports a possible interpretation of intrinsic population differences rather than a redshift effect. 
In contrast, the SALT2 color $c$ shows a very small bump/no-bump difference.

The $x_1$ difference remains robust when the analysis is repeated with SALT2 fits using a wider phase window ($[-20,+50]$ days) and with SALT3 fits, indicating that the stretch is not affected by the early bump; This supports a physical picture in which SNe~Ia with intrinsically higher $x_1$ (broader light curves, linked to younger progenitors; e.g., \citet{ Lampeitl2010}) are more likely to exhibit an early excess---suggesting that both the bump and a high $x_1$ value are independent manifestations of the same  same progenitor/explosion properties.

Host-environment trends are also significant but weaker than the light-curve-shape parameters. Local host properties ($(g-z)_{\rm local}$ and $\log_{10} (M_*/M_\odot)_{\rm local}$) show modest total significance (and persist in specific redshift ranges), suggesting that environment may affect the early-excess in SNe~Ia, consistent with previously local-environment dependencies in SN Ia samples \citep{Rigault2013,Roman2018}. Additionally, \citet{ztfdr2x1} find that high-stretch SNe~Ia are preferentially found in environments with lower global-host mass and bluer local colors; in the same sense, we find tentatively that our bump subgroup preferentially occupies lower local-mass and bluer local environments.

\subsection {Comparison to other works}

As noted in previous works \citep{Burke2022,Ye2024}, the incidence of early excesses depends strongly on the methodology adopted: the same SN can be flagged as with or without a bump according to the procedure chosen, such as comparison with other objects \citep{Yao2019}, power-law fitting \citep{Miller2020,Ye2024}, color evolution \citep{Bulla2020}, Ni-mixing models \citep{deckers2022}, or companion shock models \citep{Burke2022}. Therefore, our reported bump/no-bump catalogs should be interpreted in the context of our assumptions/cuts. 

\subsubsection{Comparison to \citet{Ye2024}}
 Since our fitting procedure is based on the work of \citet{Ye2024}, comparing our results with theirs is an important cross-check of our selection. However, there are some important differences in our samples and pipelines: 

 \begin{enumerate}[label=\roman*)]
 \item Even for overlapping SNe in their sample, based on ZTF DR1 \citep{Yao2019,Dhawan2022} and ZTF DR2, the flux measurements (zeropoints) between the two samples differ.
\item ZTF DR2 includes baseline corrections not present in their datasets.
\item We apply the ZTF DR2-recommended photometry quality cuts to remove spurious points.
\item We limit our sample to the volume-limited $z<0.06$.
 \item We apply robustness (consecutive-$N$ cut, Section \ref{sec:methodology}) to remove bump candidates that are heavily dependent on a single data point.
 \item We require the PL+G model to be statistically preferred for bump detection ($\Delta BIC > 2$), while \citet{Ye2024} don't have a fixed $\Delta BIC$ cut and include bumps with a $\Delta BIC$ as small as $-10$ or $-11$.
 \item \citet{Ye2024} no-bump sample includes SNe for which the data is not sufficient to detect an early excess, while we remove these cases from our no-bump sample.
 \end{enumerate}

Additionally, regarding SN and host-galaxy parameters:
\begin{enumerate}[label=\roman*),resume]
\item \citet{Ye2024} derive the SALT3 parameters and global host masses themselves, while we use the provided SALT2 and global host masses in ZTF DR2.
\item We include several additional parameters ($\log_{10} (M_*/M_\odot)_{\rm local}$, $(g-z)_{\rm local}$, $(g-z)_{\rm global}$, $\mathcal{F}_{r_2}$, $g-r$ color, among others; see Table \ref{tab:param_inventory}) that are not included in their analysis.
\item We do not derive Hubble residuals as done in their work.
\end{enumerate}

\citet{Ye2024} identify 17 early-excess candidates in their analysis, 11 of which have $z<0.06$ (all present in ZTF DR2 and our initial parent sample). We detect potential early-excesses in all 11 of these SNe~Ia, but discard 3 due to our robustness ``Consecutive-$N$" rule: ZTF18aaxsioa, ZTF18abkhcwl, ZTF18abvejbm. For these 3 SNe, we detect a bump with the best $N$ ($N=7)$, but we do not detect an early excess when $N=8$ in both bands (ZTF18aaxsioa and ZTF18abkhcwl) and in the $r$-band (ZTF18abvejbm). 

The remaining 8 SNe flagged in \citet{Ye2024} as having bumps under $z<0.06$ pass all our cuts and are part of our final early-excess candidate sample. These are: ZTF18aayjvve ZTF18abcflnz, ZTF18abucvbf, ZTF18aasdted, ZTF18aaslhxt, ZTF18abauprj, ZTF18abfhryc, and ZTF18abuhzfc.

\textbf{Light-curve parameter differences}: A key difference between our conclusions is that \citet{Ye2024} report no significant differences in SALT light-curve shape ($x_1$) between SNe Ia with and without a bump (ZTF DR2 uses the SALT2 version, while \citet{Ye2024} used SALT3 fits for their analysis). In contrast, in our final subgroups we find a highly significant shift to larger $x_1$ in the bump population (Table 5: $\langle x_1\rangle_{\rm bump}=0.542$ vs. $\langle x_1\rangle_{\rm nobump}=-0.602$, $\sigma_{\rm gauss}=7.91$). We also find a strong difference in the r-band secondary-maximum integrated flux $\mathcal{F}_{r_2}$ ($\sigma_{\rm gauss}=6.25$), with bumps having systematically larger $\mathcal{F}_{r_2}$ on average ($\mathcal{F}_{r_2}$ is a ZTF DR2 metric presented in \citet{ztf2max}, so it is not included in \citet{Ye2024}). In agreement with \citet{Ye2024}, we also do not find a difference in $c$ between the bump and no-bump populations.

As mentioned above, a possible explanation of this significant difference in $x_1$ is that \citet{Ye2024}’s “no strong $x_1$” statement is derived from a smaller, heterogeneous sample, and they include SNe with insufficient data in their no-bump subgroup. In contrast, our results are derived from a single, internally homogeneous DR2 sample with strict coverage gates and robustness cuts, which may reduce contamination and sharpen population contrasts. 

\textbf{Host-environment parameters}: \citet{Ye2024} detailed global host-mass comparisons show that the sign and strength of the global host mass preference vary by ZTF subset: for their \citet{Yao2019}-based analysis, bump SNe prefer lower-mass hosts at $\sim2.3\sigma$, whereas for the ZTF 2018 cosmology release \citep{Dhawan2022}, they find a small, statistically insignificant shift toward higher host masses. Overall, they conclude, there is no clear trend between bumps and global host-galaxy masses. In our DR2 sample, the host galaxy mass difference is also weak globally ($\Delta\langle \log_{10} (M_*/M_\odot)_{\rm global}\rangle=-0.246$, $\sigma_{\rm gauss}=1.74$). However, it is significant locally: we find bumps occur in lower local stellar mass regions ($\Delta\langle \log_{10} (M_*/M_\odot)_{\rm local}\rangle=-0.438$, $\sigma_{\rm gauss}=2.73$) and locally bluer $(g-z)_{\rm local}$ environments ($\Delta\langle (g-z)_{\rm local}\rangle=-0.312$ mag, $\sigma_{\rm gauss}=3.41$). This discrepancy between local and global host galaxy mass is consistent with early-excess incidence being more closely linked to local star-formation/stellar-population conditions than to global host galaxy mass \citep{Kim2018,Rigault2020}.
The $(g-z)_{\rm local}$ parameter is not analyzed in \citet{Ye2024}, as these measurements were released after their publication.

\subsubsection{Comparison to \citet{deckers2022}}
\label{subsec:discussion_deckers}

\citet{deckers2022} analyze early-time ZTF light curves with different procedure: instead of defining excesses relative to a power-law baseline, they compare observed light curves to grids of SN~Ia explosion models. They identify six SNe~Ia with detected early flux excesses.

Despite the different methodologies, several points align with our results: \citet{deckers2022} show that some flux-excess events have high-stretch values (three of six in the top 20\% of their $x_1$ distribution), although they caution that small-number statistics prevent strong conclusions. Similarly, we find that $x_1$ shows the biggest difference between the bump and no-bump populations (with much better statistics), indicating that early excesses may be connected to broader light-curve-shape diversity.

Additionally, \citet{deckers2022} find a preference toward lower-mass hosts among their bump detections (5/6 are in host galaxies below the sample’s mean mass). At the same time, our DR2 results indicate weak differences in $\log_{10} (M_*/M_\odot)_{\rm global}$, but stronger differences in $\log_{10} (M_*/M_\odot)_{\rm local}$.

Finally, \citet{deckers2022} find no strong preference for SALT2 color $c$ (three red and three blue among their six excess events), which is consistent with our result that global SALT2 $c$ does not differ significantly between the bump and no-bump populations.

\subsection{Possible Physical Interpretation}
\label{sec:physicalinterpretation}
The demographic trends found in this work can provide qualitative constraints on the origin of early-time excesses. The strongest separations between our bump and no-bump subgroups are in $x_1$ and $\mathcal{F}_{r_2}$ (larger for bumps), while the SALT2 color $c$ shows no significant difference. These trends could indicate that bumps may be related to broader light curves and secondary-maximum diversity —caused by ejecta structure, opacity/ionization evolution, or outer-layer heating—instead of dust reddening. In this context, our preference for brighter absolute magnitudes for bump SNe (at the $\sim$2--3$\sigma$ level) and strong separation in light-curve width are qualitatively consistent with recent demographic studies that find early-excess SNe to be associated with longer rise times and higher luminosities \citep{Wu2025}, who argue that these trends could reflect differences in \(^{56}\)Ni production/distribution and explosion asymmetry. Additionally, the observation that bumps may occur more frequently in locally bluer and lower local mass environments may indicate that local stellar-population conditions affect the probability of an early excess, consistent with previous evidence that the local host environment provides information about SN~Ia populations not captured by the global host parameters \citep{Rigault2013,Roman2018}.Furthermore, the fact that the $x_1$--bump association is robust against changes in the SALT fitting window and SALT version reinforces the picture that both properties could come from the same explosion physics.

Several physical scenarios have been proposed which can generate early-time excess emission, such as companion interaction in the single-degenerate scenario \citep{Kasen2010}, shallow $^{56}$Ni mixing or surface radioactive material \citep{Piro2014,Piro2016}, helium-shell detonations/double detonations \citep{Woosley1994,Fink2010,Polin2019}, and interaction with circumstellar material \citep{Moriya2023}. Since some of these models overlap in their predicted early light-curve morphologies and can be strongly sensitive to viewing angle and radiative-transfer effects, our results are best interpreted as population-level constraints instead of favoring a single physical bump model.

The bump sample also appears intrinsically more homogeneous in its light-curve properties: the standard deviations for $x_1$, $\mathcal{F}_{r_2}$, and several color/decline parameters are smaller than those of the no-bump sample (Table~\ref{tab:welch_global_allrows}). This raises the interesting possibility that early bumps do not trace explosion diversity, but instead trace a specific progenitor channel with relatively uniform explosion physics. Larger future datasets are necessary to verify this interpretation, and a physical-model analysis (beyond the scope of this paper) is required to link the demographic correlations presented here to specific progenitor channels.

\subsection{Future Improvements} \label{sec:futureimprovements}

Future work includes extending our analysis to larger samples with improved early-time cadence, such as later ZTF releases (the full ZTF survey is expected to include $\sim 8000$ spectroscopically confirmed SNe~Ia plus $\sim 25000$ additional photometrically classified SNe~Ia, \citealt{ztfdr2overview}), Vera Rubin’s LSST \citep{lsst}---where the lower cadence may require targeted follow-up to identify early excesses robustly ---and GOTTA’s planned 30-minute cadence \citep{gotta}, which will enable us to improve statistics and explore further parameter correlations in greater detail. Additionally, to translate the observed bump fraction into an intrinsic early excess, we need to model the selection function for bump detection using injection-recovery tests. Finally, to complement the PL/PL+G framework used here, future work should include physically motivated models, such as those presented in \citet{Magee2018,Magee2020,deckers2022}, fitting different $^{56}$Ni masses/distributions and explosion/outer-layer properties to the same light curves, allowing us to better distinguish between the physical scenarios which can reproduce both the early-excess morphologies and demographic trends.

\section{Conclusions} \label{sec:conclusions} 
To our knowledge, this work presents the first systematic search for early excess emission in the ZTF DR2 volume-limited sample, yielding the largest bump population derived through a coherent, homogeneous analysis to date. We have used the numerous light-curve and host-galaxy parameters in ZTF DR2 to cross-match to our bump and no-bump populations. Our main results are as follows:
\begin{itemize}
    \item Final robust catalogs: After subtype/MW-reddening/SALT2-quality cuts and robustness filtering, the final catalogs contain 42 bump candidates and 110 no-bump SNe~Ia.
    \item Most significant demographic separations: The strongest bump/no-bump differences are in SALT2 stretch $x_1$ and the secondary maximum $r$-band flux metric $\mathcal{F}_{r_2}$, both higher for the bump subgroup, at very high significance ($\sigma \approx 7.91$  for $x_1$, $\sigma \approx 6.25$ for $\mathcal{F}_{r_2}$).
    \item Robustness of the $x_1$ result: The strong $x_1$ difference between the bump and no-bump SN~Ia subsamples persists when the analysis is repeated with SALT2 fits using a wider phase window ($[-20,+50]$ days) and with SALT3 fits, suggesting that the trend is not tied to a specific light-curve fitting model or phase range.
    \item Local environment trends are present but slightly weaker: The bump subgroup is, on average, in a bluer $(g-z)_{\rm local}$ and a lower $\log_{10} (M_*/M_\odot)_{\rm local}$ environment than the no-bump subgroup ($3.41 \sigma$ and $2.73 \sigma$, respectively).
    \item SALT2 $c$ shows a minimal difference: The SALT2 color parameter $c$ appears to have an insignificant difference between bump and no-bump populations.
    \item Trends continue across redshift bins: The strongest differences between the bump and no-bump subgroups (such as $x_1$ and $\mathcal{F}_{r_2}$) are present in all redshift bins, suggesting intrinsic differences between the bump and no-bump samples.
    \item Comparison with \citet{Ye2024}  \citet{deckers2022}: From the 11 bump candidates presented in \citet{Ye2024} at $z <0.06$, we recover all 11 (plus remove 3 via our consecutive-N robustness cut). Additionally, our results are consistent with \citet{deckers2022}, while significantly increasing the statistics (bumps associated with higher $x_1$ and a in lower-mass environments).
\end{itemize}
In future papers, we will expand our analysis using physically motivated models to better connect demographic correlations to explosion physics and progenitor channels.

\normalem
\begin{acknowledgements}

This work is funded by the National Natural Science Foundation of China grant No. 12303051, the Strategic Priority Research Program of the Chinese Academy of Sciences grant No. XDB0550300, and the China Manned Space Program No. CMS-CSST-2025-A14. J.F.L. acknowledges support from the NSFC through grant No. 12588202 and from the New Cornerstone Science Foundation through the New Cornerstone Investigator Program and the XPLORER PRIZE. Software used: \texttt{Jupyter Notebook} \citep{jn},
    \texttt{matplotlib} \citep{matplotplib}, 
    \texttt{numpy} \citep{numpy},
    \texttt{pandas} \citep{reback2020pandas,mckinney-proc-scipy-2010},
    \texttt{scipy} \citep{scipy},
\end{acknowledgements}

\appendix

\section{Bump plots and bump/ no-bump tables}\label{sec:app_tables}
\input{bumps_deluxetable}
\input{nobumps_deluxetable_two_up_split_same_number}
\input{appendix_bumps_figures.tex}

\begin{figure}[!htbp]
    \centering
    \begin{subfigure}{0.49\textwidth}
        \centering
        \includegraphics[width=\linewidth]{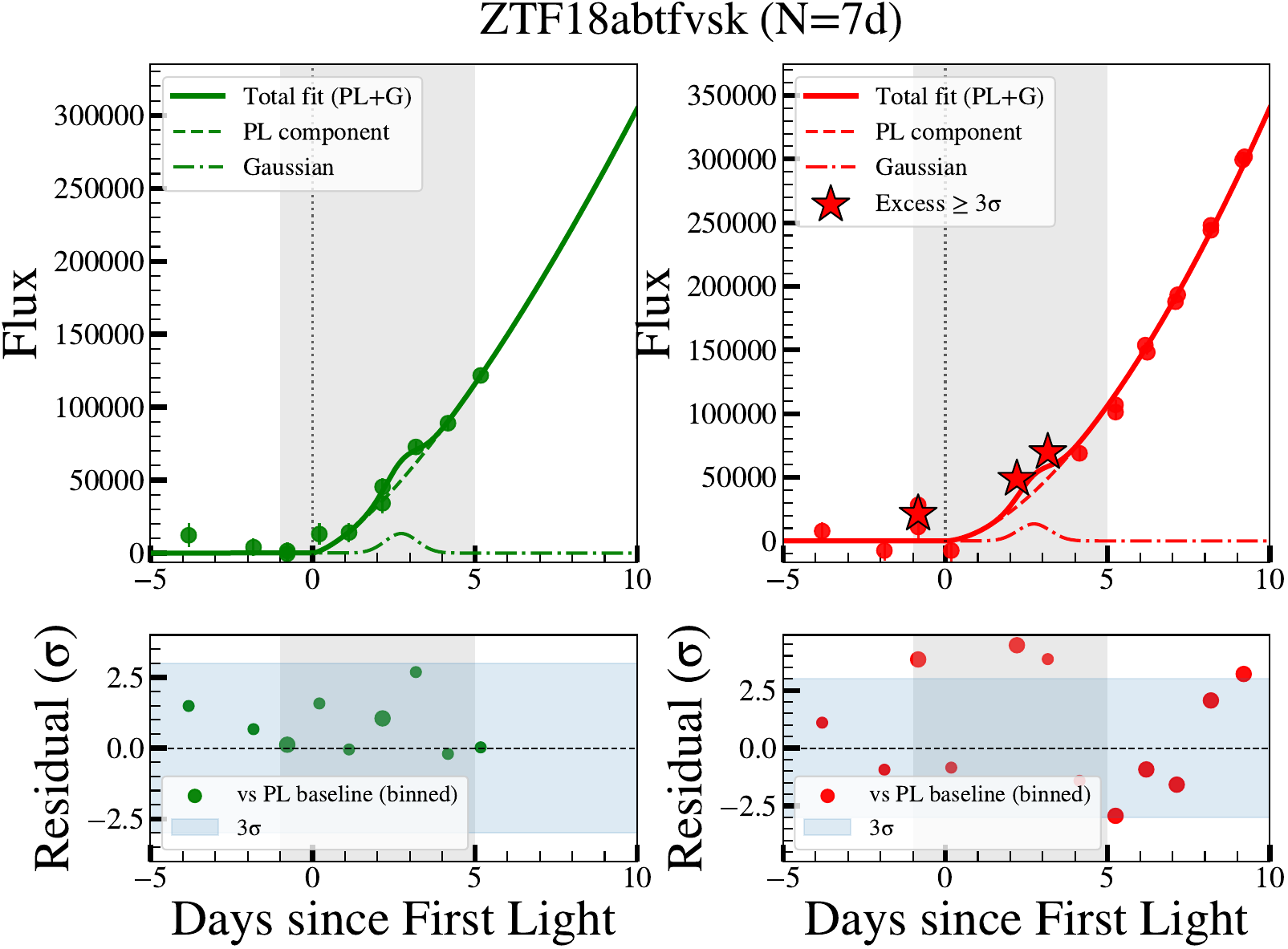}
        \caption{}
        \label{subfig:a}
    \end{subfigure}
    \hfill
    \begin{subfigure}{0.49\textwidth}
        \centering
        \includegraphics[width=\linewidth]{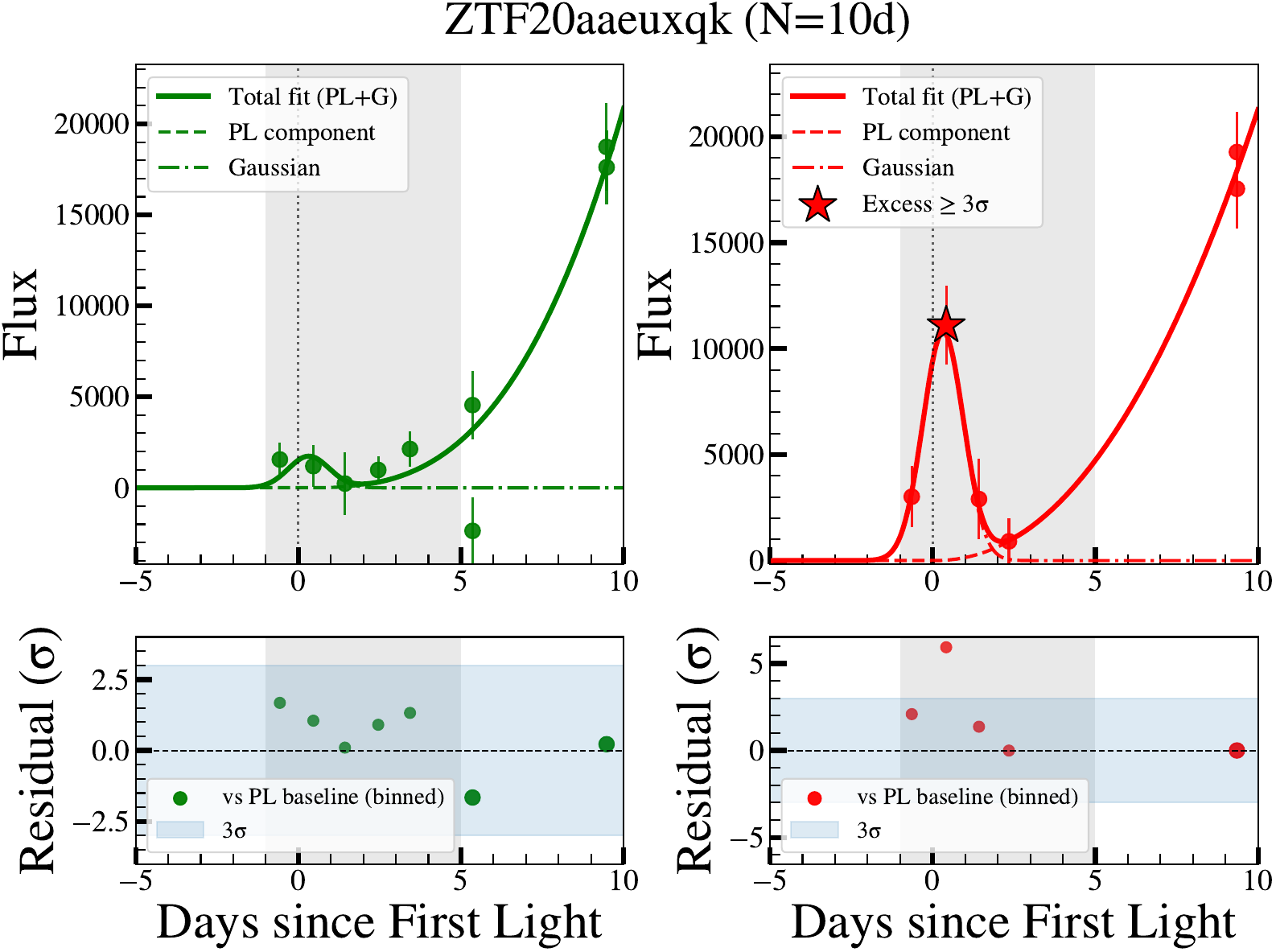}
        \caption{}
        \label{subfig:b}
    \end{subfigure}
    
    \vspace{0.5cm} 
    
    \begin{subfigure}{0.49\textwidth}
        \centering
        \includegraphics[width=\linewidth]{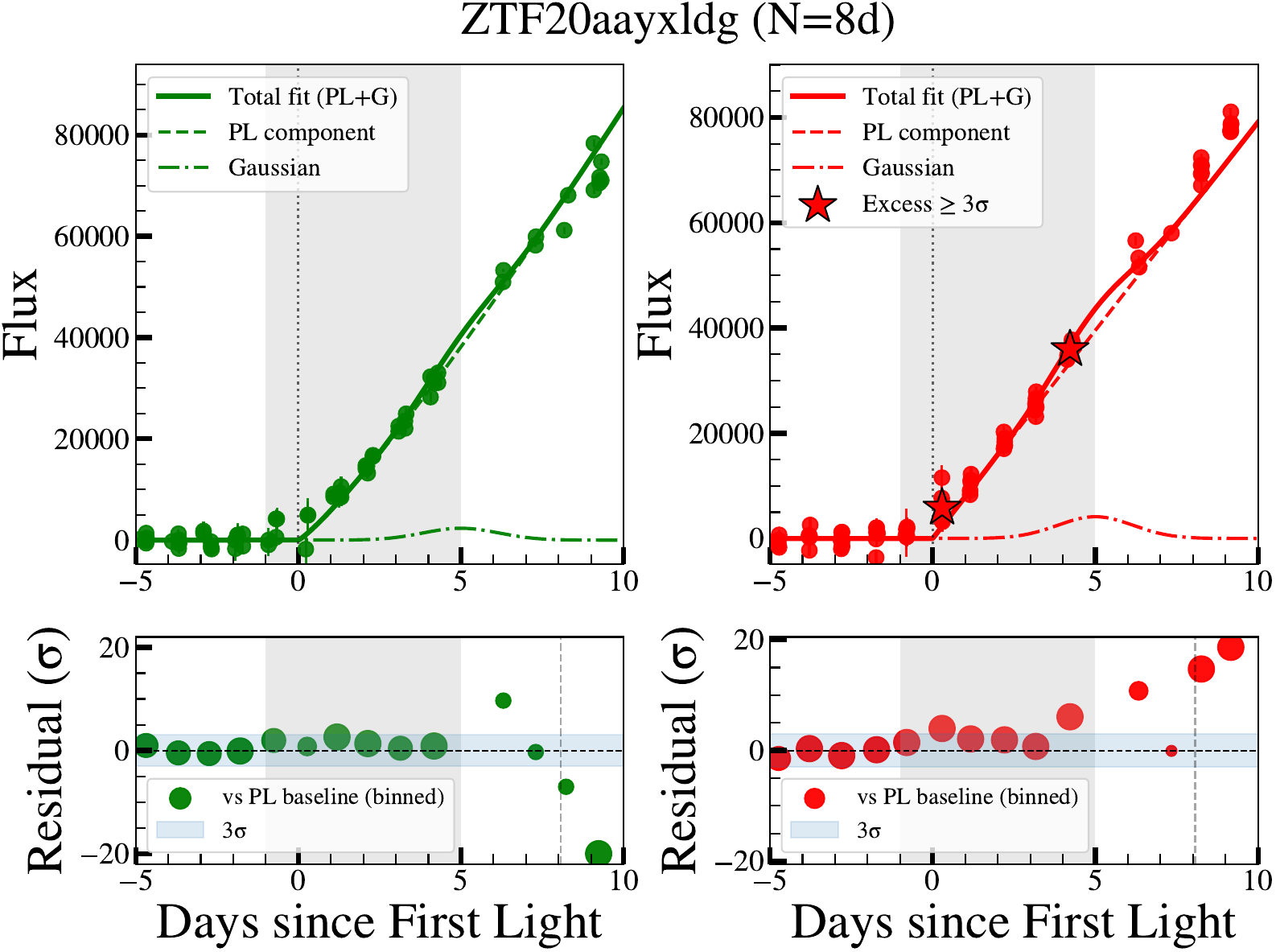} 
        \caption{}
        \label{subfig:c}
    \end{subfigure}
    \hfill
    \begin{subfigure}{0.49\textwidth}
        \centering
        \includegraphics[width=\linewidth]{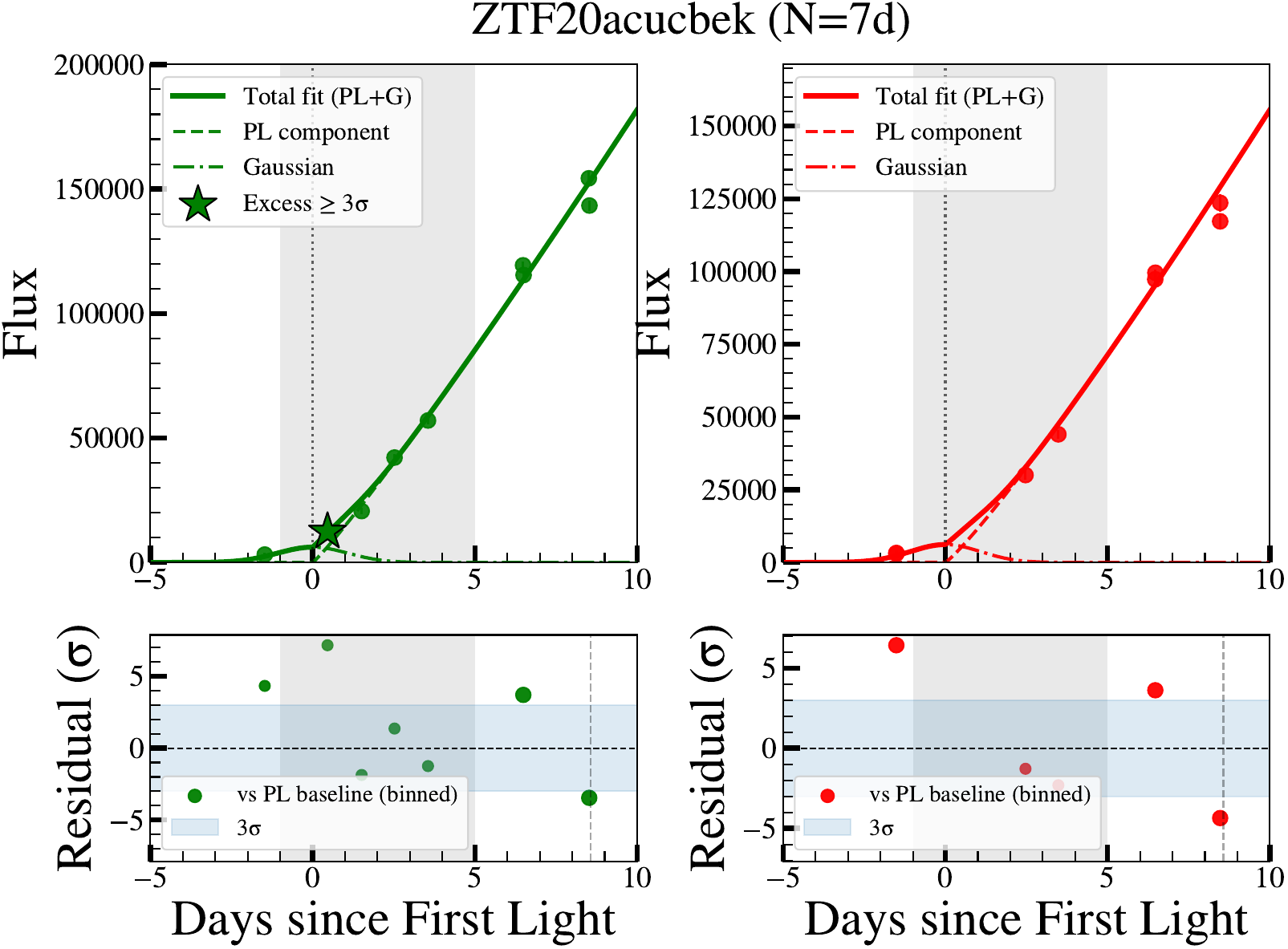} 
        \caption{}
        \label{subfig:d}
    \end{subfigure}
    
 \caption{Light-curve fits and residuals for the four bump candidates removed during visual vetting. These objects were excluded from our final sample due to indications of non-robust or contaminant-driven bumps (e.g., nuclear/host contamination, single-epoch spikes, large scatter, or a single, one-band-only excess).}

    \label{fig:removed}
\end{figure}

\bibliographystyle{raa}
\bibliography{sample701}

\end{document}

%% file: bumps_deluxetable.tex
\begin{table}[H]
\centering
\begin{minipage}{\textwidth}
\caption{Bump candidates with coordinates, redshift, and fit diagnostics.\label{tab:bumps_coords}}
\end{minipage}
\setlength{\tabcolsep}{6pt}
\small
\begin{tabular}{l r r r}
\hline\noalign{\smallskip}
ZTF name & $z$ & $N_{\rm cut}^{\rm best}$ & $\Delta{\rm BIC}$ \\
\hline\noalign{\smallskip}
ZTF18AAIYKOZ & 0.031 & 7 & 256 \\
ZTF18AASDTED & 0.018 & 7 & 888 \\
ZTF18AASLHXT & 0.055 & 7 & 40 \\
ZTF18AAYJVVE & 0.047 & 7 & 62 \\
ZTF18ABAUPRJ & 0.024 & 7 & 2540 \\
ZTF18ABCFLNZ & 0.027 & 7 & 54 \\
ZTF18ABEECWE & 0.039 & 7 & 67 \\
ZTF18ABFHRYC & 0.032 & 7 & 476 \\
ZTF18ABJVHEC & 0.057 & 7 & 8 \\
ZTF18ABPAYWM & 0.044 & 7 & 83 \\
ZTF18ABUCVBF & 0.055 & 7 & 95 \\
ZTF18ABUHYJV & 0.047 & 7 & 97 \\
ZTF18ABUHZFC & 0.038 & 7 & 40 \\
ZTF18ABUQUGW & 0.031 & 7 & 172 \\
ZTF18ACCNDXN & 0.031 & 7 & 82 \\
ZTF19AAWMQTD & 0.056 & 7 & 91 \\
ZTF19AAXFZAO & 0.048 & 7 & 84 \\
ZTF19AAYFAUM & 0.024 & 7 & 792 \\
ZTF19ABDOZNH & 0.027 & 7 & 112 \\
ZTF19ABGLMPF & 0.035 & 7 & 12 \\
ZTF19ABMYLXW & 0.045 & 7 & 27 \\
ZTF19ABVANIM & 0.060 & 7 & 29 \\
ZTF19ACCMBOE & 0.046 & 7 & 42 \\
ZTF20AAGHLKV & 0.030 & 7 & 198 \\
ZTF20AAHKGCZ & 0.041 & 7 & 153 \\
ZTF20AANYXVU & 0.047 & 7 & 36 \\
ZTF20AAOSOCT & 0.059 & 7 & 55 \\
ZTF20AAQPXTM & 0.052 & 7 & 51 \\
ZTF20AAWDJLP & 0.027 & 7 & 72 \\
ZTF20AAWQYWR & 0.050 & 7 & 16 \\
ZTF20AAZYMXI & 0.036 & 10 & 89 \\
ZTF20ABBHYXU & 0.032 & 8 & 80 \\
ZTF20ABELJUJ & 0.048 & 7 & 56 \\
ZTF20ABSVTNC & 0.033 & 7 & 44 \\
ZTF20ABSZZLP & 0.029 & 7 & 90 \\
ZTF20ABVOWVW & 0.053 & 7 & 17 \\
ZTF20ABWEEDO & 0.039 & 7 & 29 \\
ZTF20ABXGWJR & 0.040 & 7 & 35 \\
ZTF20ABYGIGY & 0.059 & 10 & 62 \\
ZTF20ACBOVRT & 0.056 & 7 & 19 \\
ZTF20ACJHHQX & 0.055 & 7 & 35 \\
ZTF20ACTDDOH & 0.015 & 7 & 108 \\
\noalign{\smallskip}\hline
\end{tabular}
\tablecomments{\textwidth}{$z$ is rounded to the nearest 0.001. $N_{\rm cut}^{\rm best}$ is the integer chosen from 7 to 13 that maximizes $\Delta{\rm BIC}$.}
\end{table}

%% file: nobumps_deluxetable_two_up_split_same_number.tex
\begin{table}[H]
\centering
\begin{minipage}{\textwidth}
\caption{No-bump candidates with redshift and fit diagnostics.\label{tab:nobumps}}
\end{minipage}
\setlength{\tabcolsep}{2.5pt}
\footnotesize
\begin{tabular}{l r r r @{\hspace{1.2em}} l r r r}
\hline\noalign{\smallskip}
ZTF name & $z$ & $N_{\rm cut}^{\rm best}$ & $\Delta{\rm BIC}$ & ZTF name & $z$ & $N_{\rm cut}^{\rm best}$ & $\Delta{\rm BIC}$ \\
\hline\noalign{\smallskip}
ZTF18AAHSHHP & 0.060 & 7 & -7 & ZTF19ABXDNZT & 0.049 & 10 & -14 \\
ZTF18AASPRUI & 0.038 & 8 & -13 & ZTF19ABXFUNP & 0.039 & 7 & -7 \\
ZTF18AAXCNTM & 0.027 & 7 & -6 & ZTF19ABZULHK & 0.050 & 7 & 1 \\
ZTF18AAYBFSD & 0.040 & 7 & -9 & ZTF19ACBJLNT & 0.057 & 11 & -7 \\
ZTF18AAZIXBW & 0.059 & 7 & -12 & ZTF19ACBPZIR & 0.055 & 10 & -1 \\
ZTF18ABDCFGZ & 0.058 & 9 & -9 & ZTF19ACEKREH & 0.035 & 11 & -8 \\
ZTF18ABDFAQI & 0.048 & 13 & 29 & ZTF19ACGQJLN & 0.049 & 9 & -9 \\
ZTF18ABDRMIN & 0.052 & 13 & 20 & ZTF19ACIHFXZ & 0.055 & 7 & -13 \\
ZTF18ABJHACK & 0.046 & 7 & 2 & ZTF19ACIHGNG & 0.058 & 10 & -9 \\
ZTF18ABLQLZP & 0.042 & 13 & 33 & ZTF19ACNJWGM & 0.034 & 7 & -10 \\
ZTF18ABNCIMO & 0.057 & 10 & 182 & ZTF19ACRDQEJ & 0.047 & 7 & 1 \\
ZTF18ABOTDEF & 0.060 & 7 & 2 & ZTF19ACWIQRC & 0.048 & 10 & -11 \\
ZTF18ABUPGQL & 0.046 & 7 & -8 & ZTF20AAEOPVV & 0.058 & 10 & -6 \\
ZTF18ABYKSWP & 0.036 & 7 & -5 & ZTF20AAFJJQJ & 0.041 & 8 & -7 \\
ZTF18ACAHUPH & 0.048 & 7 & -15 & ZTF20AAHDWZE & 0.049 & 9 & 34 \\
ZTF18ACFWMQJ & 0.056 & 9 & -9 & ZTF20AAKODIQ & 0.045 & 7 & -8 \\
ZTF18ACHBLLM & 0.047 & 10 & -6 & ZTF20AAKYOEZ & 0.041 & 7 & 0 \\
ZTF18ACHJCWV & 0.054 & 9 & 2 & ZTF20AAMKYYX & 0.050 & 7 & -6 \\
ZTF18ACMZPBF & 0.036 & 7 & -10 & ZTF20AAODKVL & 0.025 & 11 & -14 \\
ZTF18ACPFWMM & 0.049 & 9 & -2 & ZTF20AAOOJLI & 0.051 & 7 & -6 \\
ZTF18ACRDCIW & 0.052 & 7 & -3 & ZTF20AAQHTCD & 0.058 & 8 & -11 \\
ZTF18ACRDMMW & 0.050 & 7 & -8 & ZTF20AASFDFR & 0.055 & 13 & -13 \\
ZTF18ACVBIFF & 0.033 & 9 & -6 & ZTF20AAZNPOR & 0.040 & 13 & -8 \\
ZTF18ACXYARG & 0.041 & 7 & -12 & ZTF20AAZQUHC & 0.034 & 13 & -3 \\
ZTF19AABYPPP & 0.023 & 7 & -7 & ZTF20ABCQPDJ & 0.053 & 7 & -7 \\
ZTF19AAKLQOD & 0.030 & 7 & -7 & ZTF20ABCSJTX & 0.031 & 8 & -6 \\
ZTF19AANUHOX & 0.060 & 11 & -10 & ZTF20ABDSUNU & 0.055 & 7 & -4 \\
ZTF19AAPSZZY & 0.058 & 7 & -2 & ZTF20ABEEYTG & 0.043 & 7 & -12 \\
ZTF19AAQHELQ & 0.057 & 8 & -7 & ZTF20ABIPUEE & 0.057 & 7 & -16 \\
ZTF19AARHXGN & 0.033 & 9 & -8 & ZTF20ABLEIEM & 0.016 & 11 & 15 \\
ZTF19AATGWHS & 0.041 & 11 & -20 & ZTF20ABMOANS & 0.031 & 7 & -3 \\
ZTF19AAURSTB & 0.051 & 7 & -17 & ZTF20ABPYXJA & 0.043 & 10 & -12 \\
ZTF19AAWMKJA & 0.040 & 7 & -11 & ZTF20ABQCVTK & 0.053 & 7 & -7 \\
ZTF19AAWQCKX & 0.035 & 7 & -10 & ZTF20ABQMTSH & 0.025 & 7 & -8 \\
ZTF19ABASXKG & 0.039 & 7 & -8 & ZTF20ABQUPPG & 0.057 & 7 & -12 \\
ZTF19ABBVZGR & 0.050 & 7 & -2 & ZTF20ABTHXFR & 0.040 & 7 & -10 \\
ZTF19ABGJLEF & 0.058 & 7 & -12 & ZTF20ABUCJSA & 0.045 & 7 & -14 \\
ZTF19ABGPRPQ & 0.051 & 7 & -3 & ZTF20ABWAVAB & 0.050 & 10 & -16 \\
ZTF19ABITBCJ & 0.044 & 9 & -16 & ZTF20ABWCNHM & 0.047 & 7 & -8 \\
ZTF19ABLOVOT & 0.041 & 7 & -14 & ZTF20ABWKTFX & 0.027 & 12 & 73 \\
ZTF19ABMZORD & 0.058 & 7 & -3 & ZTF20ABXLHAM & 0.055 & 7 & -16 \\
ZTF19ABORNYN & 0.059 & 7 & -5 & ZTF20ABXZRQW & 0.044 & 11 & -5 \\
ZTF19ABPGGGU & 0.049 & 7 & 1 & ZTF20ACFVIAC & 0.036 & 7 & -3 \\
ZTF19ABQSVGF & 0.054 & 11 & -16 & ZTF20ACGDYCD & 0.049 & 11 & 3 \\
ZTF19ABXDNHR & 0.055 & 13 & 5 & ZTF20ACHDWMQ & 0.059 & 7 & -10 \\
\noalign{\smallskip}\hline
\end{tabular}
\tablecomments{\textwidth}{$z$ is rounded to 0.001. $N_{\rm cut}^{\rm best}$ is the integer chosen from 7 to 13 that maximizes $\Delta{\rm BIC}$.}
\end{table}
\addtocounter{table}{-1}
\begin{table}[H]
\centering
\begin{minipage}{\textwidth}
\caption{No-bump candidates with redshift and fit diagnostics (continued).}
\end{minipage}
\setlength{\tabcolsep}{2.5pt}
\footnotesize
\begin{tabular}{l r r r @{\hspace{1.2em}} l r r r}
\hline\noalign{\smallskip}
ZTF name & $z$ & $N_{\rm cut}^{\rm best}$ & $\Delta{\rm BIC}$ & ZTF name & $z$ & $N_{\rm cut}^{\rm best}$ & $\Delta{\rm BIC}$ \\
\hline\noalign{\smallskip}
ZTF20ACHZUGY & 0.056 & 9 & -11 & ZTF20ACUDKAO & 0.038 & 13 & 5 \\
ZTF20ACIKUON & 0.023 & 7 & -2 & ZTF20ACUZFPK & 0.057 & 7 & -11 \\
ZTF20ACJJKHR & 0.039 & 9 & -1 & ZTF20ACWKVGJ & 0.054 & 7 & -4 \\
ZTF20ACMZOXO & 0.029 & 8 & -9 & ZTF20ACXNCHA & 0.046 & 7 & -10 \\
ZTF20ACOJXTB & 0.056 & 8 & -8 & ZTF20ACXVVNF & 0.055 & 7 & -15 \\
ZTF20ACOYYLY & 0.048 & 7 & -10 & ZTF20ACYVZBR & 0.055 & 7 & -7 \\
ZTF20ACPCUWX & 0.055 & 7 & -10 & ZTF20ACYWEFL & 0.059 & 7 & -6 \\
ZTF20ACRQEJE & 0.059 & 7 & -11 & ZTF20ACYYAVB & 0.048 & 13 & -9 \\
ZTF20ACRUDZK & 0.039 & 7 & -11 & ZTF20ACYZSOL & 0.058 & 7 & -9 \\
ZTF20ACTSKCF & 0.029 & 9 & -12 & ZTF20ADADFFG & 0.057 & 7 & -15 \\
\noalign{\smallskip}\hline
\end{tabular}
\end{table}

%% file: appendix_bumps_figures.tex
\begin{figure}[H]
\centering
\includegraphics[width=0.48\textwidth]{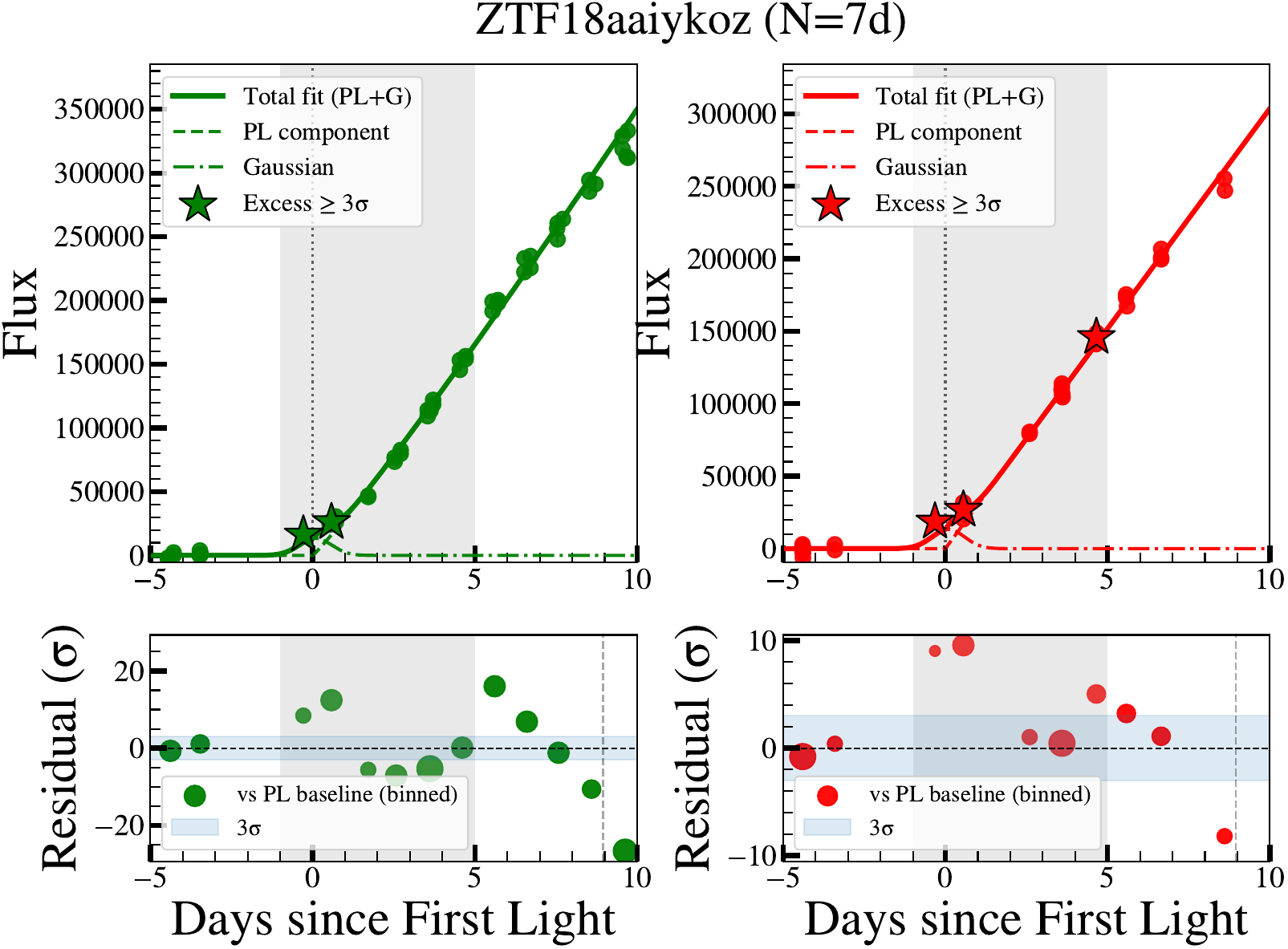}\hfill
\includegraphics[width=0.48\textwidth]{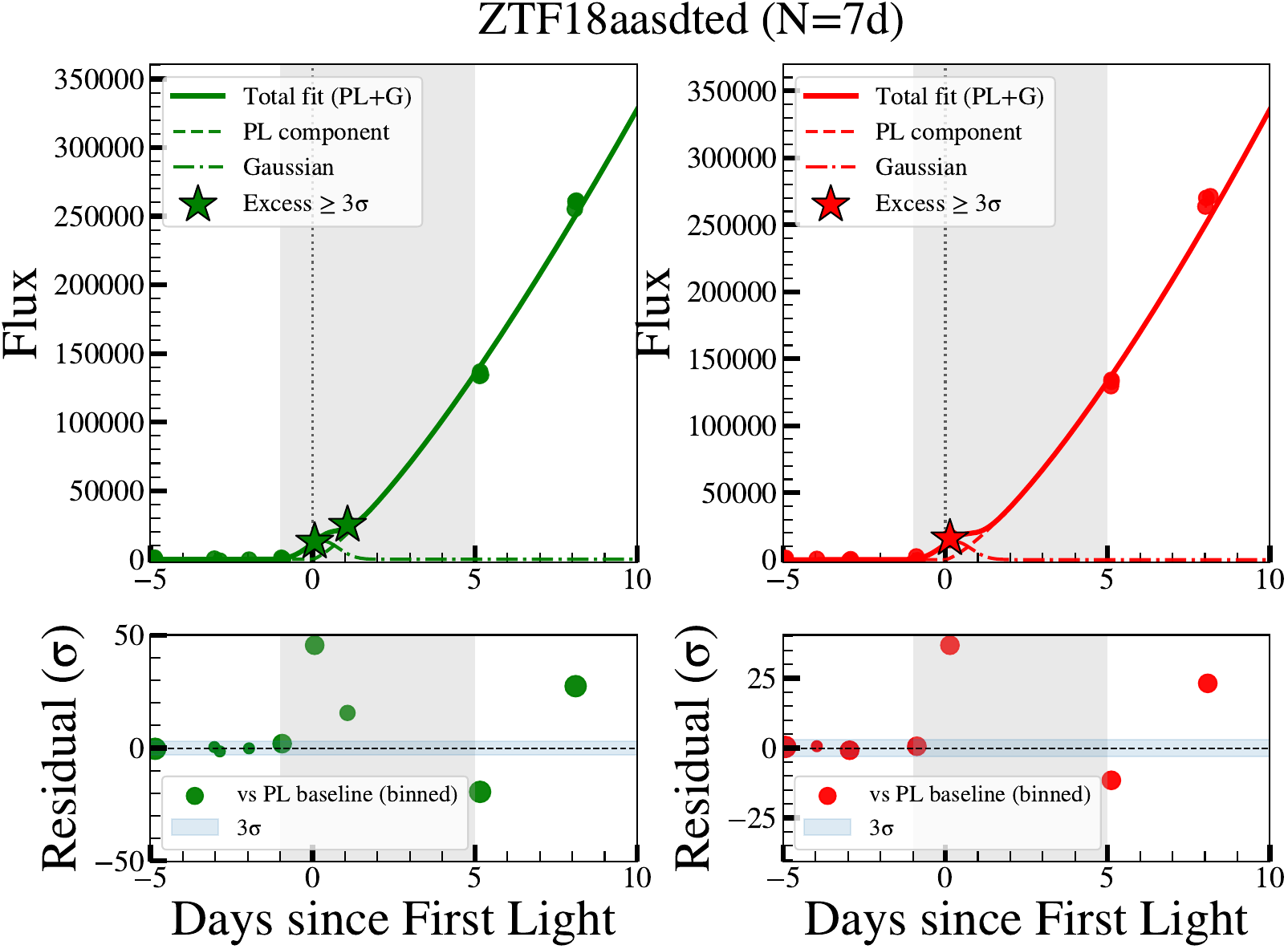}
\vspace{0.1cm}

\includegraphics[width=0.48\textwidth]{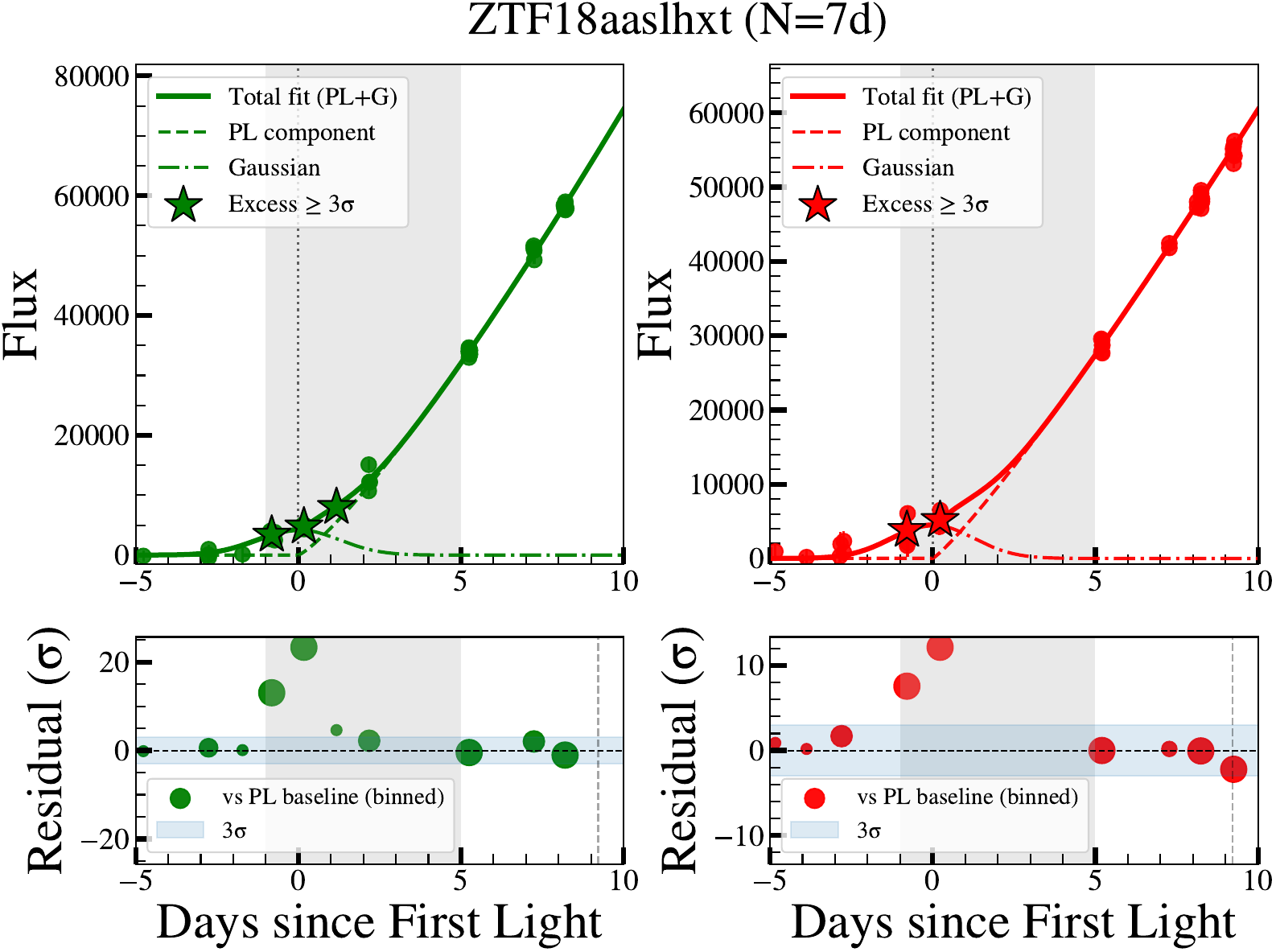}\hfill
\includegraphics[width=0.48\textwidth]{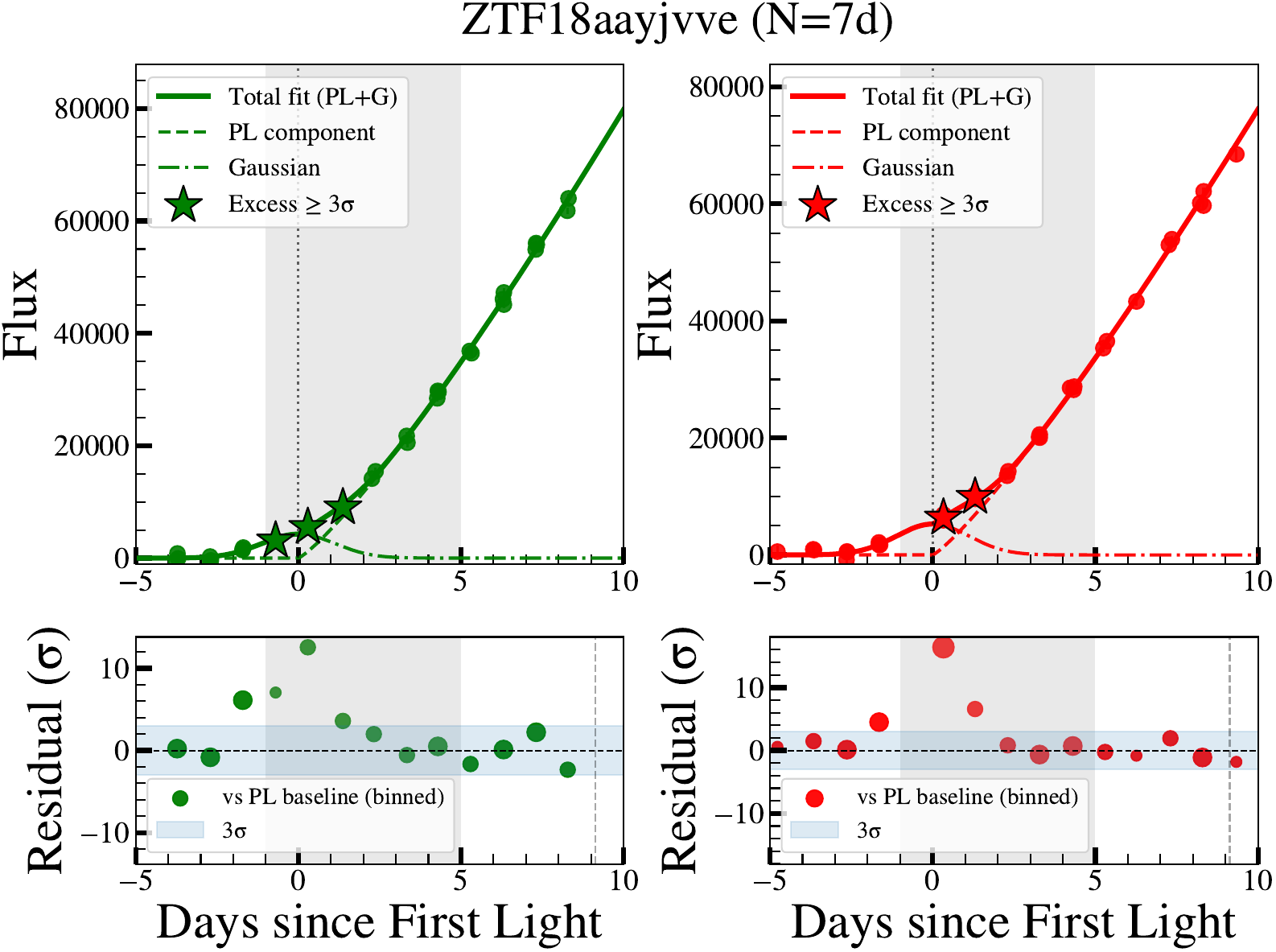}
\vspace{0.1cm}

\includegraphics[width=0.48\textwidth]{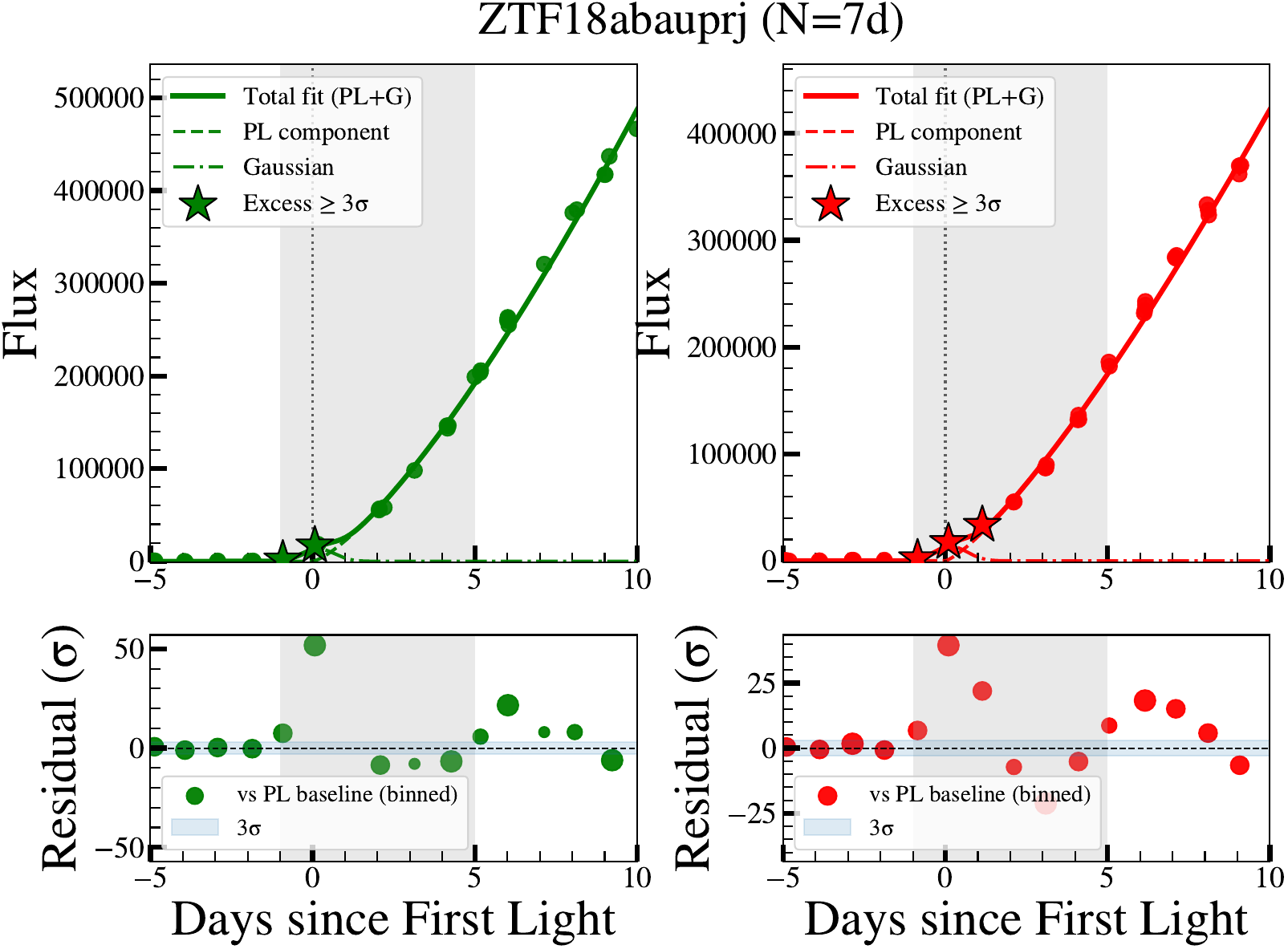}\hfill
\includegraphics[width=0.48\textwidth]{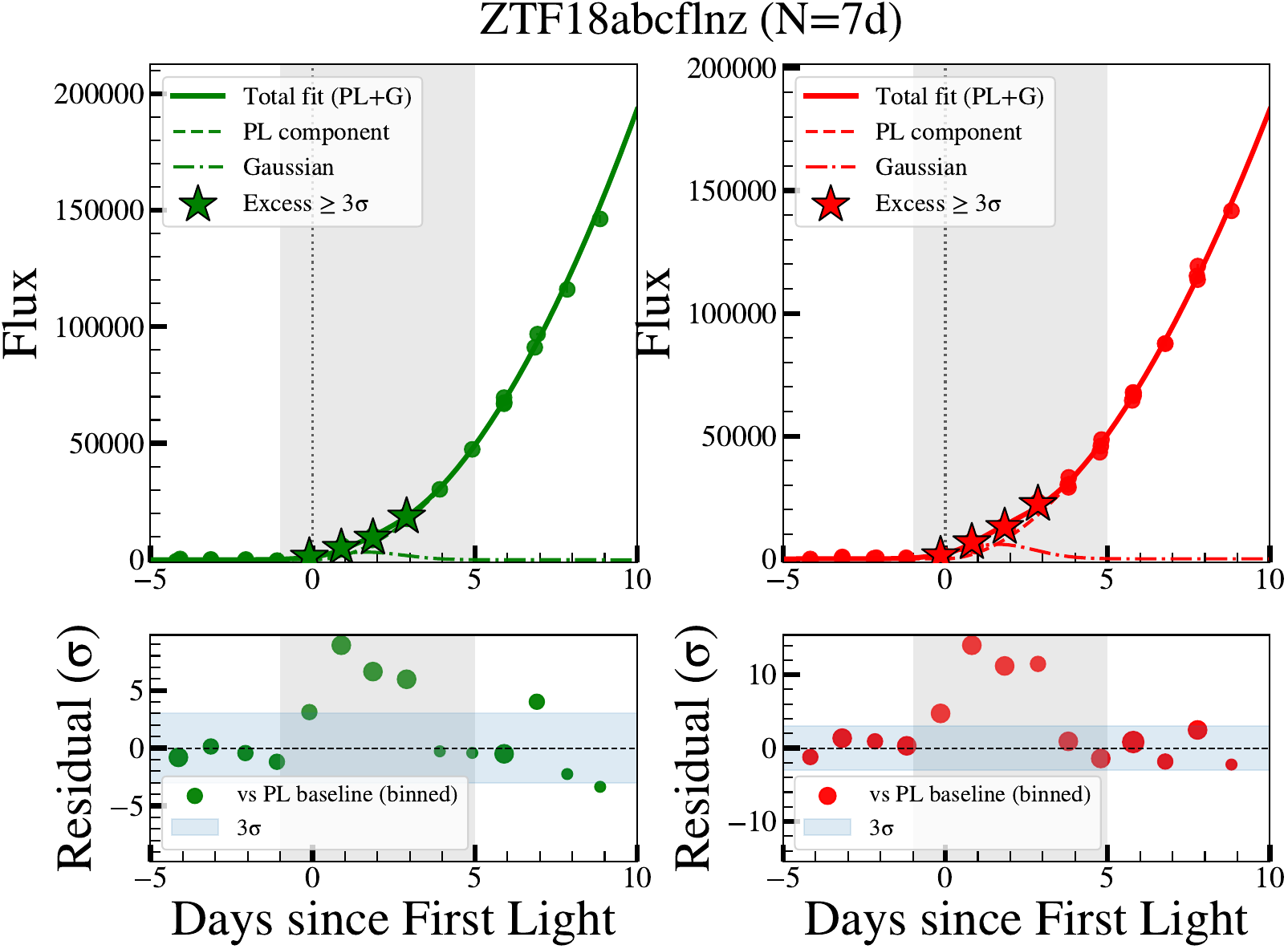}
\vspace{0.1cm}

\includegraphics[width=0.48\textwidth]{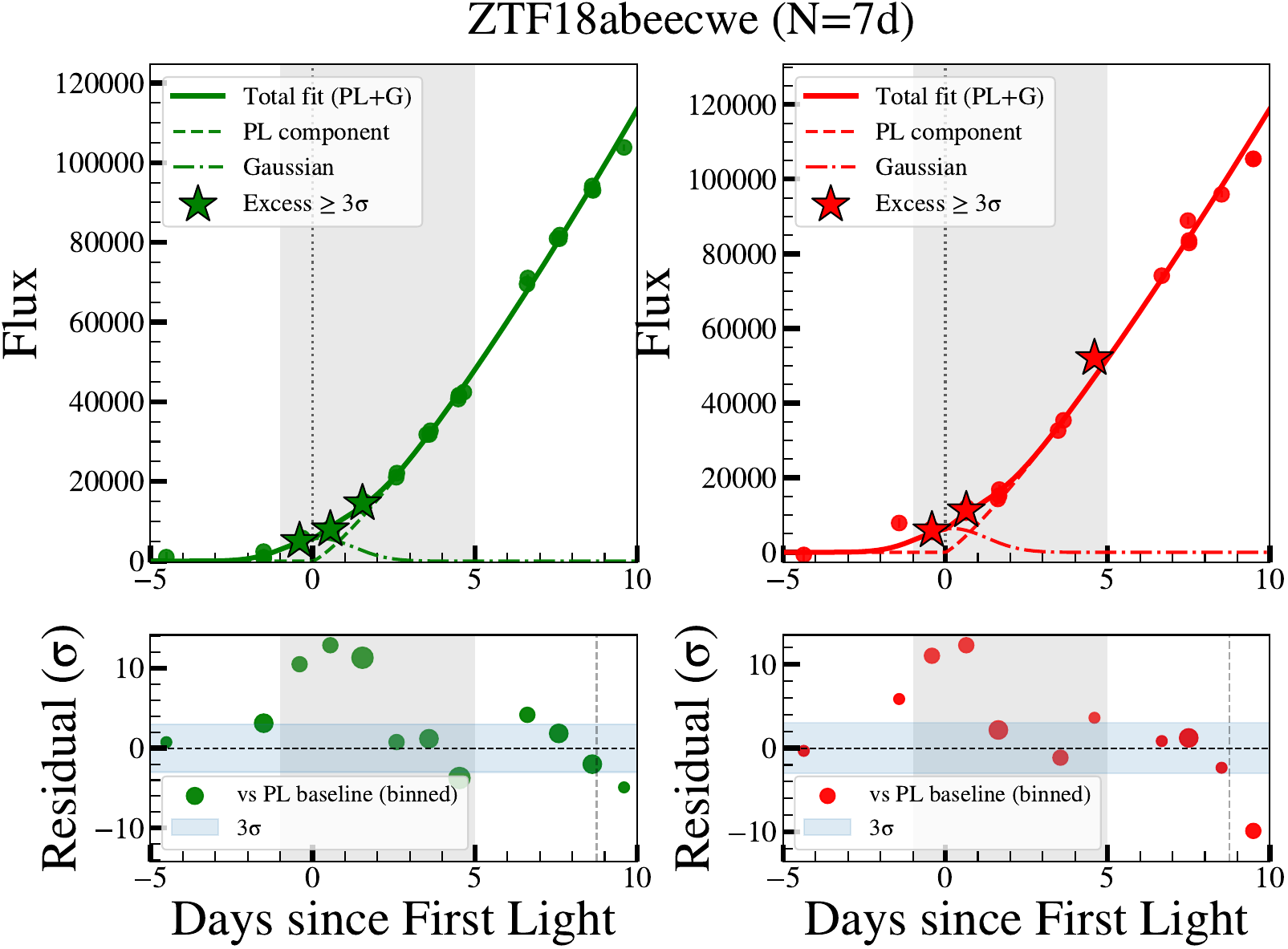}\hfill
\includegraphics[width=0.48\textwidth]{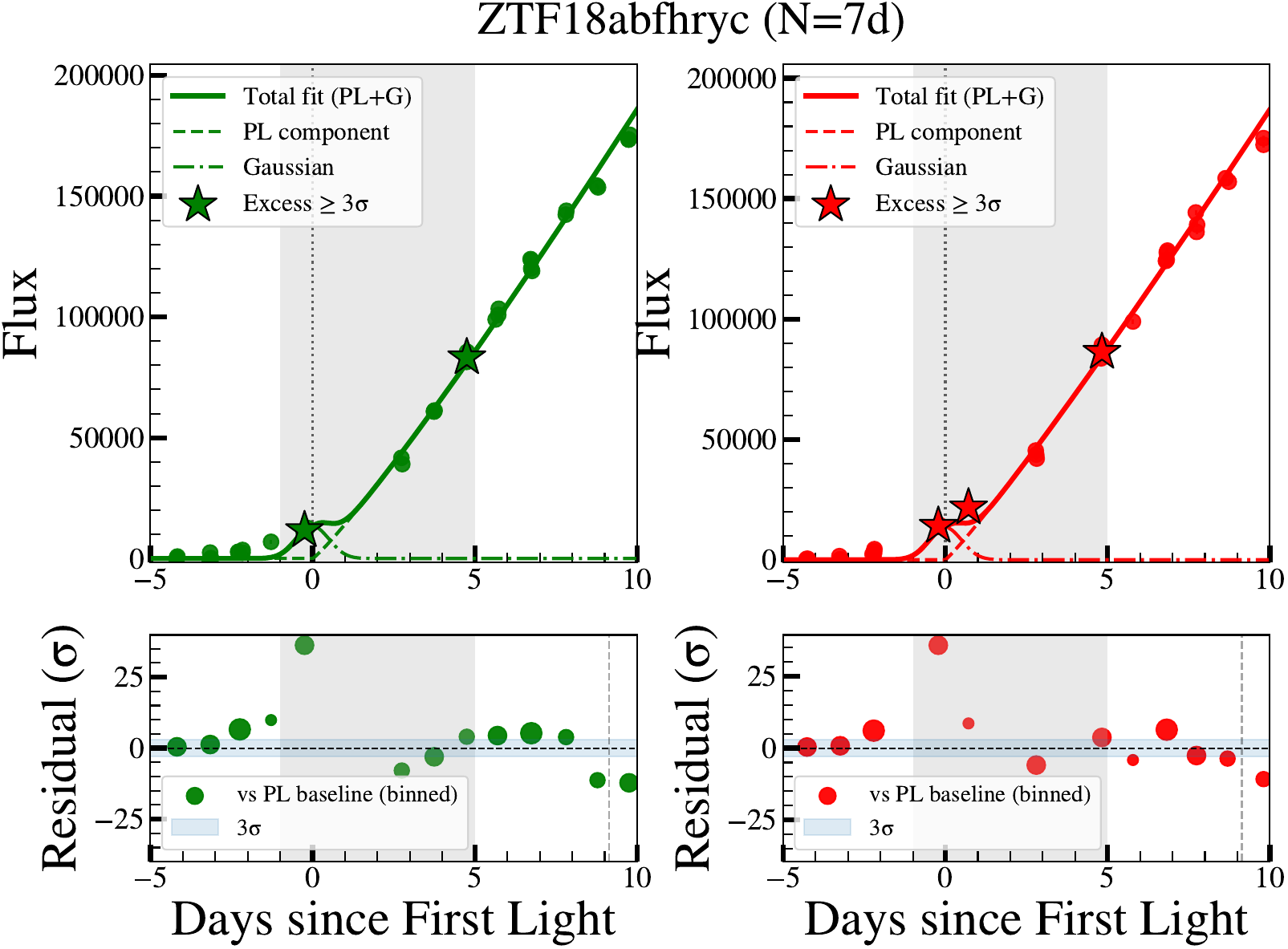}
\caption{All bump-candidate light curves.}
\label{fig:all_bumps_grid_1}
\end{figure}

\begin{figure}[H]
\centering
\includegraphics[width=0.48\textwidth]{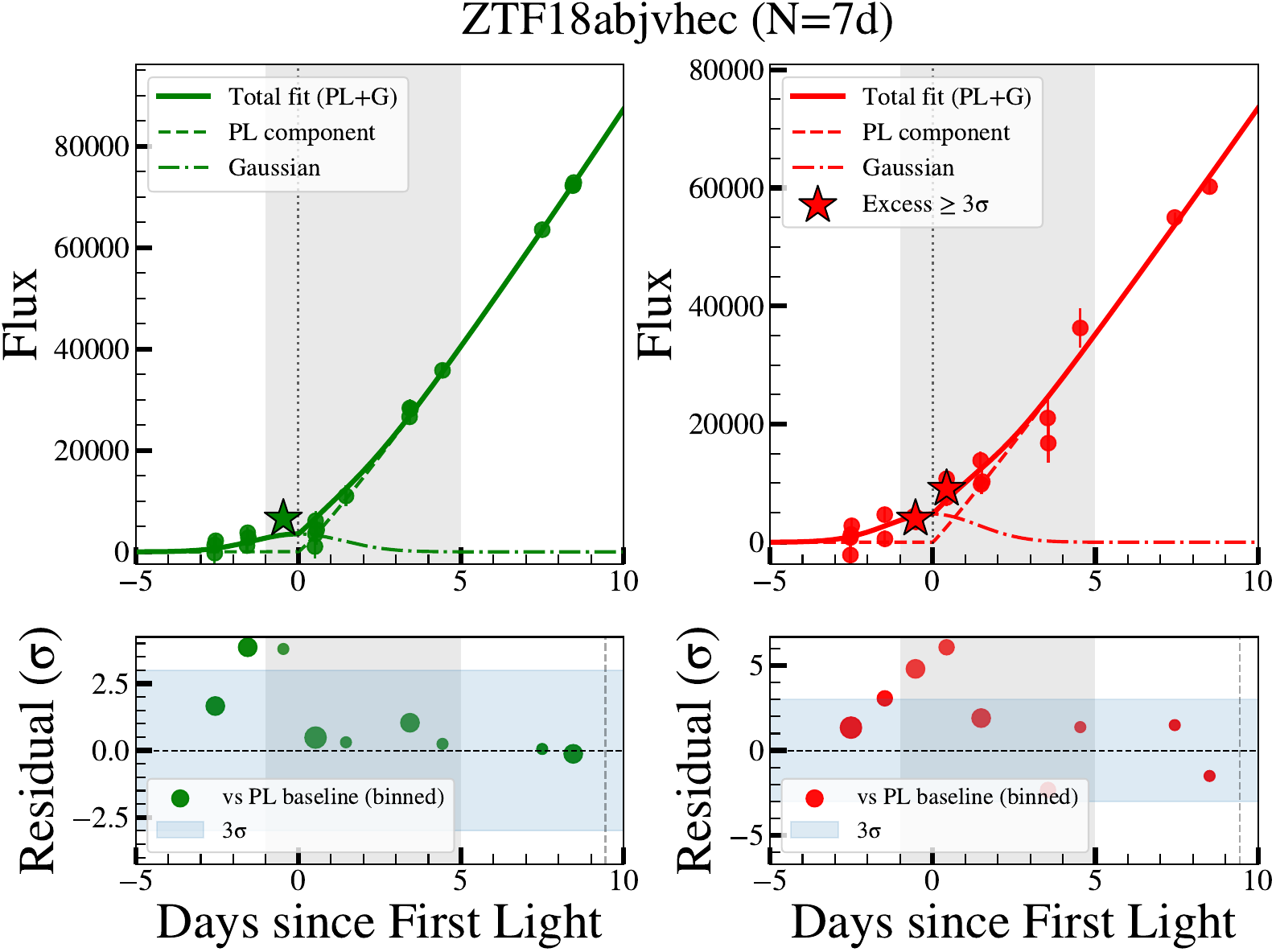}\hfill
\includegraphics[width=0.48\textwidth]{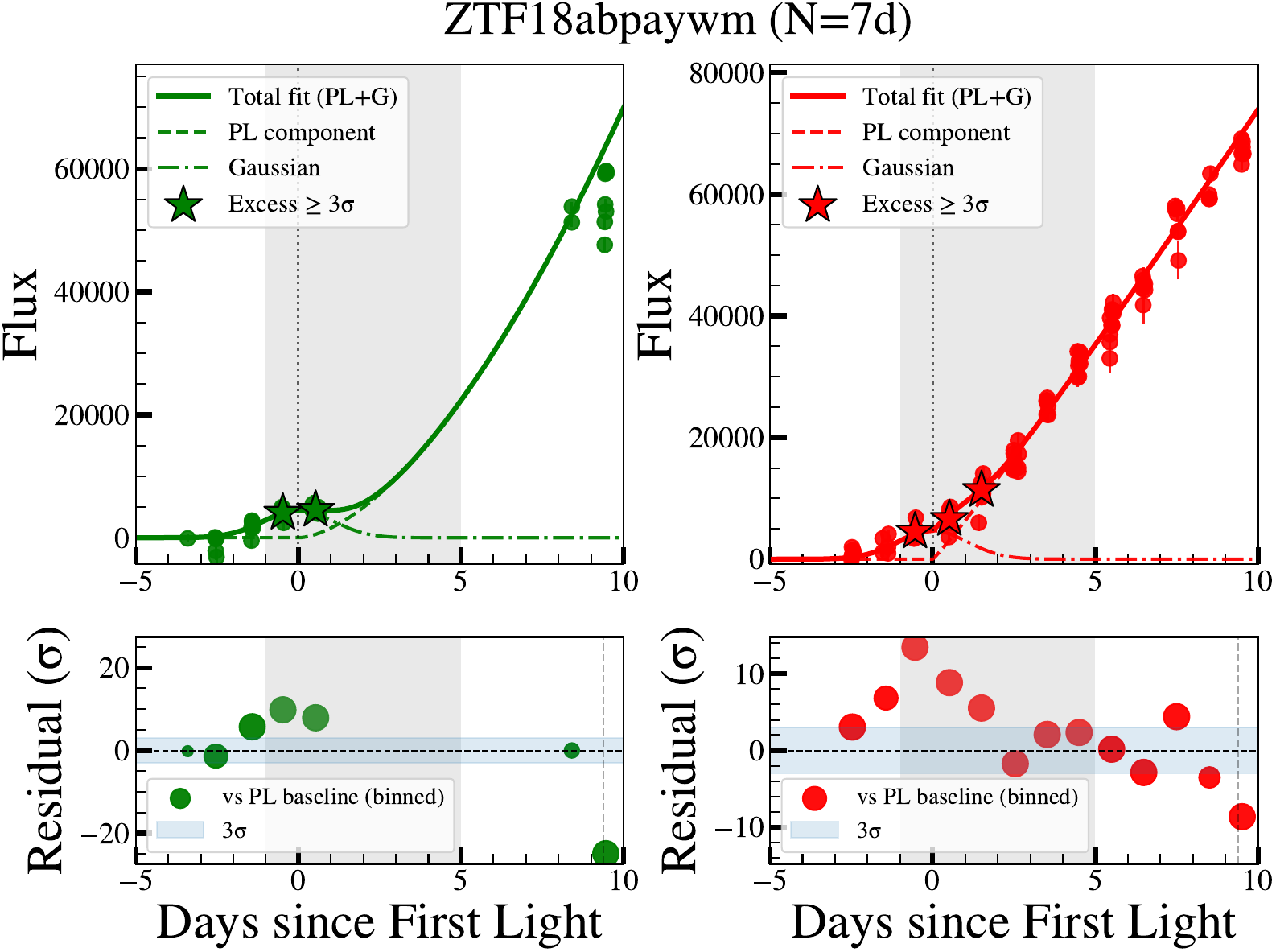}
\vspace{0.1cm}

\includegraphics[width=0.48\textwidth]{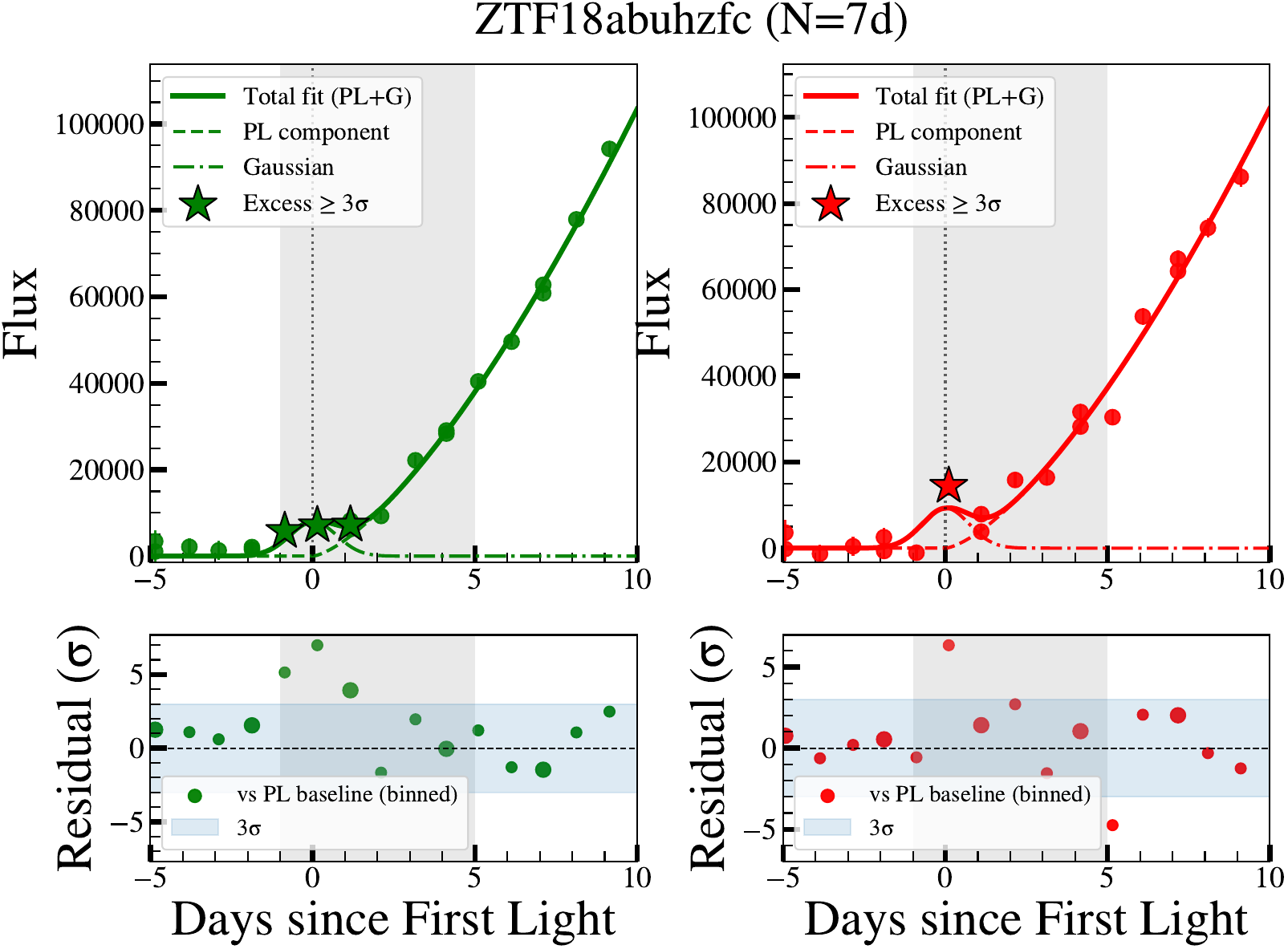}\hfill
\includegraphics[width=0.48\textwidth]{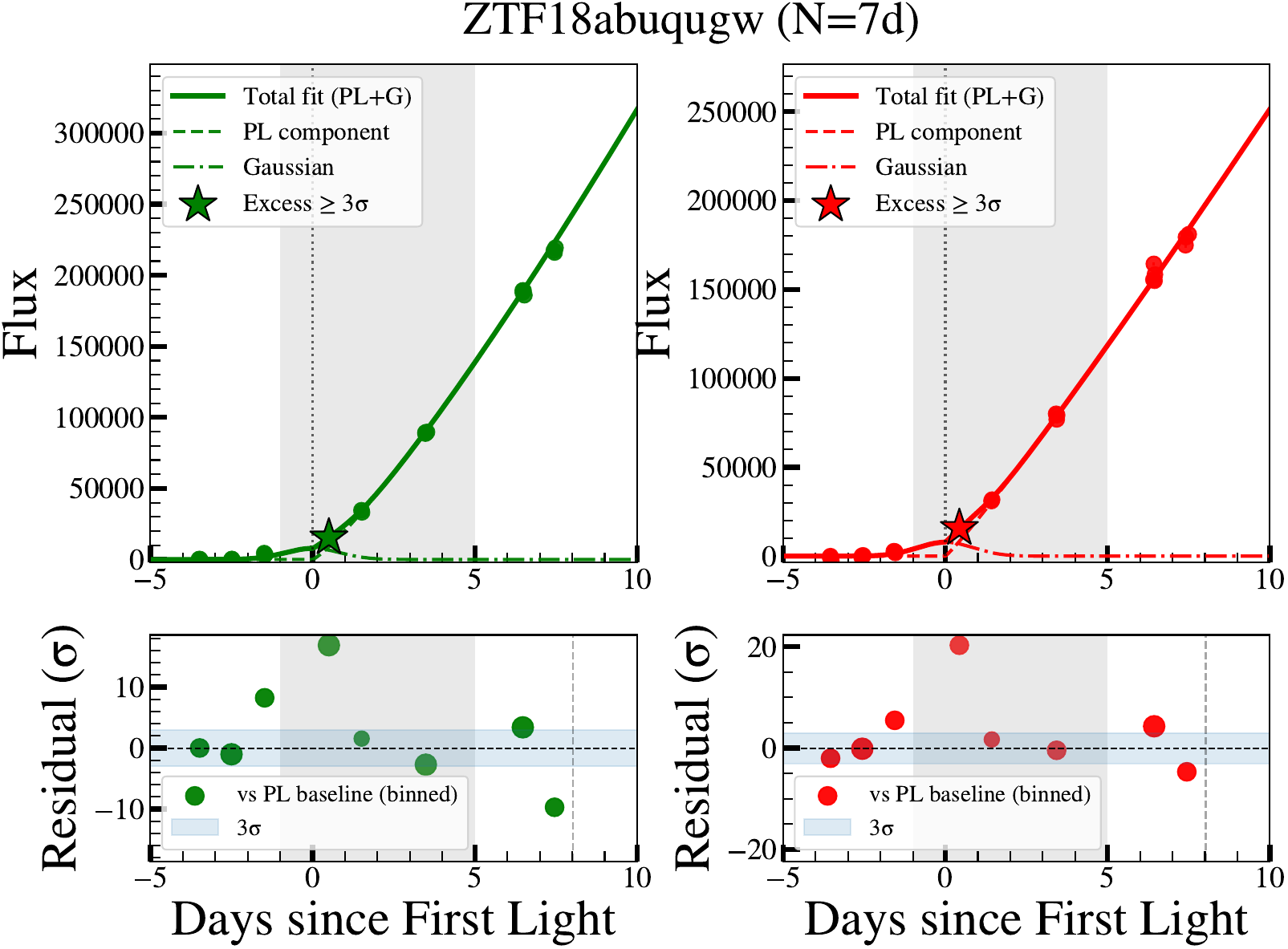}
\vspace{0.1cm}

\includegraphics[width=0.48\textwidth]{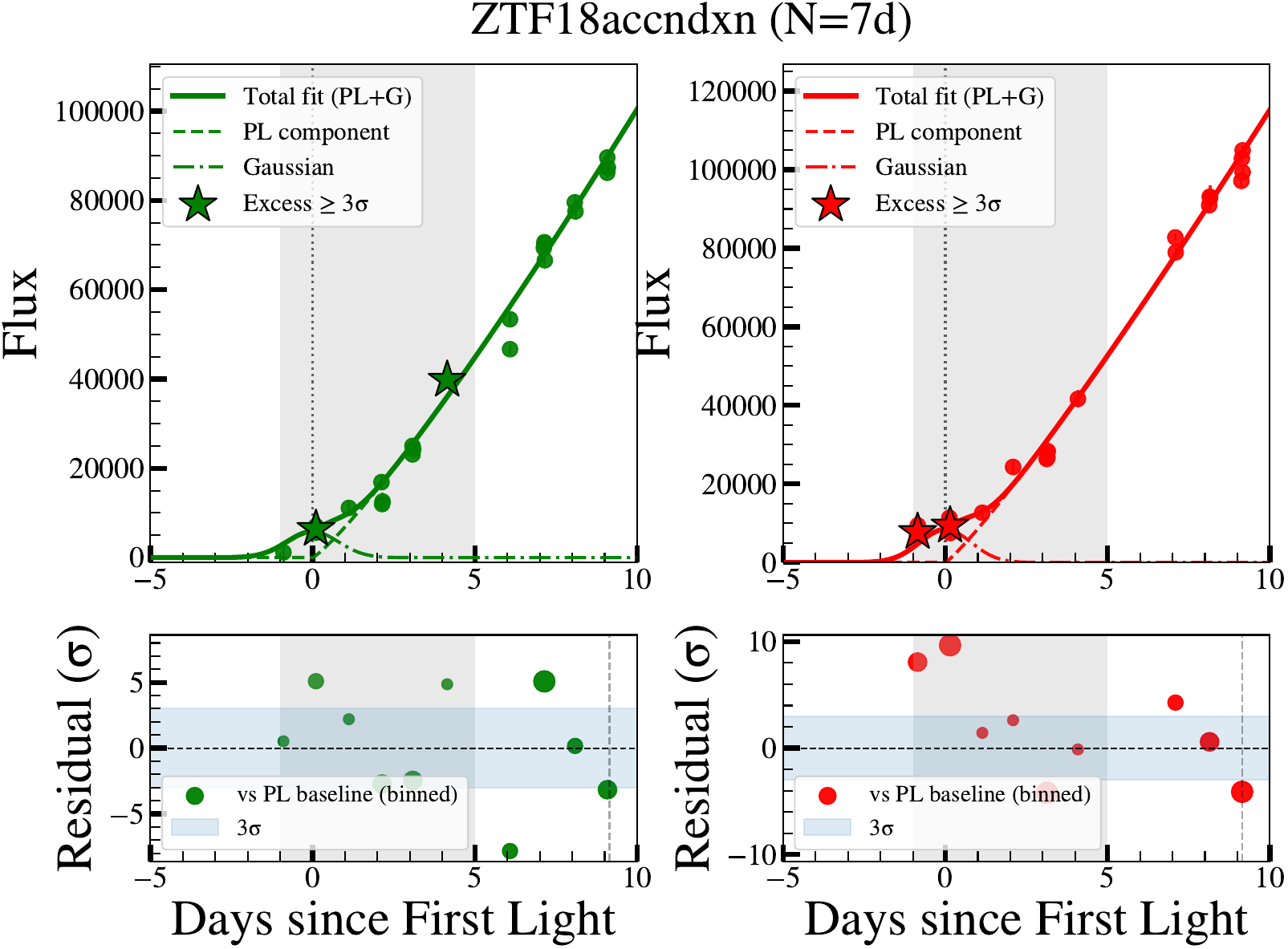}\hfill
\includegraphics[width=0.48\textwidth]{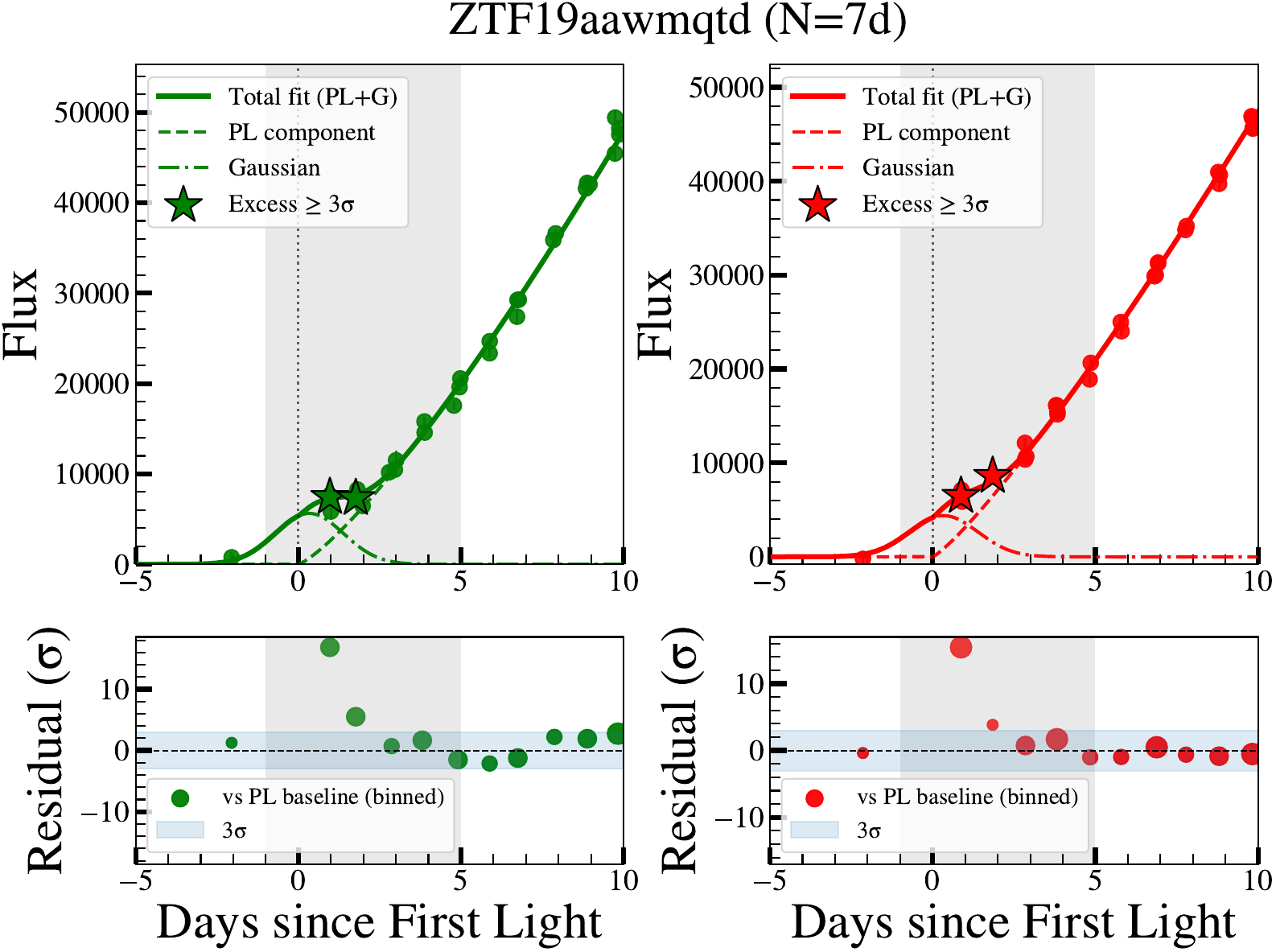}
\vspace{0.1cm}

\includegraphics[width=0.48\textwidth]{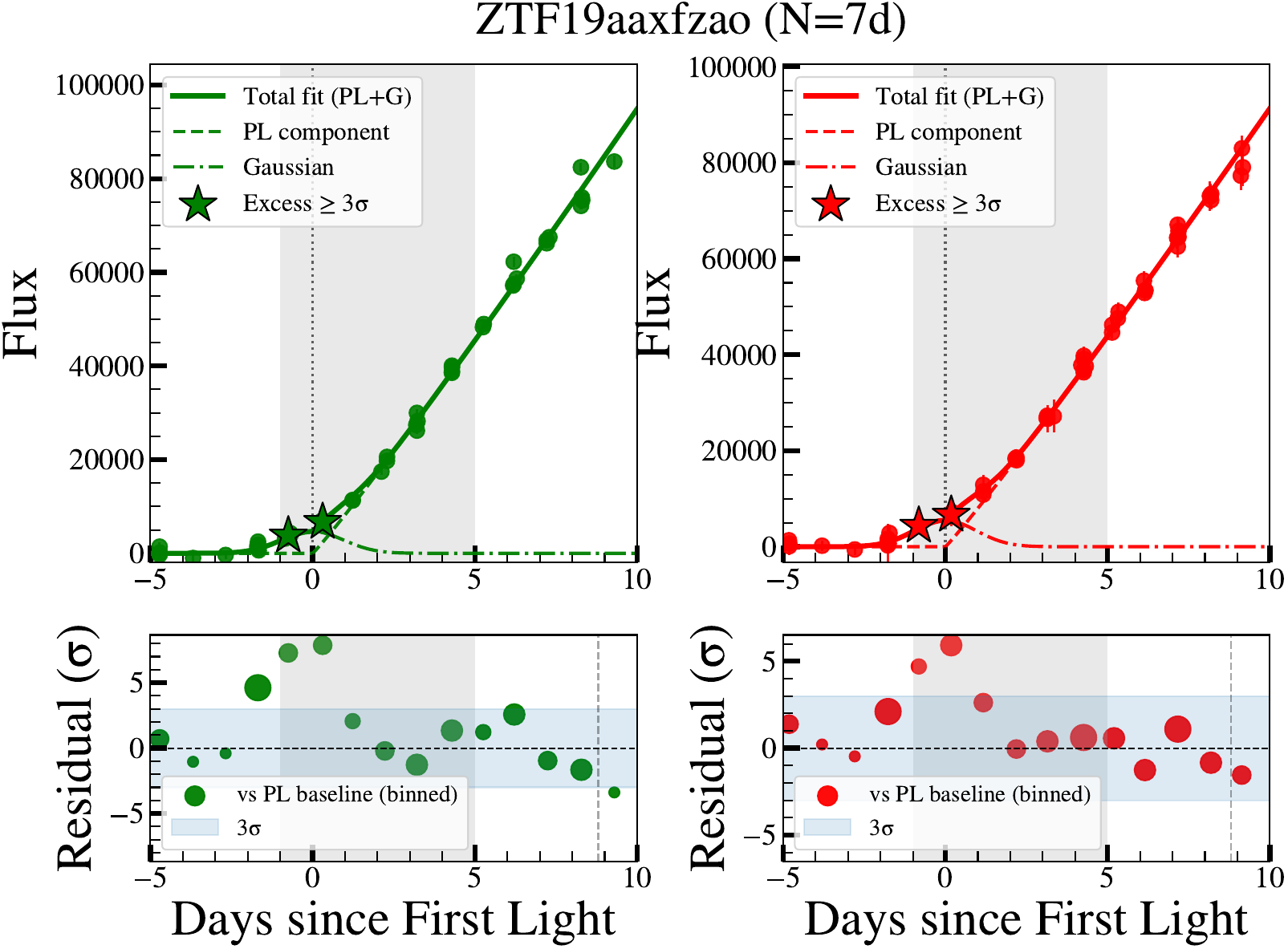}\hfill
\includegraphics[width=0.48\textwidth]{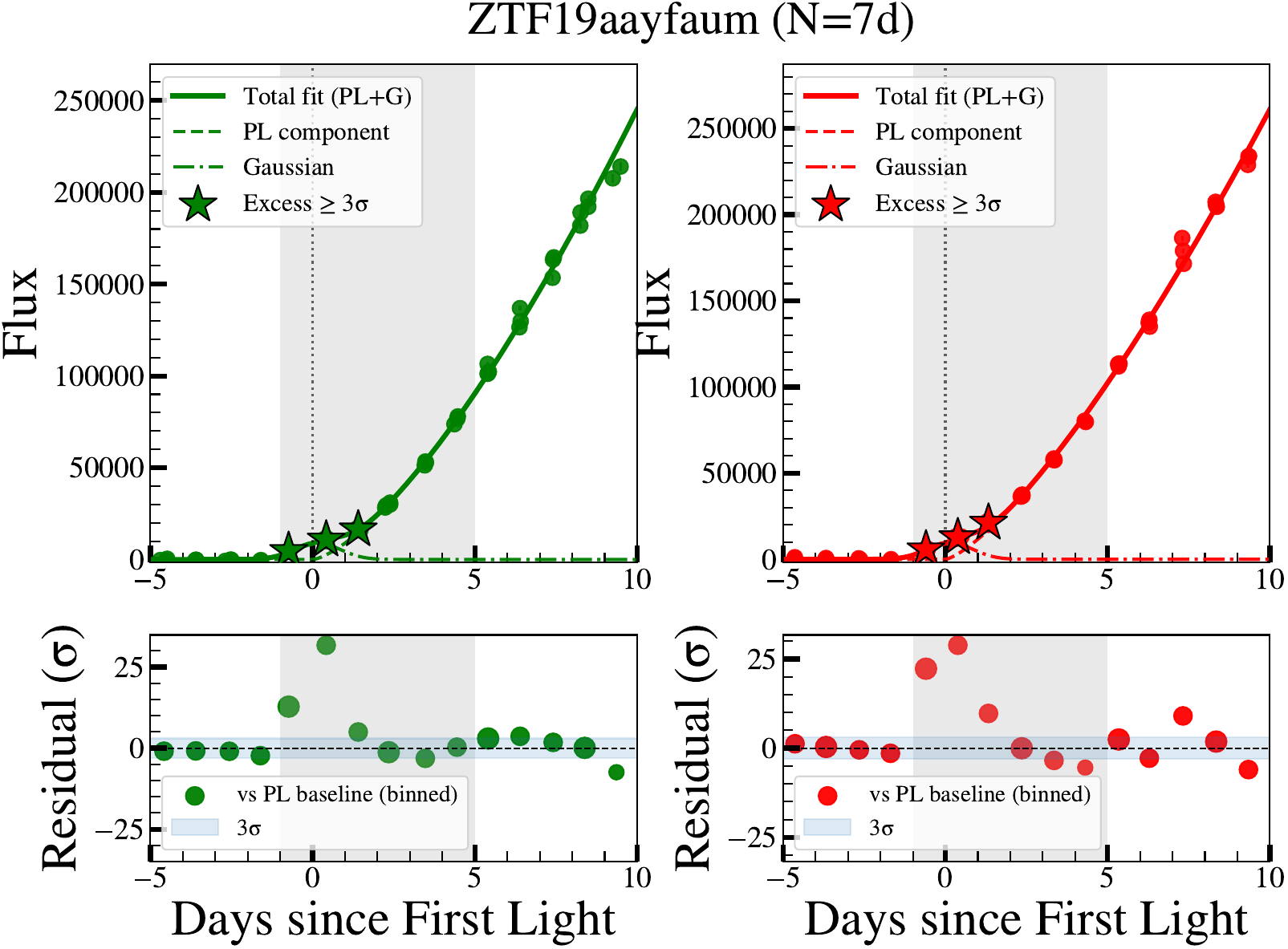}
\caption{Figure~\ref{fig:all_bumps_grid_1}. Continued.}
\end{figure}

\begin{figure}[H]
\centering
\includegraphics[width=0.48\textwidth]{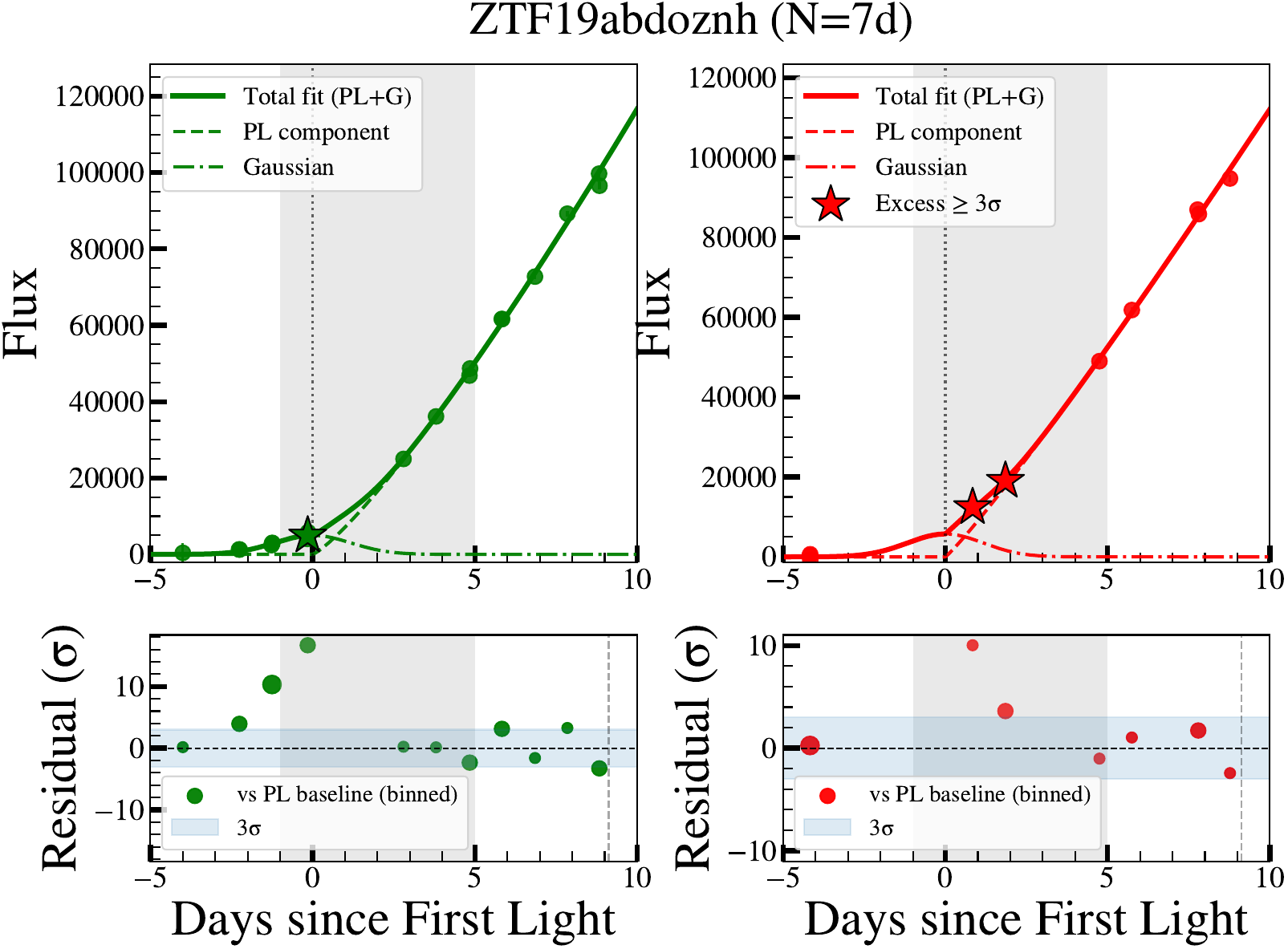}\hfill
\includegraphics[width=0.48\textwidth]{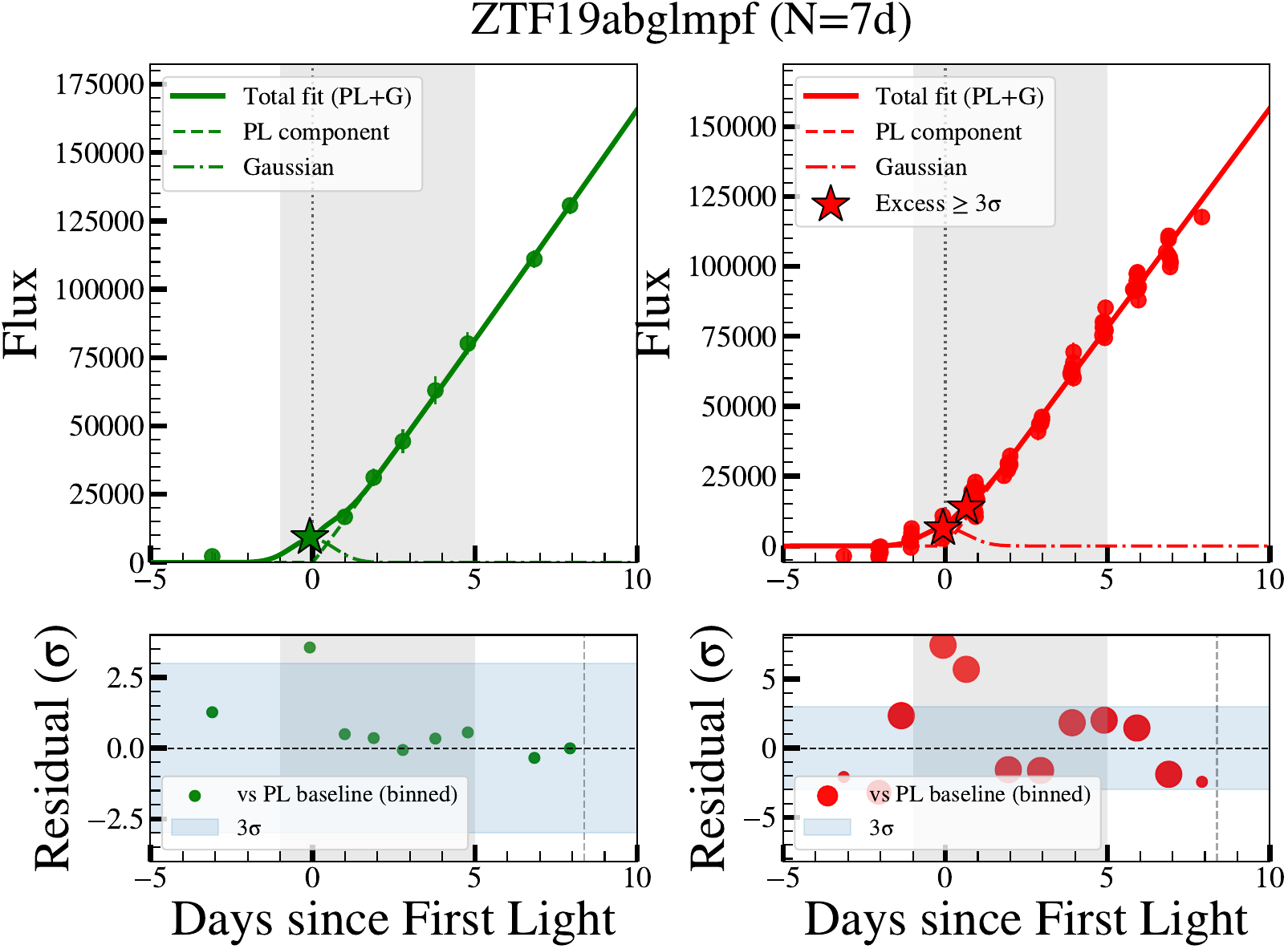}
\vspace{0.1cm}

\includegraphics[width=0.48\textwidth]{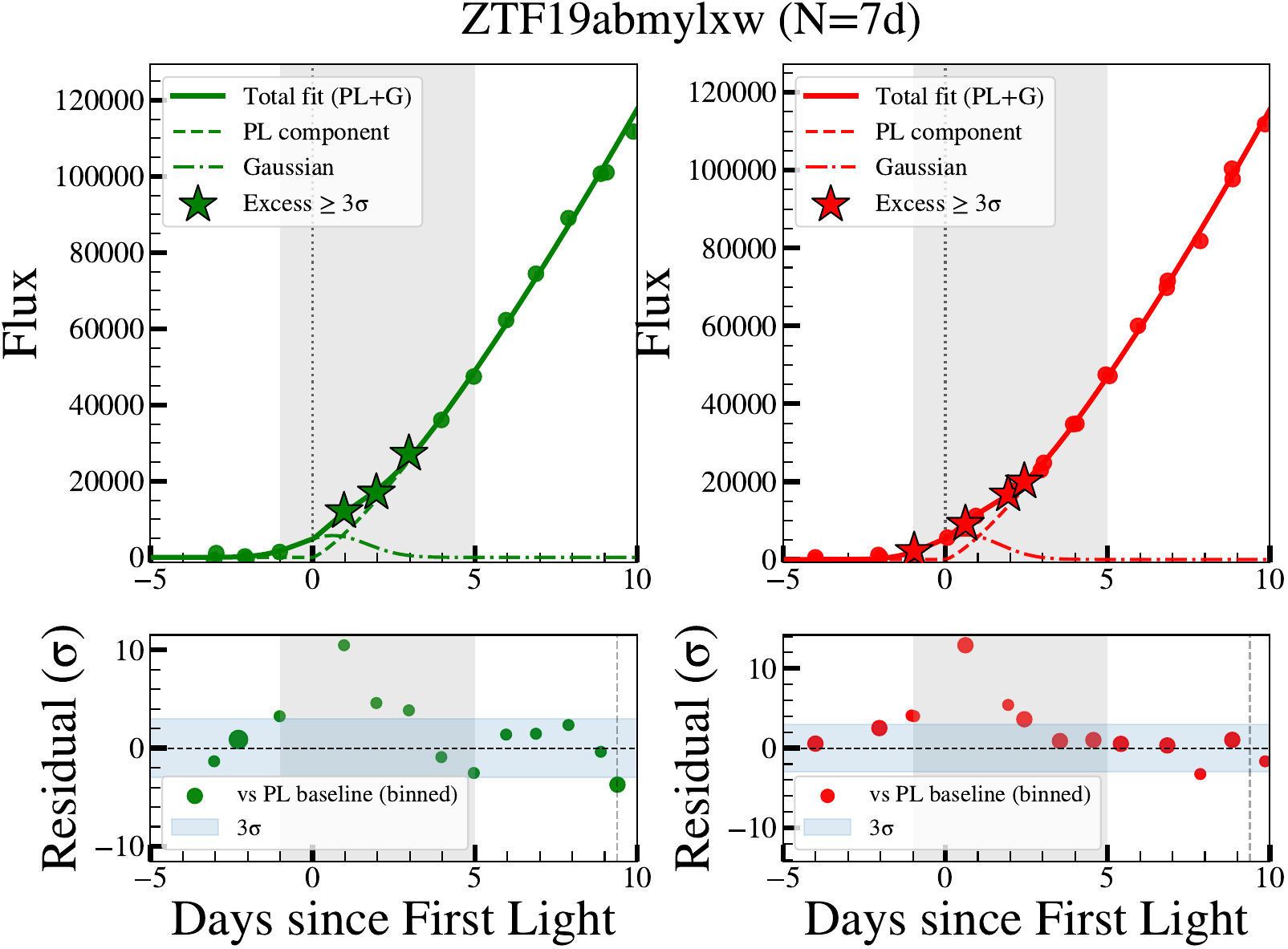}\hfill
\includegraphics[width=0.48\textwidth]{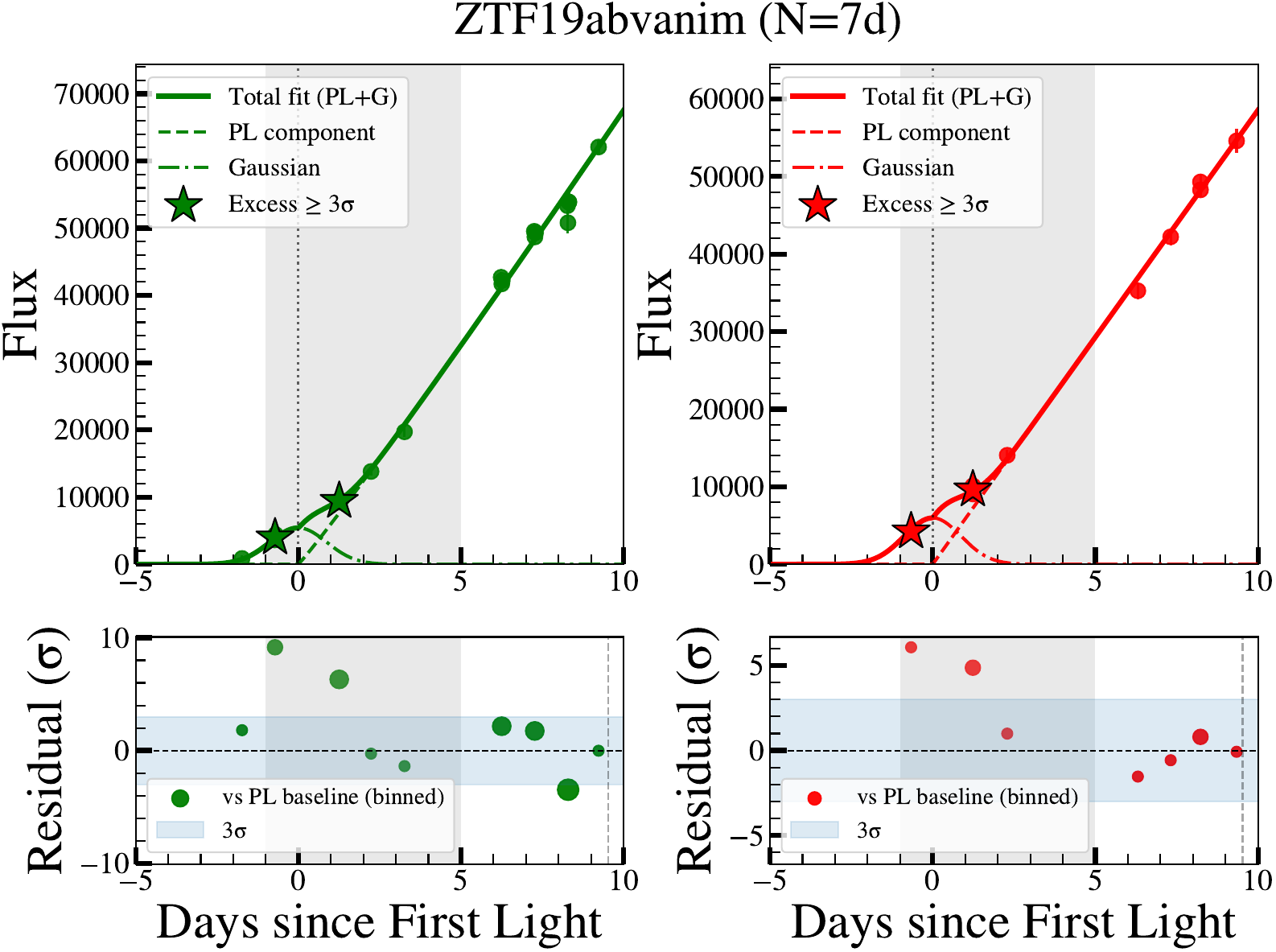}
\vspace{0.1cm}

\includegraphics[width=0.48\textwidth]{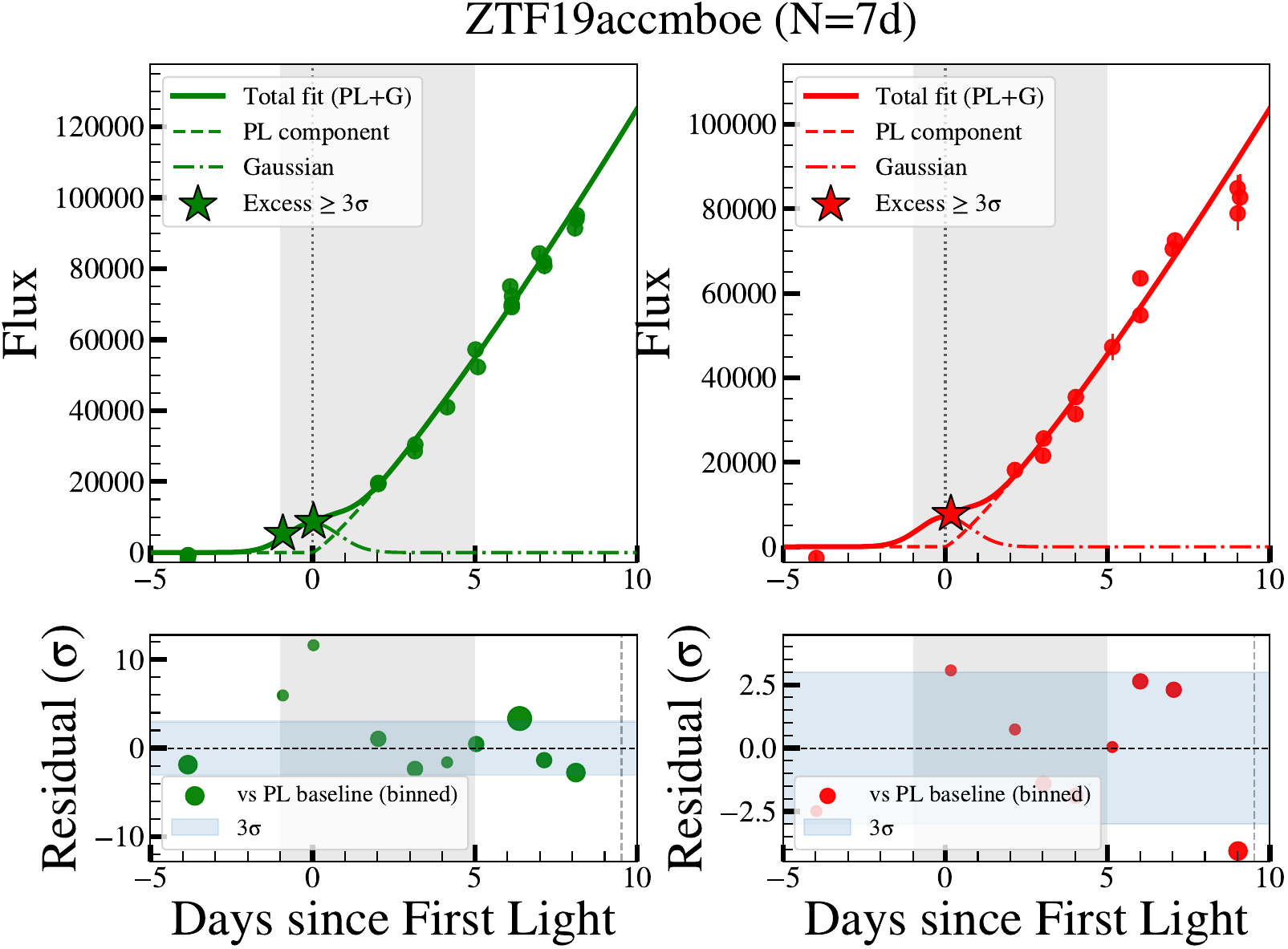}\hfill
\includegraphics[width=0.48\textwidth]{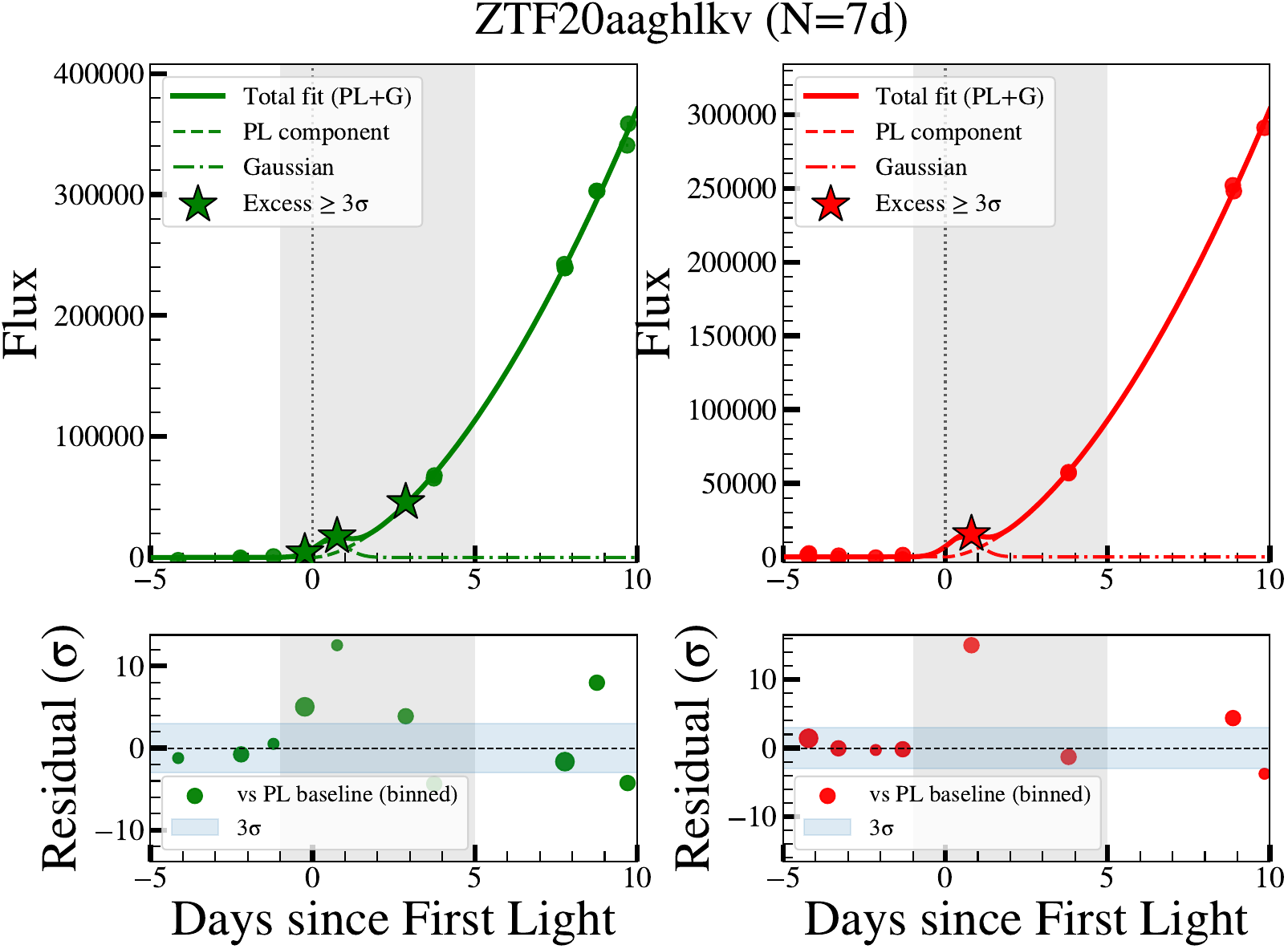}
\vspace{0.1cm}

\includegraphics[width=0.48\textwidth]{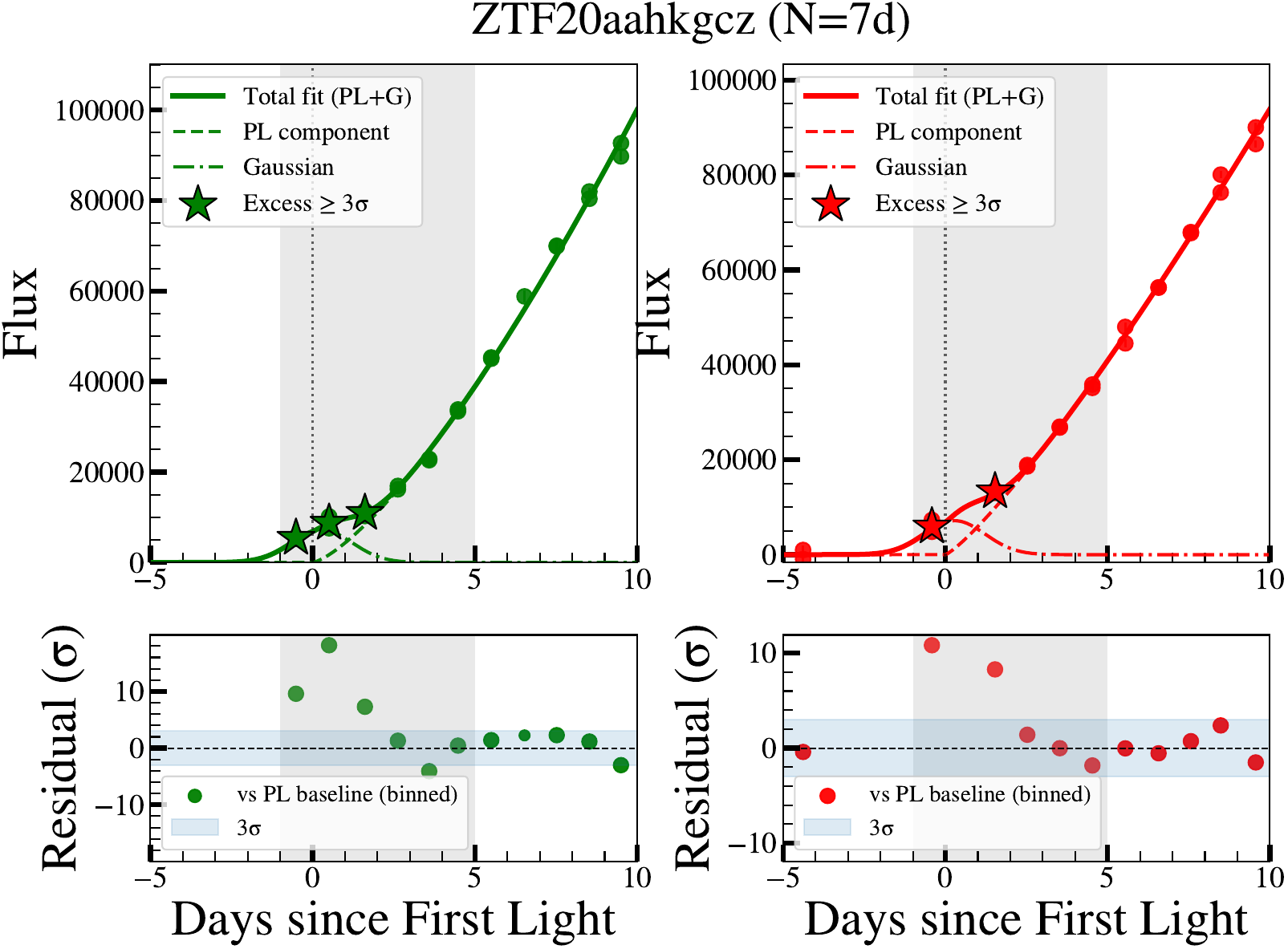}\hfill
\includegraphics[width=0.48\textwidth]{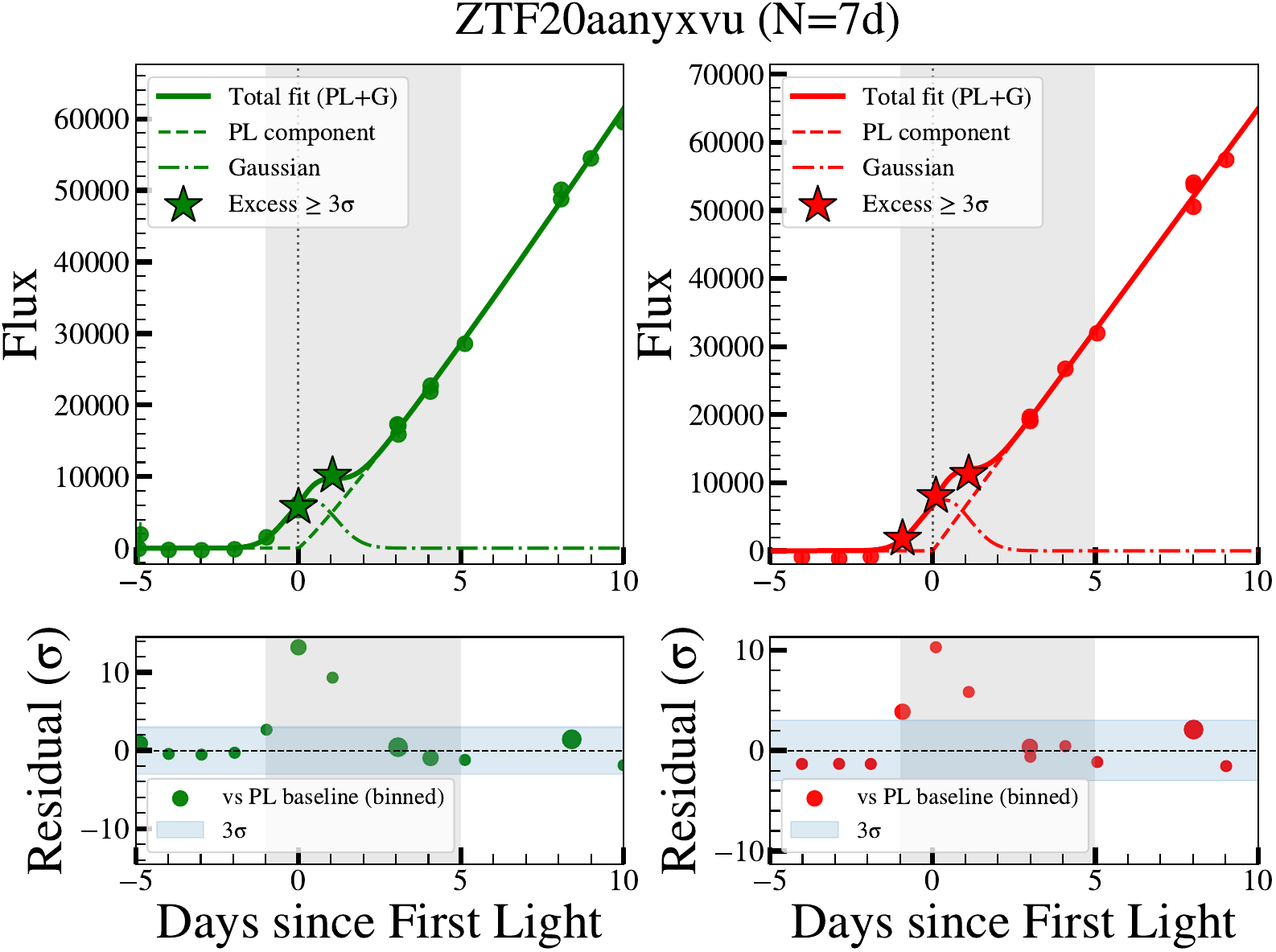}
\caption{Figure~\ref{fig:all_bumps_grid_1}. Continued.}
\end{figure}

\begin{figure}[H]
\centering
\includegraphics[width=0.48\textwidth]{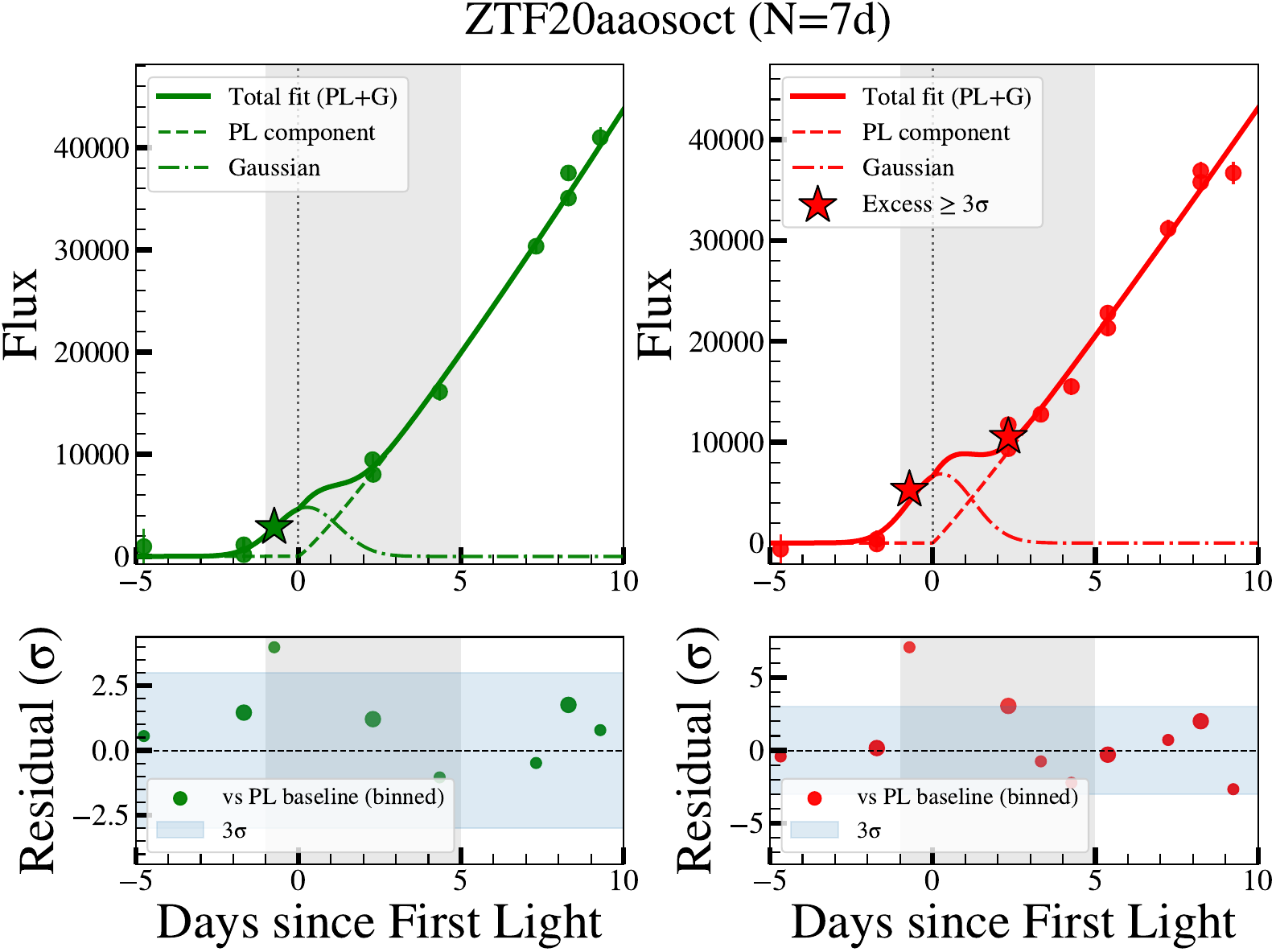}\hfill
\includegraphics[width=0.48\textwidth]{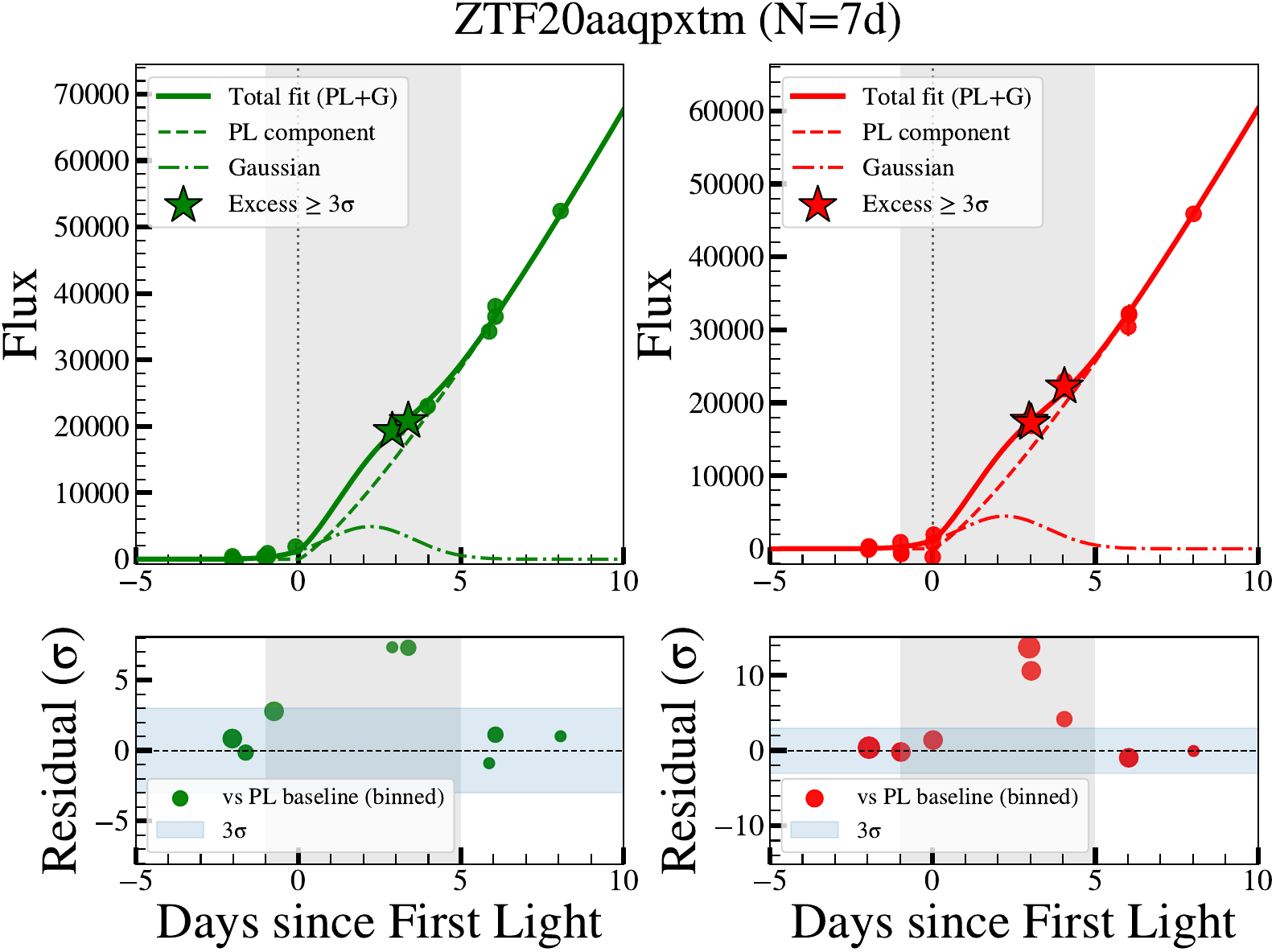}
\vspace{0.1cm}

\includegraphics[width=0.48\textwidth]{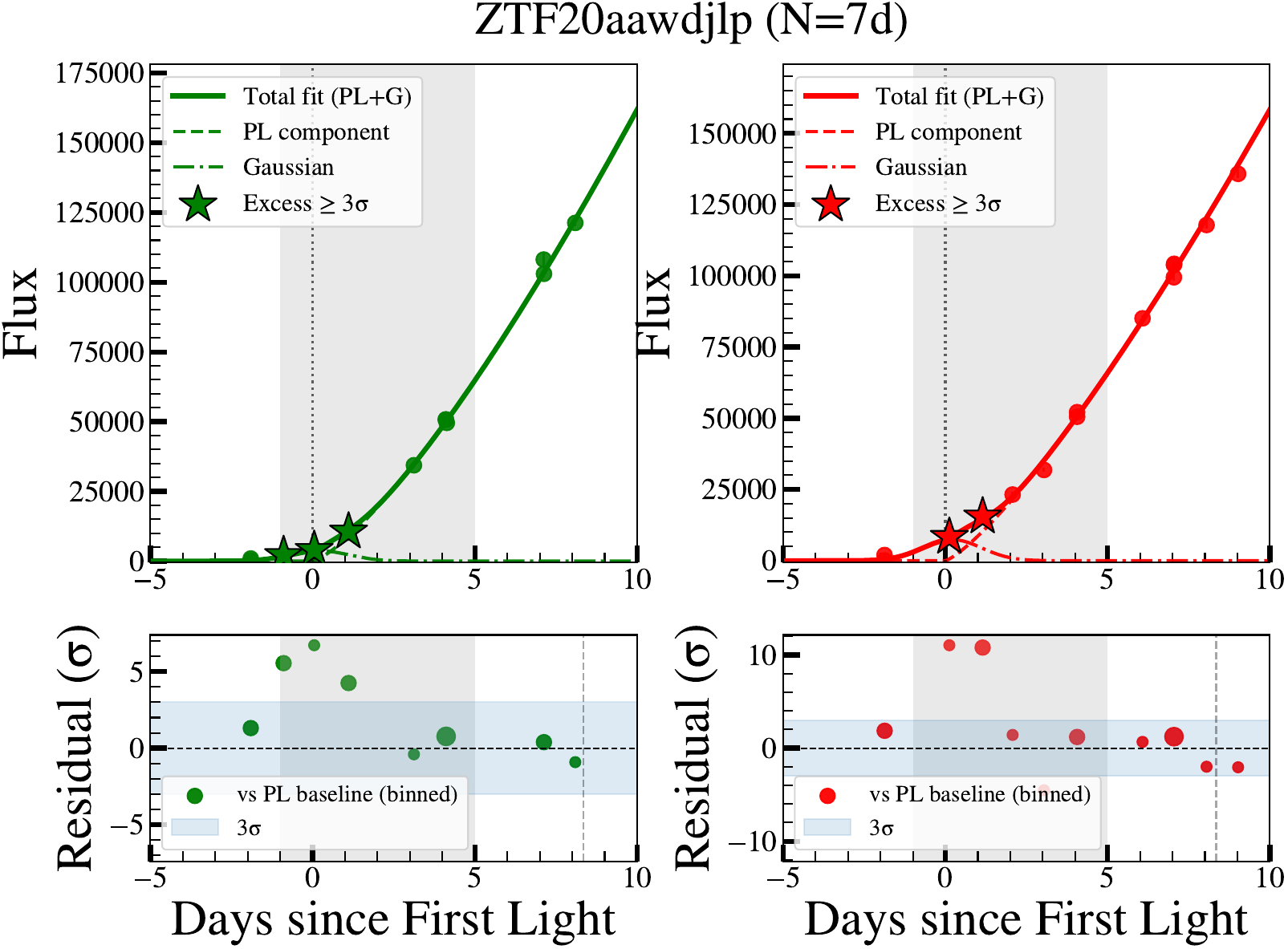}\hfill
\includegraphics[width=0.48\textwidth]{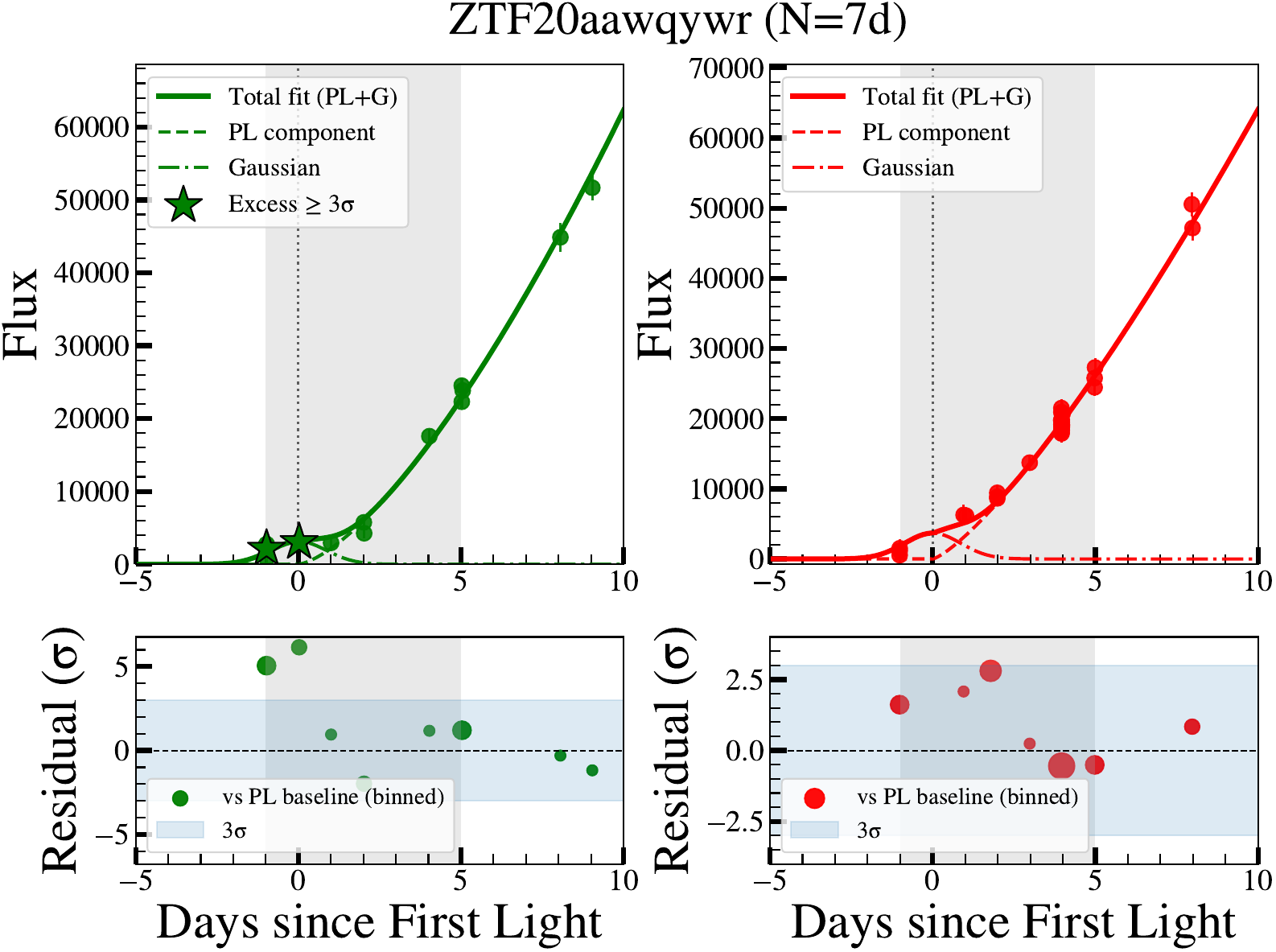}
\vspace{0.1cm}

\includegraphics[width=0.48\textwidth]{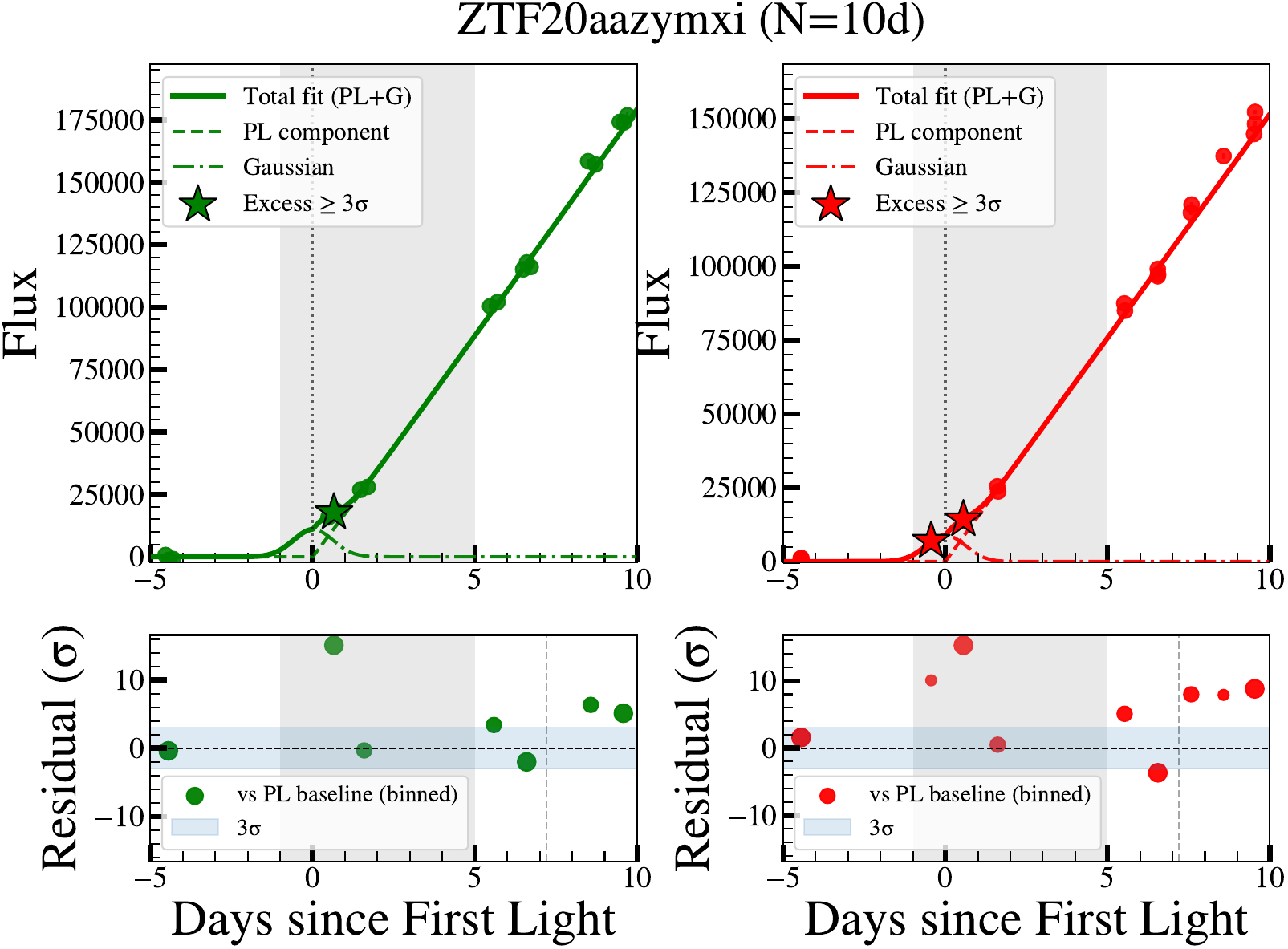}\hfill
\includegraphics[width=0.48\textwidth]{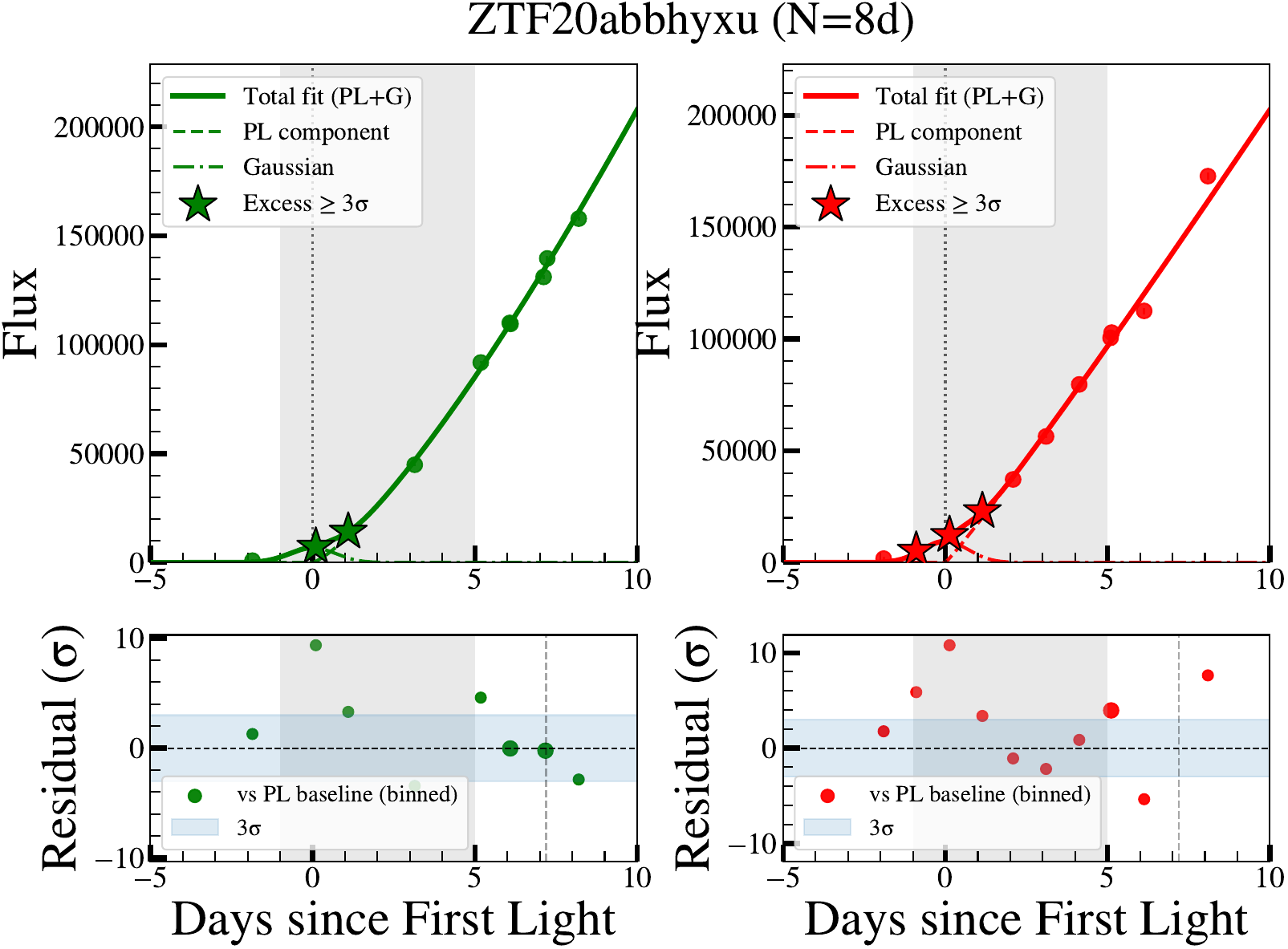}
\vspace{0.1cm}

\includegraphics[width=0.48\textwidth]{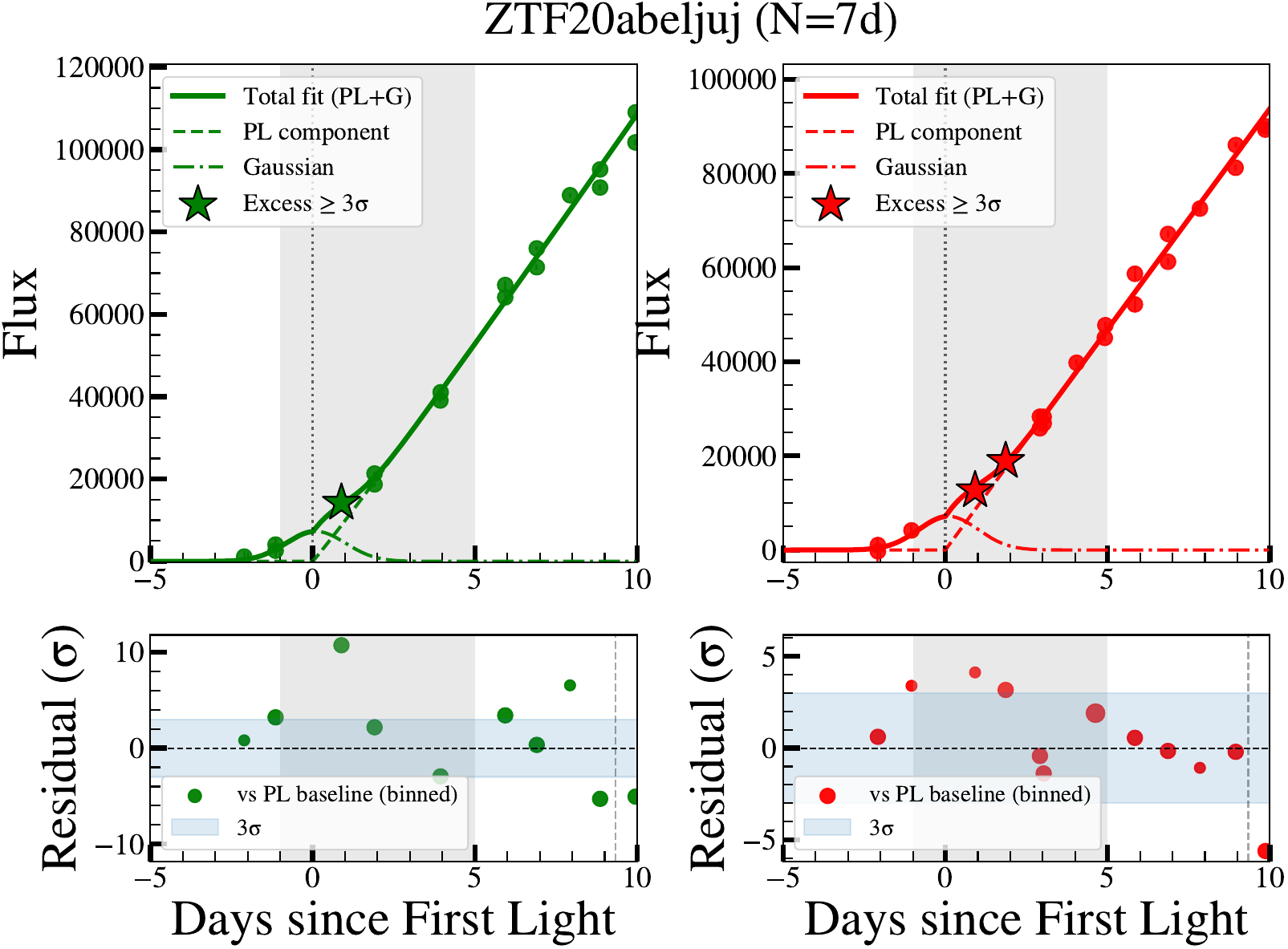}\hfill
\includegraphics[width=0.48\textwidth]{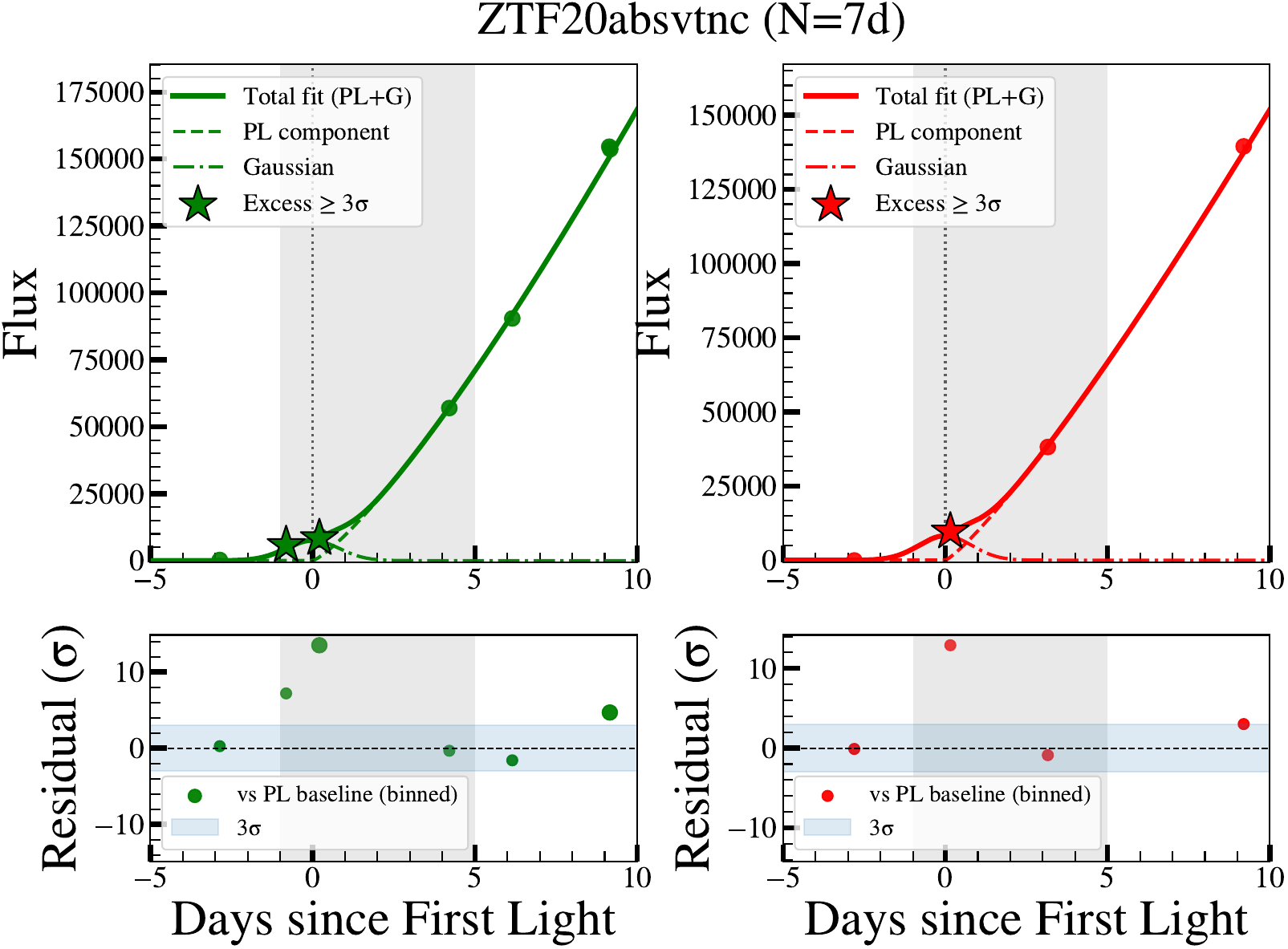}
\caption{Figure~\ref{fig:all_bumps_grid_1}. Continued.}
\end{figure}

\begin{figure}[H]
\centering
\includegraphics[width=0.48\textwidth]{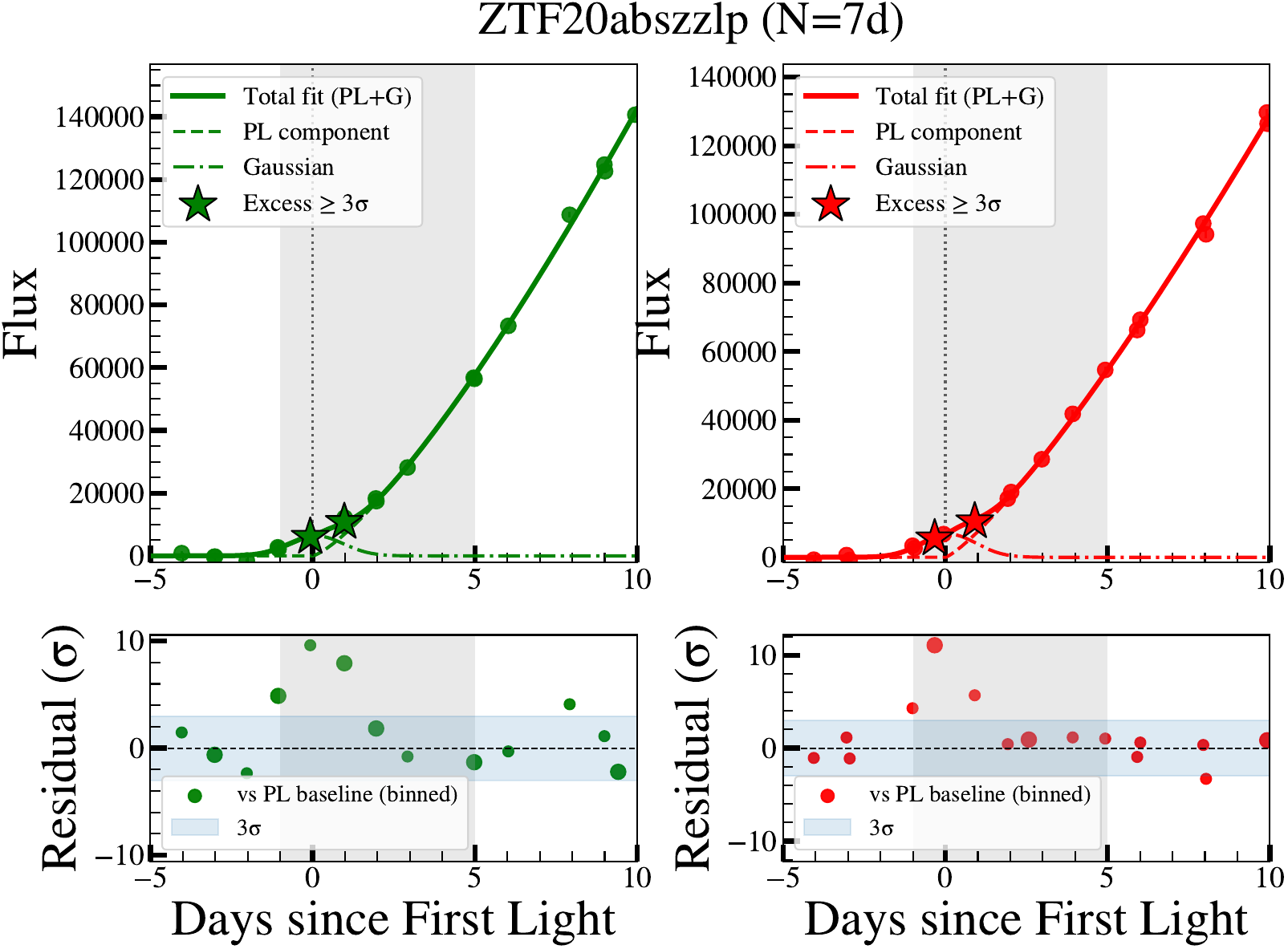}\hfill
\includegraphics[width=0.48\textwidth]{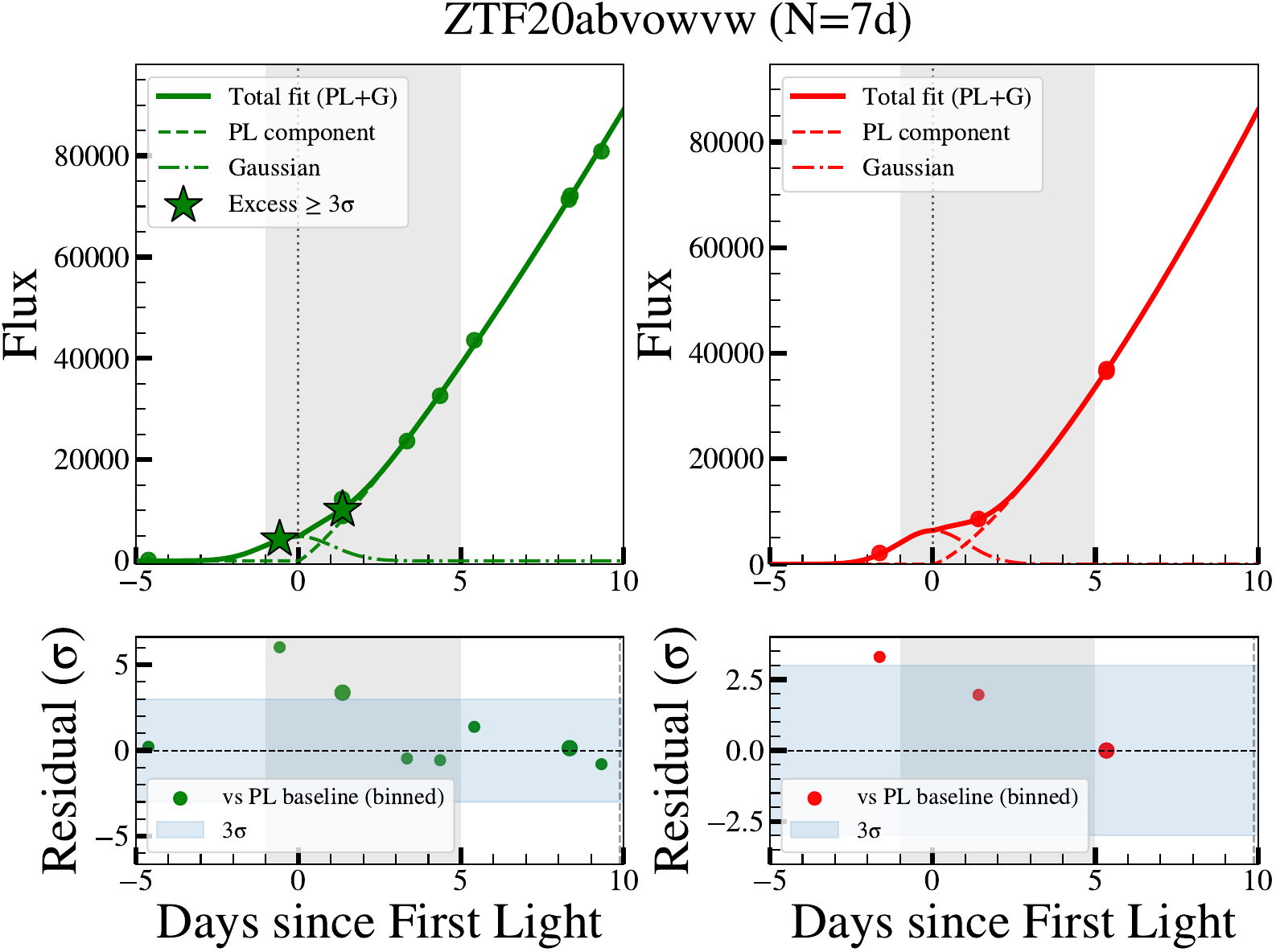}
\vspace{0.1cm}

\includegraphics[width=0.48\textwidth]{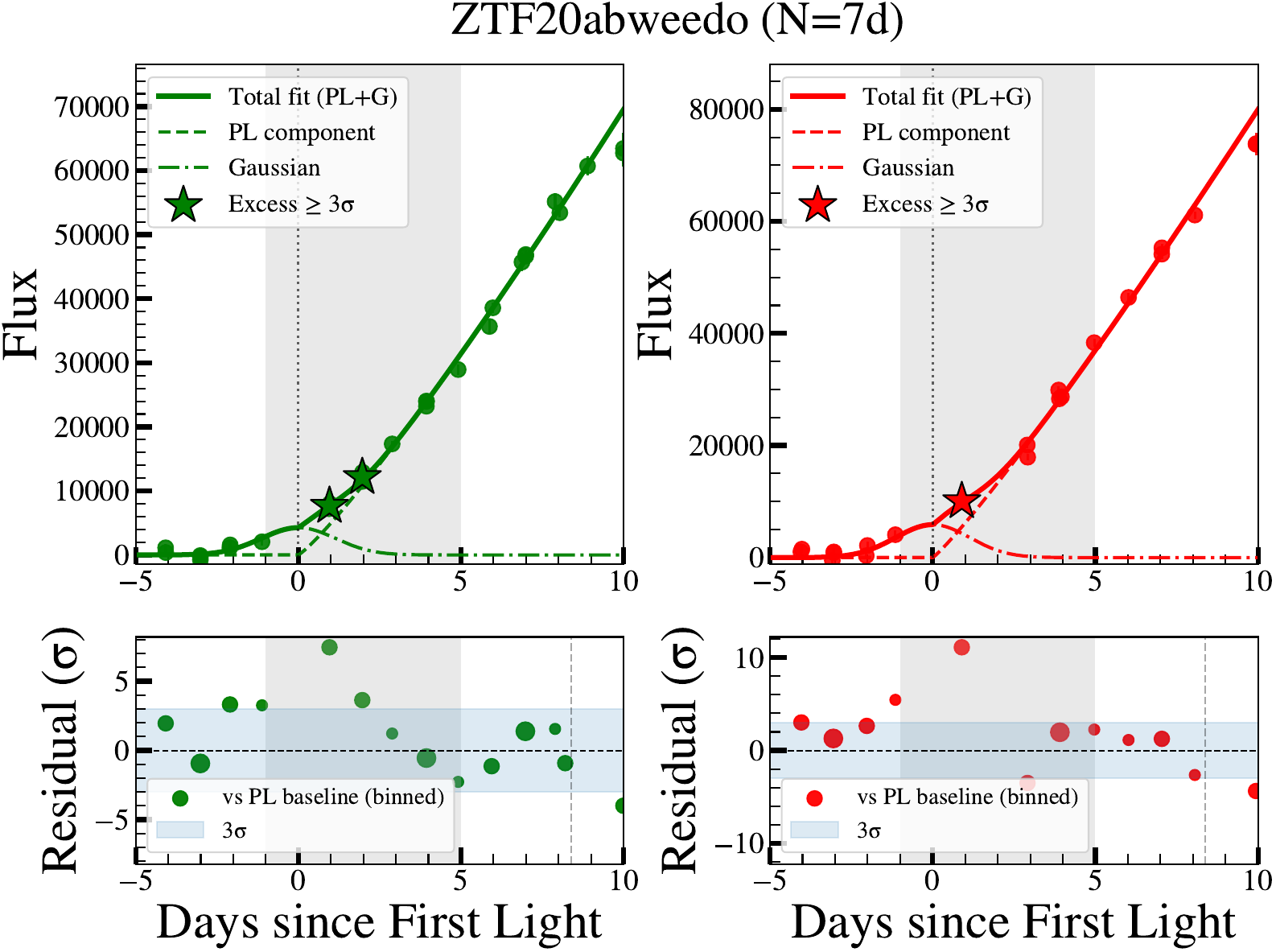}\hfill
\includegraphics[width=0.48\textwidth]{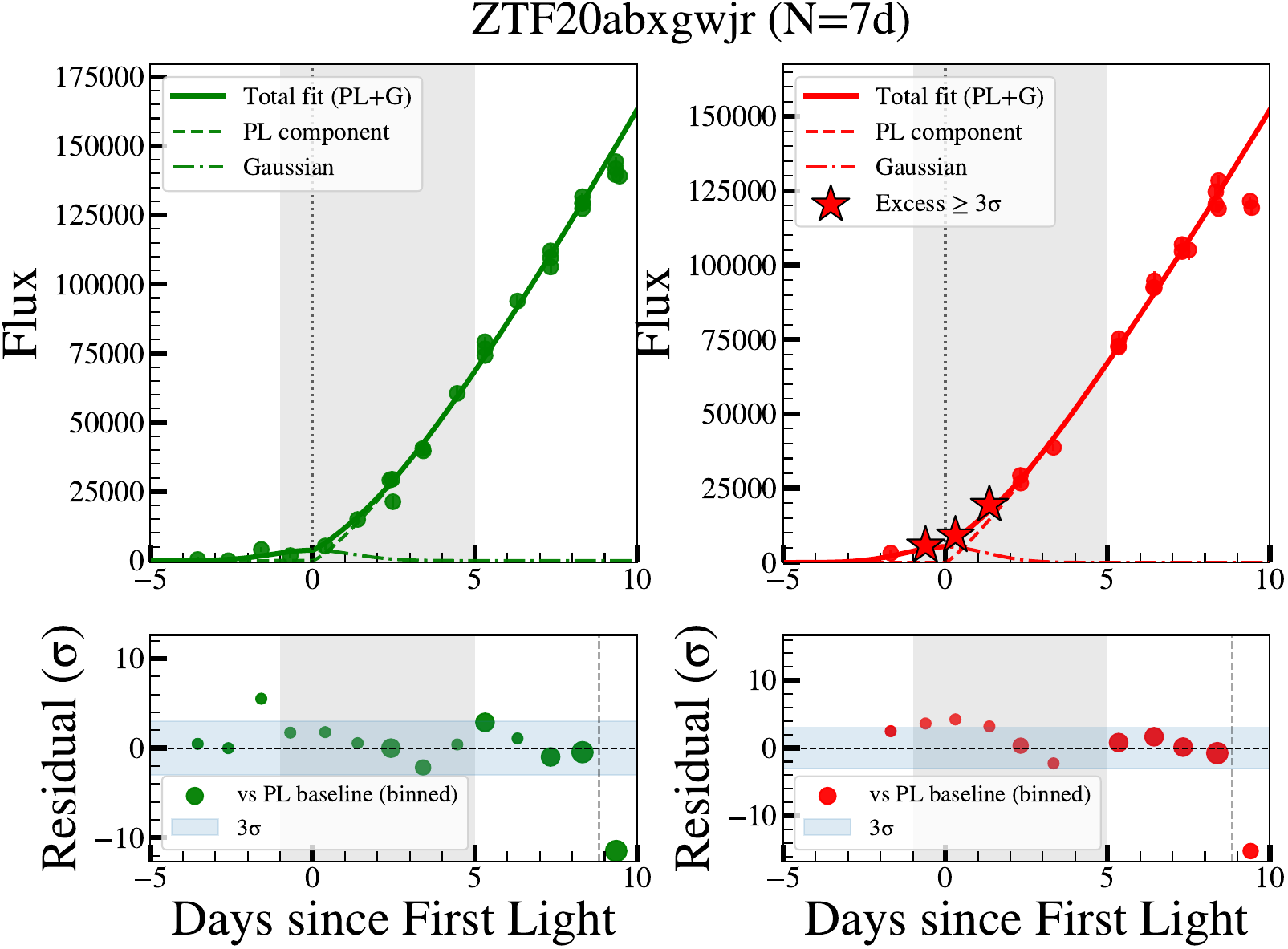}
\vspace{0.1cm}

\includegraphics[width=0.48\textwidth]{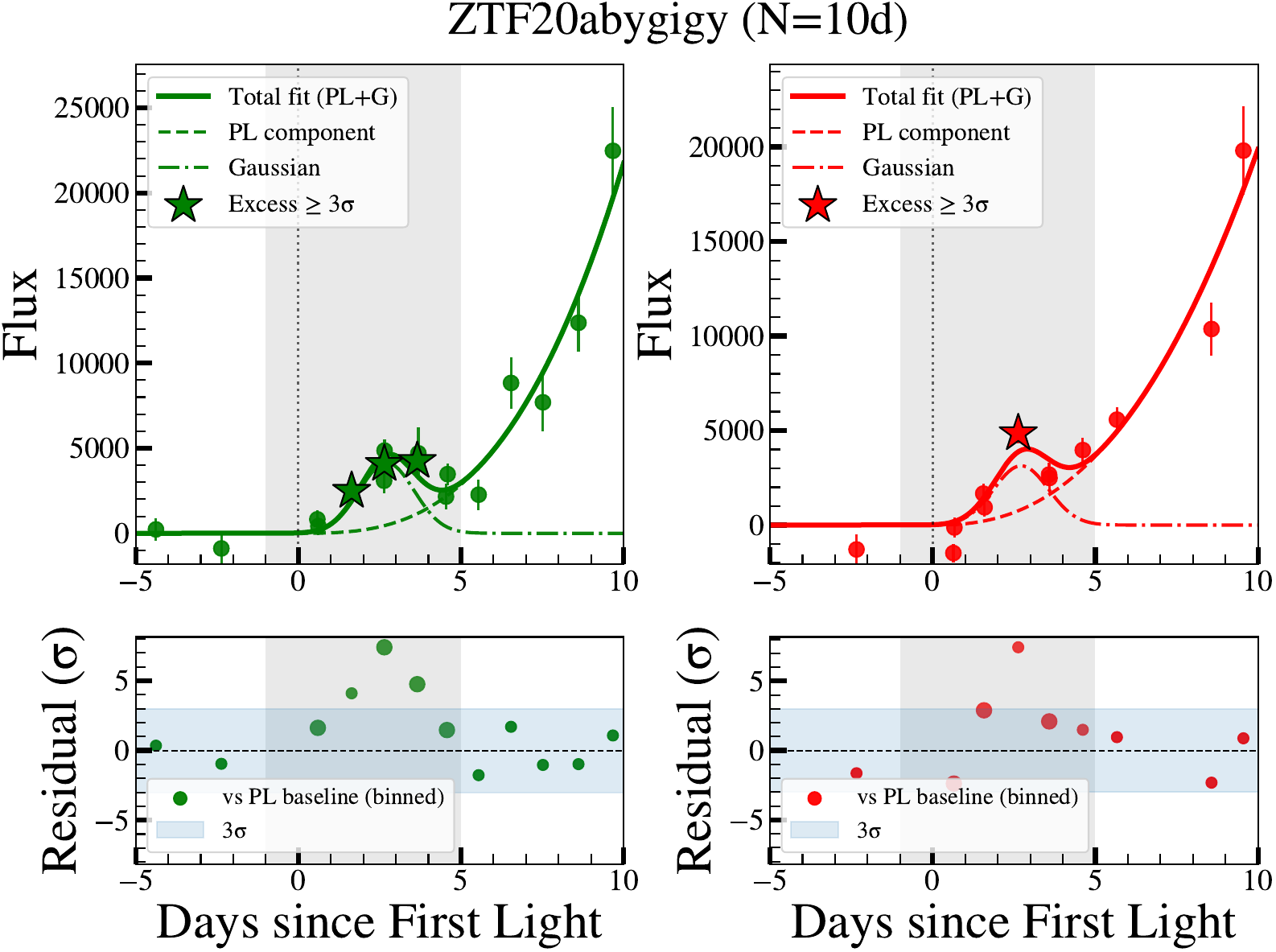}\hfill
\includegraphics[width=0.48\textwidth]{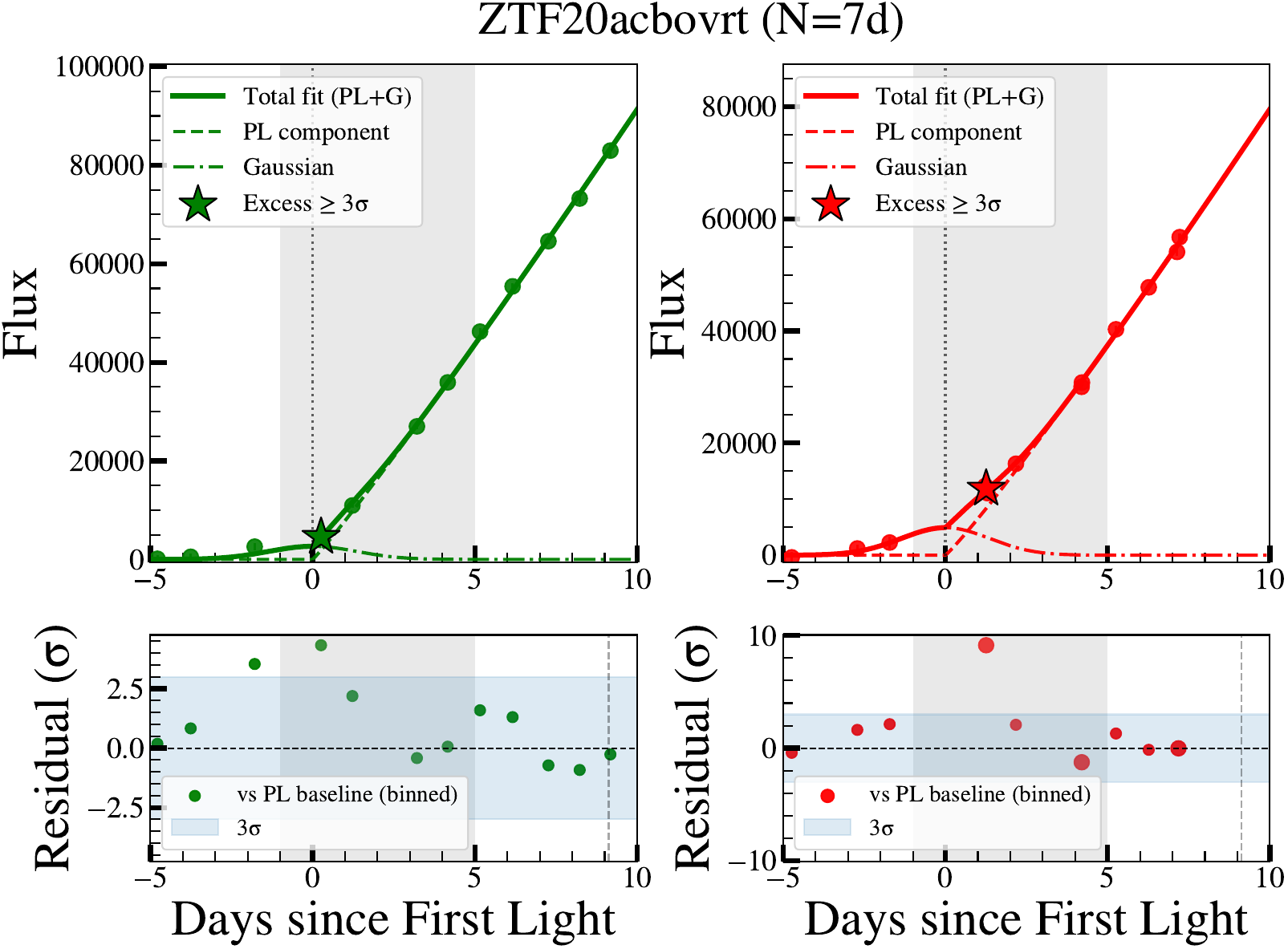}
\vspace{0.1cm}

\includegraphics[width=0.48\textwidth]{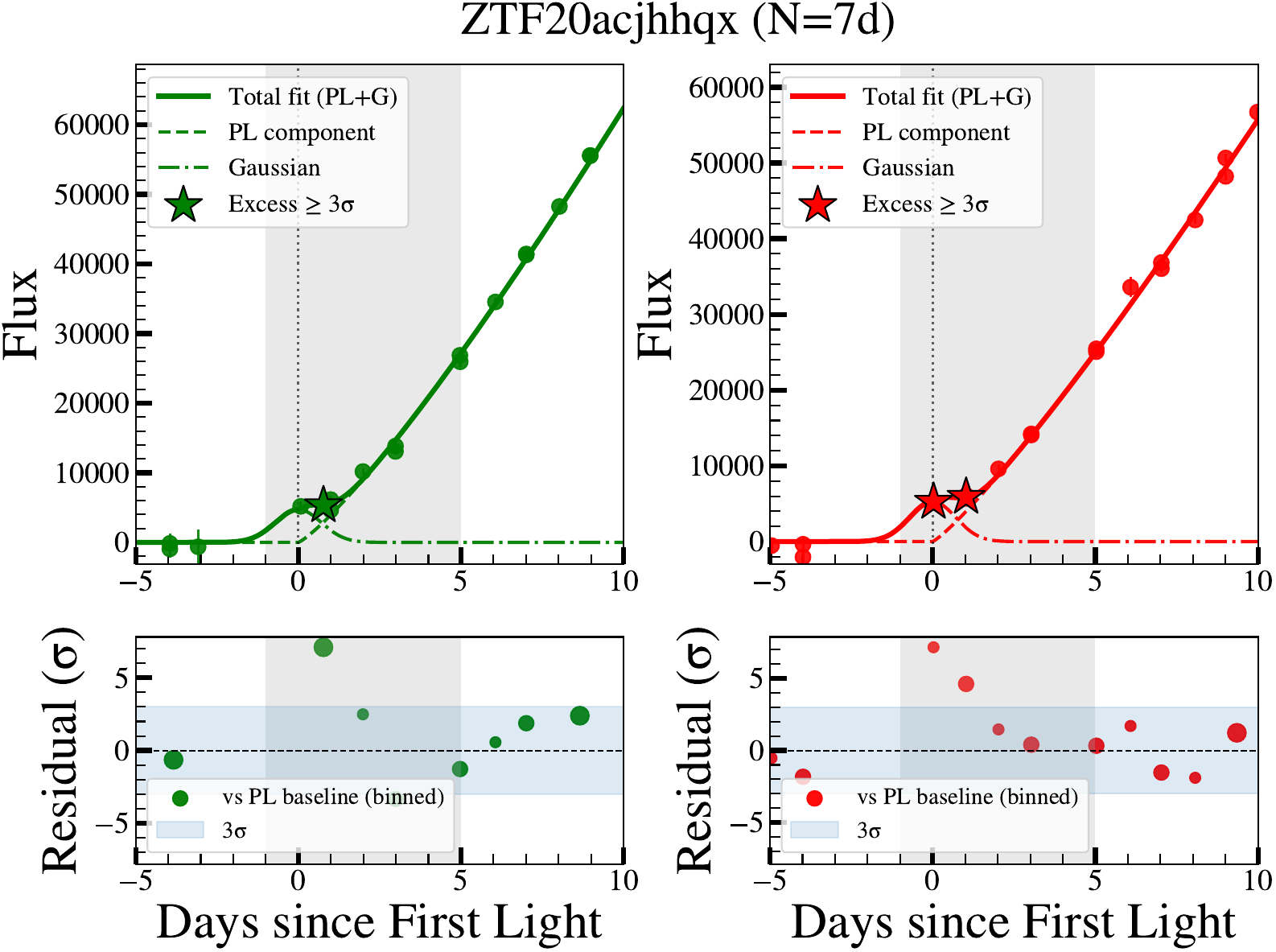}\hfill
\includegraphics[width=0.48\textwidth]{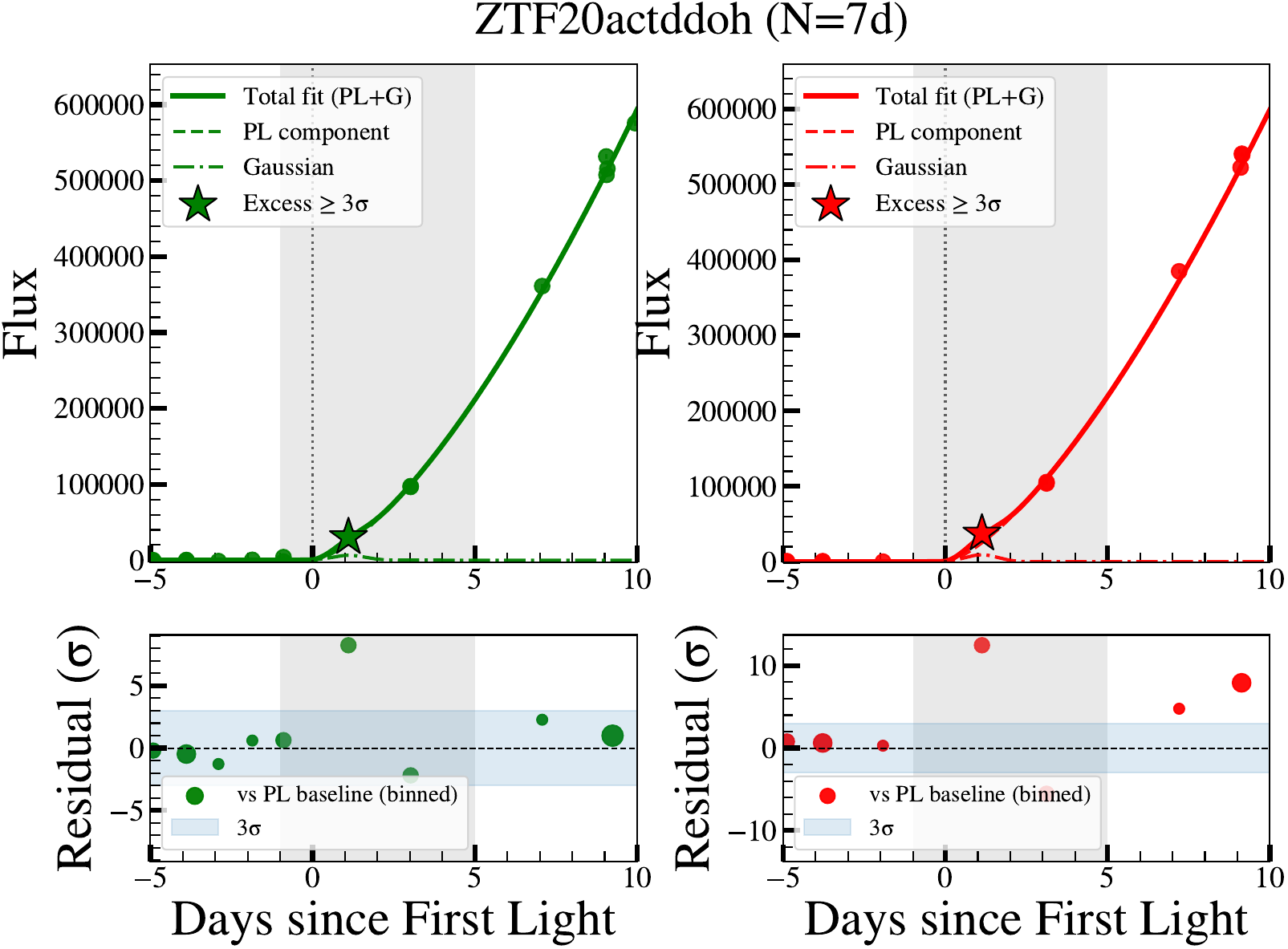}
\caption{Figure~\ref{fig:all_bumps_grid_1}. Continued.}
\end{figure}
